\documentclass[prX,twocolumn,showpacs,preprintnumbers,amsmath,amssymb,longbibliography ]{revtex4-1} 

	\usepackage[utf8]{inputenc}
	\usepackage[T1]{fontenc}
	\PassOptionsToPackage{hyphens}{url} 
	\usepackage[allcolors=blue, colorlinks=true, pdfborder={0 0 0}]{hyperref}

	\usepackage{url}

	\makeatletter\newcommand\multiabel[1]{\quad \refstepcounter{equation}(\theequation)\ltx@label{#1}}\makeatother

\usepackage{graphicx}
\usepackage{xcolor}
\usepackage{enumitem}
\usepackage{empheq}
\usepackage{bbold}
\usepackage{dsfont}

\pdfpageattr{/Group << /S /Transparency /I true /CS /DeviceRGB>>} 

\def\thesection{\arabic{section}}
\def\thesubsection{\arabic{section}.\arabic{subsection}}

\makeatletter
\renewcommand{\p@subsection}{}
\renewcommand{\p@subsubsection}{}
\makeatother

\renewcommand{\vec}{\mathbf}

\usepackage{ulem}
\definecolor{darkgreen}{rgb}{0,0.5,0} 
\definecolor{violet}{rgb}{0.5,0,0.5}
\definecolor{orange}{rgb}{0.8,0.5,0.2}
\definecolor{gray}{rgb}{0.3,0.3,0.3}
\definecolor{green}{RGB}{50,177,65}

\begin{document}


\title{Topologically robust zero-sum games and Pfaffian orientation -- How network topology determines the long-time dynamics of the antisymmetric Lotka-Volterra equation}

\author{Philipp M. Geiger$^{1}$}
\author{Johannes Knebel$^{1}$}
\author{Erwin Frey$^{1}$}

\email[]{frey@lmu.de}
\date{\today}

\affiliation{
$^{1}$Arnold Sommerfeld Center for Theoretical Physics and Center for NanoScience, Department of Physics, Ludwig-Maximilians-Universit\"at M\"unchen, Theresienstrasse 37, 80333 M\"unchen, Germany}

\begin{abstract}
To explore how the topology of interaction networks determines the robustness of dynamical systems, we study the antisymmetric Lotka-Volterra equation (ALVE).   
The ALVE is the replicator equation of zero-sum games in evolutionary game theory, in which the strengths of pairwise interactions between strategies are defined by an antisymmetric matrix 
such that typically some strategies go extinct over time.
Here we show that there also exist topologically robust zero-sum games, such as the rock-paper-scissors game, for which all strategies coexist for all choices of interaction strengths. 
We refer to such zero-sum games as coexistence networks and construct coexistence networks with an arbitrary number of strategies.
By mapping the long-time dynamics of the ALVE to the algebra of antisymmetric matrices, we identify simple graph-theoretical rules by which coexistence networks are constructed. 
Examples are triangulations of cycles characterized by the golden ratio $\varphi = 1.6180...$, cycles with complete subnetworks, and non-Hamiltonian networks. 
In graph-theoretical terms, we extend the concept of a Pfaffian orientation from even-sized  to odd-sized networks. 
Our results show that the topology of interaction networks alone can determine the long-time behavior of nonlinear dynamical systems, and may help to identify robust network motifs arising, for example, in  ecology.
\end{abstract}



\maketitle

\twocolumngrid
\section{Introduction}

\begin{figure*}[htb!]
\centering
\includegraphics[width=0.95\textwidth]{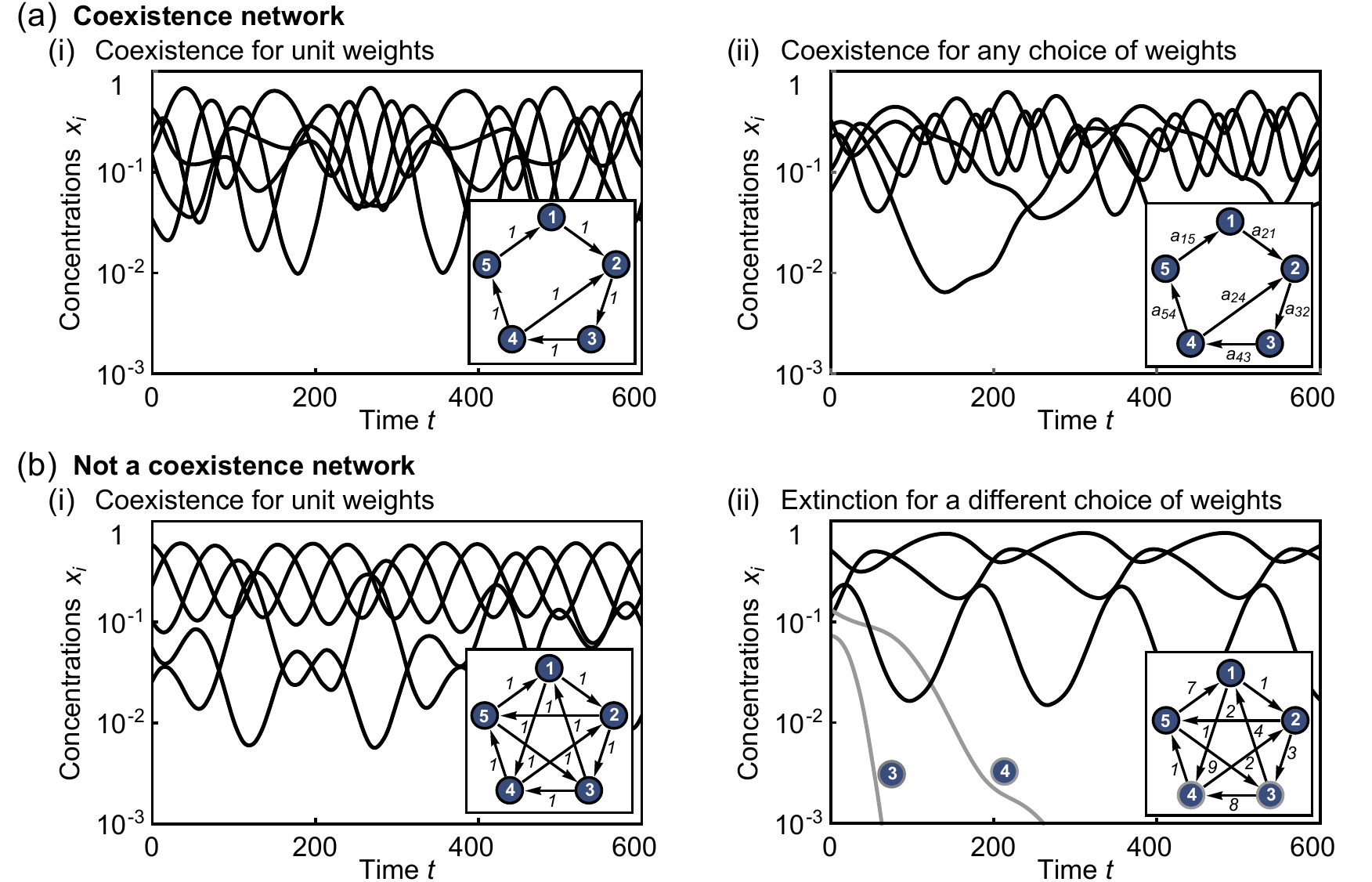}
\caption{(Color online)
\textbf{
Coexistence networks in the antisymmetric Lotka-Volterra equation (ALVE)~\eqref{eq:ALVE}.} 
The long-time dynamics of the ALVE are independent of the initial conditions and two scenarios are possible for a state $i$ (equivalently strategy): 
Either the state concentration vanishes ($x_i\to 0$ as $t\to \infty$; extinction and depletion) or it remains bounded away from 0 for all times ($x_i\geq Const> 0$ for all $t$; survival and condensation). 
Survival and extinction of states depend only on the weighted network defined by the network topology and weights.
(a) Trajectories of the ALVE for a directed cycle of 5 nodes with an interior edge from node 4 to 2 (see insets). All states coexist. This coexistence does not only occur for unit weights (i), but \textit{for all choices} of weights on that network topology (ii). 
Such network topologies are called \textit{coexistence networks} or \textit{topologically robust zero-sum games}. 
(b) The vast majority of networks are not coexistence networks; here shown for the rock-paper-scissors-lizard-spock game (network topology of five states with two in-going and two out-going links for every node); see insets. 
(i) For unit weights, all states coexist, but states 3 and 4 go extinct for differently chosen weights (ii). Thus, coexistence depends on the rates.
}
\label{fig:ALVE_topological_stability}
\end{figure*}

\begin{figure*}[th!]
\centering
\includegraphics{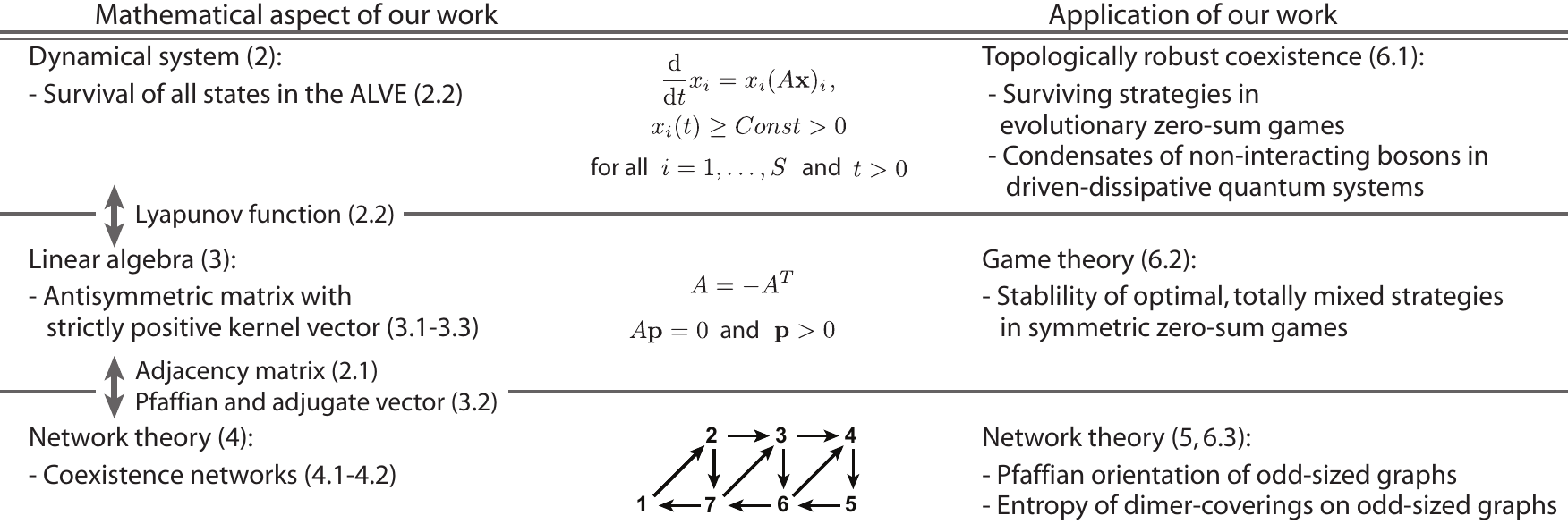}
\caption{
\textbf{Organisation and application of this work.}
The mathematical aspects are presented in Sections~\ref{sec:ALVE},~\ref{sec:algebra}, and~\ref{sec:coexistence_networks}. 
We discuss the application of our work in Section~\ref{sec:discussion}.
In Section~\ref{sec:examples} we discuss topology and properties of two exemplary coexistence networks.
In each row, the application corresponds to the mathematical aspect.
}
\label{fig:visualabstract}
\end{figure*}

\textit{Robustness of dynamical systems.}
The temporal behavior of models arising in nonequilibrium statistical physics are often adequately described in terms of nonlinear dynamical systems. 
How the qualitative long-time behavior of a dynamical system depends on the initial conditions, on the interaction of different degrees of freedom, and on the coupling parameters that determine the interaction strengths remain central questions in theoretical physics and applied mathematics~\cite{Drazin1992, Hirsch2013, Strogatz2014, Mobilia2016}.
In this work, we focus on the robustness of the qualitative long-time behavior of dynamical systems against arbitrary changes of the coupling parameters~\cite{Dai2012, Dai2015, Rohr2014, Allesina2012, Gao2016, Szolnoki2009, Intoy2015, Tu2017, Allesina2008}.

\textit{Antisymmetric Lotka-Volterra equation.} 
The antisymmetric Lotka-Volterra equation (ALVE)~\eqref{eq:ALVE} is a well-suited dynamical system to study robustness properties. 
Besides its applications in the fields of quantum physics~\cite{Vorberg2013, Vorberg2015}, population dynamics~\cite{Reichenbach2006, Volterra1931, Goel1971, May1972, Frey2010, Dobrinevski2012, Berr2009}, chemical kinetics~\cite{Itoh1971, DiCera1988, DiCera1989}, and plasma physics~\cite{Zakharov1974, Manakov1975}, the ALVE describes the dynamics of general zero-sum games in evolutionary game theory~\cite{Akin1984, Chawanya2002, Knebel2015}.
At the heart of evolutionary game theory stands the idea that interactions between different phenotypes or behavioral programs are encoded by interacting strategies in a population of individuals. 
For an evolutionary zero-sum game, the gain of one strategy equals the loss of another in a pairwise interaction. Thus, the game's payoff matrix is antisymmetric.

The ALVE describes the dynamics of the fraction of individuals playing a certain strategy in an infinitely large, well-mixed population and follows as the replicator equation for zero-sum games~\cite{Akin1984, Hofbauer1998, Knebel2015, Dobrinevski2012}.
Whether a strategy survives or goes extinct depends on both the network topology of the zero-sum game and the interaction strengths, but is independent of the initial conditions~\cite{Chawanya2002, Knebel2015, Sadeghi2018}.

\textit{Coexistence networks.}
We investigated the conditions on a zero-sum game under which all strategies coexist \textit{for all choices} of interaction strengths. 
Consider, for example, the rock-paper-scissors zero-sum game in which each of the three strategies dominates one strategy and is also dominated by another one ($1\to 2\to 3\to 1$; the game's network topology is a directed cycle of three nodes)~\cite{Hofbauer1998, Szabo2007, Frachebourg_2_1996, Reichenbach2006, Park2017, Frey2010, Dobrinevski2012, Berr2009}.
Notably, all strategies coexist in an interacting population irrespective of the chosen interaction strengths, which characterize the dominance relations through the entries of the antisymmetric payoff matrix~\cite{Hofbauer1998}. 
Since coexistence of all states depends only on the game's network topology, but not on the choice of interaction strengths, we refer to the rock-paper-scissors game as a \textit{topologically robust zero-sum game} or a \textit{coexistence network}; see Figure~\ref{fig:ALVE_topological_stability}(a).

 \textit{How to determine coexistence networks?}
The existence of coexistence networks as well as their characterization are non-trivial because strategies typically go extinct for some choice of interaction strengths~\cite{Knebel2013}.
On the one hand, complete networks such as the rock-paper-scissors-lizard-spock game~\cite{Knebel2013, Vukov2013, Hawick2011, Kang2013, Cheng2014, Park2017}
are not coexistence networks; see Figure~\ref{fig:ALVE_topological_stability}(b).
On the other hand, cycles with an odd number of strategies, in which every strategy dominates exactly one strategy and is dominated by another ($ 1\to 2\to \dots \to 2n-1\to 1 $), are coexistence networks~\cite{Zia2011,Durney2011}. 
Coexistence networks with an arbitrary number of strategies beyond odd-sized cycles have not been characterized thus far.

\textit{This work: construction of coexistence networks.}
Here, we construct coexistence networks that are zero-sum games for which all strategies coexist irrespective of the chosen interaction strengths. 
By connecting the long-time dynamics of the ALVE to an algebraic problem of antisymmetric matrices, we employed graph-theoretical tools to (i) characterize coexistence networks, (ii) establish an algorithm to construct coexistence networks, and (iii) analyze properties of specific coexistence networks.
We find coexistence networks that are certain triangulations of odd-sized cycles whose topology is characterized by the golden ratio $\varphi = 1.6180...$, odd-sized cycles with a complete subnetwork on half of the network's nodes, and non-Hamiltonian networks.
By linking our graph-theoretical results to the ALVE, we propose an evolutionary zero-sum game, with which one can \textit{dynamically measure} the golden ratio.

\textit{Applications to game theory and statistical physics.}
Besides evolutionary game theory, our results have a broad range of applications through the established link to the algebra of antisymmetric matrices and graph theory.
From an algebraic point of view, we construct antisymmetric matrices whose kernel is one-dimensional and strictly positive for all choices of matrix entries. 
Applied to game theory, we find those symmetric zero-sum games that have a Nash equilibrium in which all strategies are played with non-vanishing probability and remain so for any change of payoff in the zero-sum game.
In other words, our analysis facilitates the study of symmetric zero-sum games whose optimal strategy is both totally mixed and stable~\cite{Fisher92, Brandl17, Brandt09, Arsham95}.

From a statistical physics point of view, we propose a concept to compute the entropy of dimer molecules adsorbed on graphs of odd size.
To compute the entropy of systems in which dimers completely fill the nodes of regular lattice graphs of even size, Kasteleyn, Fisher, and Temperley introduced techniques to count the number of closed-packing configurations~\cite{Kasteleyn1961, Temperley1961, Fisher1961} (in graph theory, so-called perfect matchings of an even-sized graph). 
It was shown that certain even-sized graphs can be oriented in a so-called Pfaffian orientation such that the dimer entropy can be computed either numerically efficiently or analytically~\cite{Kasteleyn1963,Wu2006,Thomas2006}. 
Following our results on coexistence networks, we suggest the equivalent concept to orient graphs of odd size, which facilitates the counting of closed-packing dimer configurations that leave one node of the graph uncovered. 
Overall, in graph-theoretical terms, we extend the concept of a Pfaffian orientation of even-sized to odd-sized graphs.

\textit{Organization of this manuscript (see Figure~\ref{fig:visualabstract}).}
In Section~\ref{sec:ALVE}, we introduce the mathematical framework of the ALVE~\eqref{eq:ALVE}. Our work is motivated by the coexistence of all strategies and coexistence networks, which are qualitative properties of the dynamics of the ALVE. These dynamical properties are illustrated in Figure~\ref{fig:ALVE_topological_stability}, which provides the starting point of this work. Furthermore, the connection between coexistence of all strategies and coexistence networks to algebraic properties of the defining antisymmetric payoff matrix is established. It is the central idea of this work to map the question about the dynamical system to an algebraic problem, analyze and solve the algebraic problem, and to interpret the obtained results for the dynamical system, but also to exemplify further applications for game theory and statistical physics.
In Section~\ref{sec:algebra}, we characterize algebraic properties of antisymmetric matrices in terms of a graph-theoretical interpretation by exploiting the notion of the Pfaffian of an antisymmetric matrix. Readers, who are familiar with the determinant-like Pfaffian, may skip over this part of the text; Figure~\ref{fig:Pfaffian} sketches the computation of the Pfaffian in terms of the graph-theoretical interpretation of antisymmetric matrices and provides the background for a first read. 
All of our results are obtained through the graph-theoretical interpretation of antisymmetric matrices, and their corresponding Pfaffian and adjugate vector; see Figure~\ref{fig:coexistence_networks}. In particular, the characterization of perfect matchings of the network representation and its subnetworks lies at the heart of our work.
In Section~\ref{sec:coexistence_networks}, we exploit this approach to construct Hamiltonian coexistence networks as generalizations of odd-sized, directed cycles. 
Furthermore, a numerical survey of coexistence networks with up to 9 nodes is presented.
In Section~\ref{sec:examples}, two specific classes of network topologies are studied in detail. 
Our main results on coexistence networks obtained in Sections~\ref{sec:coexistence_networks} and~\ref{sec:examples} find applications in the context of the ALVE, which are exemplified for topologically robust zero-sum games in evolutionary game theory and topologically robust quantum networks for non-interacting bosons in driven-dissipative systems; see Section~\ref{sec:discussion}.
Because our results relate to general antisymmetric matrices, further applications are immediate. We exemplify one application for symmetric zero-sum games in the field of game theory, and discuss in detail the dimer problem for odd-sized graphs in statistical physics as a second application. 
This work is summarized and concluded in Section~\ref{sec:conclusion}.

\section{Long-time dynamics of the ALVE and coexistence networks}
\label{sec:ALVE}
%

\subsection{Definition of the ALVE and weighted networks} 
\label{sec:ALVE_definition}

\textit{Definition of the ALVE.}
The antisymmetric Lotka-Volterra equation is defined for a system of $S$ dynamical variables, which we refer to as states (or strategies in the context of evolutionary game theory). 
The concentration or mass in state $i$ is denoted as $x_i$ and the vector of state concentrations is denoted as $\vec{x}(t) = (x_1(t), \dots, x_S(t))$. 
These masses evolve through a system of nonlinearly coupled ordinary differential equations of first order in time:
\begin{align}\label{eq:ALVE}
\frac{\text{d}}{\text{d} t} x_i(t) =x_i(t)\sum_{j=1}^S a_{ij}x_j(t)\ , 
\end{align}
for all $i = 1, \dots, S$. 
The matrix $A = \{a_{ij}\}_{i,j}\in \mathbb{R}^{S \times S}$ is \textit{antisymmetric} (or skew-symmetric), that is $a_{ij} = -a_{ji}$. 
The vector of initial masses is assumed to be strictly positive and normalized, such that $\vec{x}(t=0) \eqqcolon \vec{x}_0$ lies in the open $(S-1)$-simplex $\Delta_{S-1}$ ($x_{0,i} >0$ for all $i = 1, \dots, S$ and $\sum_{i = 1}^S x_{0, i} = 1$).
For brevity, the time variable $t$ is dropped in most of the following derivations.

The antisymmetric matrix $A$ defines the set of control parameters of the ALVE (and defines the zero-sum game). It specifies how mass is exchanged between the $S$ states through pairwise interactions. Mass in state $i$ changes through interaction with state $j$ as $a_{ij} x_i x_j$. A negative matrix entry $a_{ij}<0$ means that mass is transported from state $i$ to $j$. Thus, at the same time $j$ gains this mass through $ -a_{ij} x_j x_i = a_{ji} x_j x_i$ and $a_{ji}>0$. 
A vanishing off-diagonal entry $a_{ij} = a_{ji} = 0$ implies that no mass can be exchanged between states $i$ and $j$. It is those zero entries of the antisymmetric matrix that are most relevant to our work on coexistence networks.
Since no other interactions are defined by the ALVE~\eqref{eq:ALVE}, the total mass is conserved over time ($\frac{\text{d}}{\text{d} t} \sum_{i=1}^S x_i = 0$).
Consequently, the ALVE~\eqref{eq:ALVE} defines a trajectory bound to the open simplex, that is, $\vec{x}(t)\in \Delta_{S-1}$ for all times~\cite{Hofbauer1998}. 
If the dynamics were initialized on the boundary of the simplex, $\vec{x}_0\in \partial\Delta_{S-1} = \overline{\Delta}_{S-1}\backslash\Delta_{S-1}$, they would remain restricted to the boundary.

The natural question about the long-time behavior of a state concentration $x_i$ is whether it remains bounded away from 0, whether it approaches 0, or whether it expresses any other qualitatively different behavior (such as, for example, a heteroclinic orbit).
All properties of the ALVE that we summarize below (coexistence and condensation) and that are central results of this work (topological robustness and coexistence networks), can be traced back to the quadratic interaction structure of the ALVE~\eqref{eq:ALVE} and the antisymmetry of $A$.

\textit{Interpretation of the antisymmetric matrix $A$ as a weighted network.} 
For our analysis of topological robustness and coexistence networks, we interpret the antisymmetric matrix $A$ as the antisymmetric adjacency matrix of a weighted network (also referred to as the skew-adjacency matrix of a weighted directed graph), see Figure~\ref{fig:ALVE_topological_stability} (insets). States in the ALVE correspond to nodes of the weighted network and entries of the antisymmetric matrix $A$ characterize the links between nodes.
This mapping to graph theory enables us to separate the discussion of the network topology (direction of links) from the weights of the network (weights of links).

In general, a network (or directed graph) $\mathcal{N}$ consists of a set of labeled nodes (or vertices) $V(\mathcal{N}) = \{1,...,S\}$ and a set of links (or directed edges) each of which connects two nodes $E(\mathcal{N}) = \{ (1 \to 2), \dots, (i \to j), \dots\}$~\cite{West2001}. 
Every antisymmetric matrix $A \in \mathbb{R}^{S\times S}$ gives rise to a \textit{weighted network} $\mathcal{N}(A)$ (or weighted directed graph or interaction network) of size $S$ and with vertex-set $V(\mathcal{N}(A)) = \{1,...,S\}$. 
The weighted edge set of $\mathcal{N}(A)$ is obtained from the positive matrix entries as $w(i \to j) \coloneqq a_{ji}$ if $a_{ji}>0$.
On the other hand, if $a_{ji} = 0$, nodes $i$ and $j$ are not connected. Naturally, two nodes are connected by at most one link and self-loops do not appear, such that we deal with simple networks here.
The \textit{network topology} of $A$ is recovered by discarding the weights of the links, but keeping their direction. In other words, the network topology is the oriented graph without weights. 

Conversely, from a simple, weighted network~$\mathcal{N}$ with $S$ nodes, the antisymmetric adjacency matrix $A(\mathcal{N})\in \mathbb{R}^{S\times S}$ is obtained by defining for every edge $i \to j$ with weight $a_{ji} > 0$ the matrix entries $A(\mathcal{N})_{ji} = - A(\mathcal{N})_{ij} = a_{ji} $.
Thus, the sign of an entry in the adjacency matrix corresponds to the direction of the edge (positive weight for incoming link, negative for outgoing link), and the absolute value denotes to the magnitude of the weight.

\subsection{Qualitative long-time dynamics: condensation of some states and coexistence of all states}
\label{sec:ALVE_long_time}

The long-time behavior of the ALVE~\eqref{eq:ALVE} shows two qualitatively different scenarios: condensation of some states and coexistence of all states.

\textit{Condensation and depletion of some states.}
Depending on the weights of the interaction network, a state concentration either vanishes for long times ($x_i(t)\to 0$ as $t\to \infty$), in which case $i$ is referred to as a \textit{depleted state} (``depletion'' or ``extinction''), or it remains bounded away from zero for all times ($x_i(t)\geq Const>0$ for all times $t$), in which case $i$ is called a \textit{condensate} (``condensation'' or ``survival''). 
Whether a state is a condensate or becomes depleted is independent of the initial conditions $\vec{x}_0$ and depends only on the antisymmetric matrix~$A$; see Section~S1.a of the Supplementary Material \cite{Supplement} and~\cite{Chawanya2002, Knebel2015} for details.
However, the details of the dynamics within the surviving condensate states depend both on $\vec{x}_0$ and $A$.
Condensation and depletion of states constitute one central feature of the ALVE.

\textit{Coexistence of all states.}
A situation in which none of the $S$ states becomes depleted is referred to as \textit{coexistence of all states}~\cite{Durney2011, Zia2011, Knebel2013, Knebel2015}, see Figure~\ref{fig:ALVE_topological_stability}. 
All states are condensates in this case. 
In mathematical terms, the trajectory of state concentrations stays away from the boundary of the $(S-1)$-simplex by a finite distance for all times. 
In the context of evolutionary game theory, coexistence of all states in the ALVE translates to an evolutionary zero-sum game in which none of the strategies goes extinct. 
Despite the interactions between the agents of the population playing different strategies, all strategies remain coexisting for all times.
To obtain coexistence of all states in the ALVE, either the network topology needs to be chosen carefully, or the weights on a given network topology need to be adjusted, or both network topology and weights need to be adjusted (to fine-tuned values or in a broader regime).
Coexistence of all states is only possible for strongly connected networks.
A network is strongly connected if for all pairs of nodes $i$ and $j$ there is a directed path connecting $i$ to $j$ and, vice versa, a directed path connecting $j$ to $i$. 
In other words, only if every state can indirectly exchange mass with any other state is coexistence of all states possible; see Section~S1.b of the Supplementary Material \cite{Supplement}.
In this work, we do not consider unconnected networks because unconnected network components do not interact with each other and, thus, can be treated separately.
The simplest strongly connected network is a directed cycle as discussed below.

A concise characterization of coexistence is obtained as follows: coexistence of all states for long times follows \textit{if and only if} the kernel of the antisymmetric matrix $A$ is strictly positive. Recall that the kernel (or nullspace) of a matrix $A$, $\text{Ker}(A)$, consists of all vectors $\vec{p}$ with $A \vec{p} = 0$.
We call a kernel element $\vec{p}\in \text{Ker}(A)$ a \textit{strictly positive kernel element} if $p_i>0$ for all $i=1, \dots, S$ (denoted as $\vec{p}>0$) and the kernel strictly positive (denoted as $\text{Ker}(A)>0$) if it contains a strictly positive kernel element.
The details of this characterization can be found in~\cite{Knebel2013, Knebel2015}; we present a concise summary in the following.

\textit{A strictly positive kernel of $A$ implies coexistence of all states.} 
To show this direction, let $\vec{p}$ be an element of the kernel of $A$. 
It follows that the collective quantity defined as the Kullback-Leibler divergence (or, equivalently, relative entropy) of the kernel element $\vec{p}$ to the trajectory~$\vec{x}$,
\begin{align}\label{eq:ALVE_conserved_quantity}
D(\vec{p}||\vec{x}(t)) = \sum_{\substack{i=1\\(p_i\neq 0)}}^S p_i\log \left( \frac{|p_i|}{x_i(t)}\right)\ ,
\end{align}
is conserved under the dynamics of the ALVE~\eqref{eq:ALVE} due to the pairwise interaction structure and antisymmetry of matrix $A$ as one confirms straightforwardly ($\frac{\text{d} }{\text{d} t} D(\vec{p}||\vec{x}) = \sum_{i} (A\vec{p})_ix_i = 0$).
Furthermore, independent kernel elements of $A$ give rise to independent conserved quantities, and the dimension of the kernel, $\dim(\text{Ker}(A))$, determines how many such conserved quantities of form $D$ exist; see~\cite{Knebel2013} for details.
If the kernel of $A$ contains a strictly positive kernel element, the relative entropy~\eqref{eq:ALVE_conserved_quantity} is conserved and positive for all times: $0<D(\vec{p}||\vec{x}(t)) = D(\vec{p}||\vec{x}(0))<\infty$. Thus, none of the state concentrations vanishes (otherwise, $D$ would diverge in contradiction to the boundedness of $D$).
In other words, if the kernel of $A$ is strictly positive, all states coexist.

\textit{Coexistence of all states implies a strictly positive kernel of $A$.}
To show the reverse direction, one may (i) exploit an algebraic property of antisymmetric matrices, and (ii) connect this algebraic property to the long-time dynamics of the ALVE via a suitable collective quantity that has the same form as the conserved quantity in Equation~\eqref{eq:ALVE_conserved_quantity}; see 
Section~S1.a of the Supplementary Material \cite{Supplement} for more details and~\cite{Chawanya2002, Knebel2015}.
Taken together, coexistence of all species occurs in the ALVE~\eqref{eq:ALVE} if and only if the kernel of the antisymmetric matrix~$A$ is strictly positive.

\textit{Steady state concentrations.} 
As shown above, a strictly positive kernel ensures coexistence of all states independent of the initial conditions. When the strictly positive kernel of the matrix $A$ is one-dimensional, the temporal average of the state concentrations, $\langle\vec{x}\rangle_t = \frac{1}{t}\int_0^t \mathrm{d}s\ \vec{x}(s)$, converges to the unique kernel element in the open simplex~\cite{Knebel2015}. In mathematical terms, if $\text{dim}(\text{Ker}(A)) = 1$ and $\vec{p}\in \text{Ker}(A)$ with $\vec{p}\in \Delta_{S-1}$, then $\langle\vec{x}\rangle_t\to \vec{p}$ as $t\to \infty$.
If, however, the dimension of the strictly positive kernel is greater than 1 (that is, $\dim (\text{Ker}(A)) \geq 2$ or, in other words, $\text{Ker}(A)$ is degenerate), a comparably simple characterization of the dynamics in terms of $\langle\vec{x}\rangle_{t\to \infty}$ has not yet been obtained; see~\cite{Knebel2015} for details. 
For this reason, we focus our discussion of coexistence networks primarily on networks with a one-dimensional kernel.

\subsection{Topologically robust coexistence and coexistence networks}
\label{sec:topological_robustness}
In this manuscript, we ask which network topologies admit coexistence of all states for all choices of weights and initial conditions; see Figure~\ref{fig:ALVE_topological_stability} for an illustration.
For the rock-papers-scissors-lizard-spock network topology, coexistence of all states depends on the choice of weights; see Figure~\ref{fig:ALVE_topological_stability}(b).
However, for the network topology of a directed cycle of five states supplemented with the directed link in the inside of the cycle as depicted in Figure~\ref{fig:ALVE_topological_stability}(a), coexistence of all states is observed for all choices of weights.
We refer to the latter network topology as a \textit{topologically robust zero-sum game} or a \textit{coexistence network} because coexistence of all states in the ALVE~\eqref{eq:ALVE} is robust against arbitrary changes of the weights on that network topology.
In other words, coexistence of all states depends only on the network topology, but not on the specific values of the weights.

\textit{Algebraic characterization of coexistence networks.}
Having established the equivalence of coexistence of all states in the ALVE and a strictly positive kernel of the antisymmetric matrix, coexistence networks can be algebraically characterized as follows.
A network is a \textit{coexistence network} if its antisymmetric adjacency matrix has a strictly positive kernel element for all choices of weights that do not change the underlying network topology, that is, keeping the direction of links and not adding links to or removing links from the network. 
In algebraic terms, this question of a strictly positive kernel amounts to determining the conditions on an antisymmetric matrix under which its kernel remains strictly positive for all choices of the non-zero matrix entries (as long as their sign is kept).
An antisymmetric matrix whose kernel is one-dimensional is a coexistence network if, for all choices of weights, all entries of the kernel vector have the same sign, such that the unique normalized kernel vector is always strictly positive.
The mapping between the entries of a matrix and its kernel elements is, in general, not straightforwardly answered for arbitrary matrices. 
For antisymmetric matrices, however, analytical progress is possible as we show in this manuscript; see Section~\ref{sec:algebra}.
Before proceeding, we illustrate these algebraic insights with directed cycles of odd and even length as examples. 

\textit{Examples of directed cycles: algebraic method for coexistence networks.} 
As an example for a coexistence network, consider the rock-paper-scissors network topology (that is, mass can only flow in the directed cycle $1\to 2\to 3\to 1 $ such that $a_{13}, a_{21}, a_{32}>0$). The antisymmetric adjacency matrix $A_\text{3-cycle}$ is given by: 
\begin{align}
A_\text{3-cycle} = 
\begin{pmatrix}
0 & -a_{21} & a_{13}  \\
a_{21} & 0 & -a_{32}  \\
-a_{13} & a_{32} & 0  
\end{pmatrix}\ ,
\end{align}
with kernel $\text{Ker}(A_\text{3-cycle}) = \{(a_{32}, a_{13}, a_{21})\}$ %
\footnote{Note that we do not distinguish notationally between column and row vectors in this manuscript if the meaning is clear from the context.}.
Consequently, the kernel of the rock-paper-scissors network topology is strictly positive and all states coexist in the ALVE for all times \textit{for all choices} of weights as long as the rock-paper-scissors network topology is not altered.

The above observation can be generalized to cycles of odd size, that is, to network topologies in which mass can only flow in the cycle $ 1\to 2\to 3\to \dots \to S-1 \to S \to 1$ for $S$ odd (such that $a_{1,S}, a_{21}, a_{32}, \dots, a_{S, S-1}>0$) with antisymmetric adjacency matrix $A_\text{odd-cycle}$:
\begin{align}\label{eq:odd_cycle}
A_\text{odd-cycle} = 
\begin{pmatrix}
0 & -a_{21} & 0  &  \dots & a_{1, S}  \\
a_{21} & 0 & -a_{32}  &  \dots & 0\\
\vdots & \ddots & \ddots  & \ddots &   \vdots\\
0 & 0 & \dots  & 0 &   -a_{S, S-1}\\
-a_{1,S} & 0 & \dots  & a_{S, S-1} &   0\\
\end{pmatrix}\ .
\end{align}
Its kernel is obtained as $\text{Ker}(A) = \{(a_{32}a_{54}\cdots a_{S, S-1}, a_{43}a_{65}\cdots a_{1, S}, \dots, a_{21}a_{43}\cdots a_{S-1, S-2})\}$ (see Section~\ref{sec:main_adjugate_vector}), which is again strictly positive for all choices of weights on the cyclic network topology.
Thus, cycles of odd size are coexistence networks.

In contrast, directed cycles of even size have a non-trivial kernel only if the weights $a_{ij}$ are fine-tuned. 
The determinant (in suiting labeling) is given by $\text{Det}(A_{\text{even-cycle}})  = (a_{21}a_{43} \dots a_{S,S-1} - a_{32}a_{54}\dots a_{1,S})^2$, which is zero only for specific choices of weights; see Section~\ref{sec:algebra} and Section~S2.c of the Supplementary Material \cite{Supplement}. Thus, for cycles of even size, the occurrence of a strictly positive kernel depends on the choice of matrix entries. Cycles of even size are not coexistence networks.
In the following, we further explore this possibly counter-intuitive behavior between even and odd dimension. The discussion of odd-sized directed cycles as examples of coexistence networks is extended in Section~\ref{sec:coexistence_networks} to Hamiltonian networks, which have a directed cycle through all nodes, and generalized to non-Hamiltonian networks (network topologies without a cycle).

\section{Kernel of an antisymmetric matrix and graph theory}
\label{sec:algebra}

\textit{Overview of this section.}
In this section, we characterize the kernel of an antisymmetric matrix in terms of graph-theoretical properties of its network representation. 
To this end, we introduce the \textit{Pfaffian}, which is a determinant-like function defined for antisymmetric matrices, and its interpretation in terms of perfect matchings. By using the notion of the Pfaffian, kernel elements of antisymmetric matrices can be computed via the \textit{adjugate vector}; see Section~\ref{sec:main_adjugate_vector}. This computation enables us to construct coexistence networks; see Section~\ref{sec:coexistence_networks}.

\subsection{Kernel and Pfaffian of antisymmetric matrices}
\label{sec:Pfaffian_Intro}

We start with recalling some general spectral properties of antisymmetric matrices. 
The determinant of an antisymmetric $- A^T = A\in\mathbb{R}^{S\times S}$ is always zero when $S$ is odd, since $\mathrm{Det}(A) = \mathrm{Det}(-A^T) = (-1)^S \mathrm{Det}(A)$. Thus, the kernel (or nullspace) of an odd-sized antisymmetric matrix is at least one-dimensional and, thus, always nontrivial. We refer to the kernel of an antisymmetric matrix as \textit{degenerate} if the kernel dimension is greater than 1 ($\dim (\text{Ker}(A)) \geq 2$).
It is straightforward to show that the non-zero eigenvalues of an antisymmetric matrix $A$ always occur as pairs of purely imaginary complex conjugate numbers, such that $\mathrm{Det}(A) \geq 0$. In other words, all antisymmetric matrices have an even number of nonzero eigenvalues, implying that their kernel is even-dimensional if the matrix is even-sized, whereas the kernel is odd-dimensional if the matrix is odd-sized.

To further characterize the kernel of an antisymmetric matrix $A$, it is suitable to introduce the concept of the Pfaffian, $\text{Pf}(A)$.
The Pfaffian can be thought of as a determinant-like function tailored to antisymmetric matrices. The square of the Pfaffian equals the determinant of $A$, $\text{Pf}(A)^2 = \text{Det}(A)$~\cite{Muir1882, Kasteleyn1967, Wimmer2011} indicating that the Pfaffian of an antisymmetric matrix can carry a sign as opposed to its determinant.
The sign of the Pfaffian is central to our analysis of coexistence networks. 
\begin{figure}[bth!]
\centering
\includegraphics{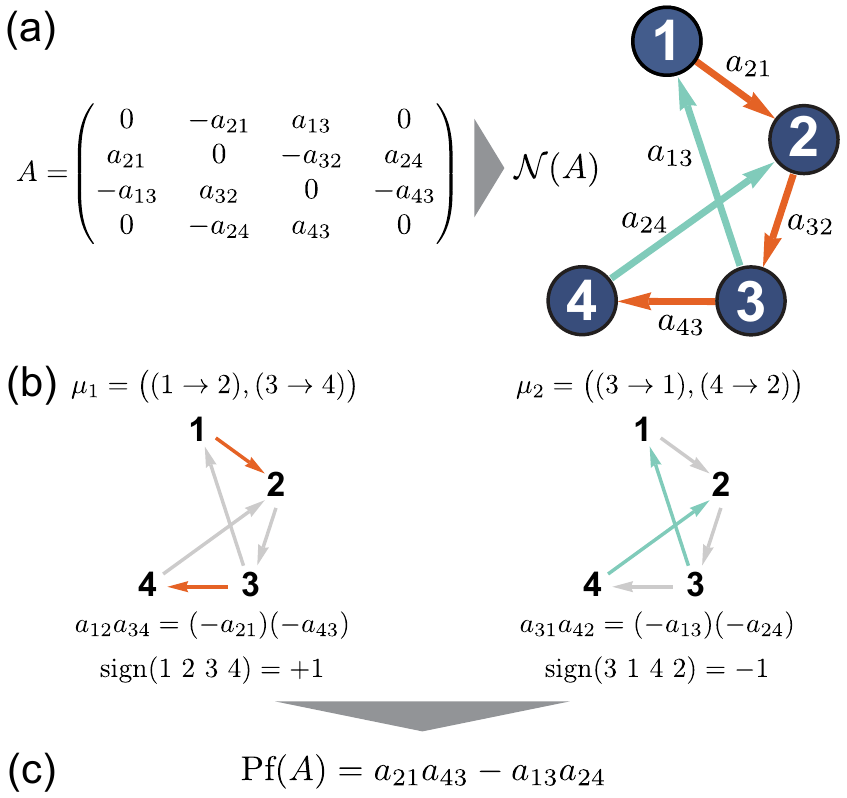}
\caption{(Color online)
\textbf{Graph-theoretical definition of the Pfaffian.}
(a) The antisymmetric matrix $A$ and the corresponding, pretzel-like, weighted network $\mathcal{N}(A)$.
(b) $\mathcal{N}(A)$ has two perfect matchings, $\mu_1 = \big((1\to 2),(3\to 4)\big)$ and $\mu_2 = \big((3\to 1), (4 \to 2)\big)$. Each of the perfect matchings gives rise to a summand in the Pfaffian of $A$. 
Note that the graph-theoretical definition of the Pfaffian only includes negative matrix entries. Thus, the difference of the signs of the two summands arises from the network topology (via the permutations of the respective perfect matchings).
The sign of a perfect matching depends on the number of transposition needed to permute the indices of the matching's partition to the ordered partition $(1, 2, 3, 4)$. 
(C) The Pfaffian of $A$ is the sum over the contributions stemming from all perfect matchings; see Equation~\eqref{eq:Pfaffian_matchings}. 
All signs in the final sum of the graph-theoretical definition of the Pfaffian are determined by the network topology alone.
 }
\label{fig:Pfaffian}
\end{figure}
%

\subsection{Graph-theoretical definition of the Pfaffian}
\label{sec:Pfaffian_graph}

Because the Pfaffian of an antisymmetric matrix is central to our analysis, we now present its graph-theoretical definition; see Equation~\eqref{eq:Pfaffian_matchings} below and the illustration in Figure~\ref{fig:Pfaffian}. 
Following the above statements about antisymmetric matrices, the Pfaffian of odd-sized antisymmetric matrices is always zero.
For even-sized antisymmetric matrices, the Pfaffian is typically defined through a combinatorial formula, which we present in 
Section~S2.a of the Supplementary Material \cite{Supplement} for completeness. 
However, an intuitive understanding of the Pfaffian that is suited for our purposes is obtained via the weighted network $\mathcal{N}(A)$ by considering all of its perfect matchings. 
The graph-theoretical interpretation of the Pfaffian has already been appreciated and applied in statistical physics to compute the entropy of systems in which dimer molecules are placed on regular lattice graphs~\cite{Kasteleyn1961, Temperley1961, Fisher1961,Wu2006,Thomas2006}. We discuss the significance of our results in this context in Section~\ref{sec:Pfaffian_orientation}.

\begin{figure*}[htb!]
\centering
\includegraphics{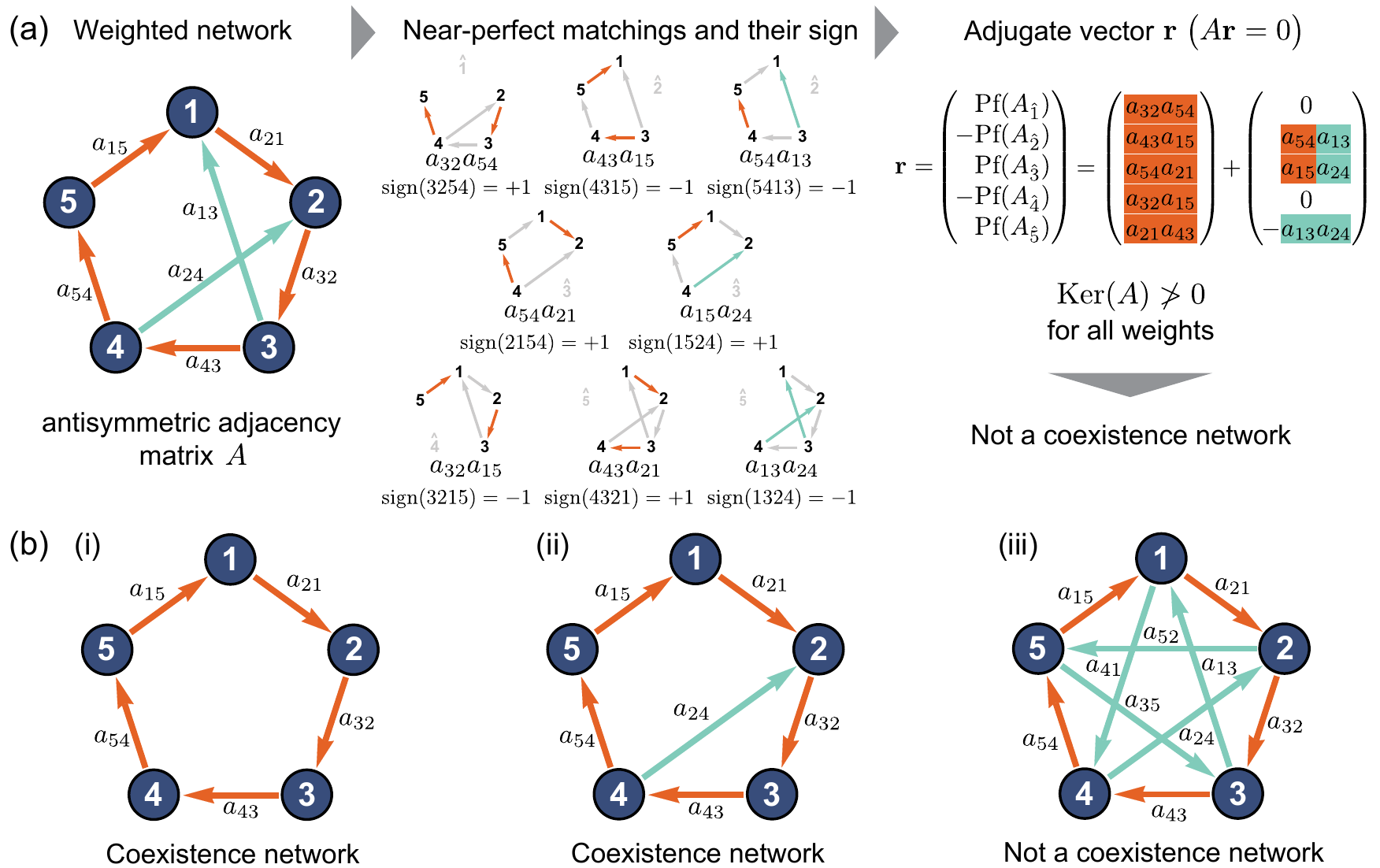}
\caption{(Color online)
\textbf{Algebraic characterization of coexistence networks via the adjugate vector of $A$.}
(a) Algebraic determination of coexistence networks; here illustrated for a cycle of five nodes (orange edges) with the two additional interior edges $(4, 2)$ and $(3, 1)$ (cyan edges). The network is factor-critical.
The entries of the adjugate vector are calculated via the near-perfect matchings of the network and their signs (here, eight near-perfect matchings in total); see Figure~\ref{fig:Pfaffian} for details of the computation. 
The near-perfect matching $\big((3\to 1),(4\to 2)\big)$ creates the negative summand ($-a_{13}a_{24}$). Thus, the weights can be chosen such that the kernel of $A$ is not strictly positive. Therefore, this network is not a coexistence network. Note that by setting one of the two or both cyan matrix entries to zero, the resulting network topology is a coexistence network.
(b)(i) Cycles of odd length are coexistence networks because their adjugate vector is always strictly positive, whereas cycles of even length are not.
(ii) The cycle with five nodes and the additional interior edge $(4, 2)$ is a coexistence network as in inferred from the adjugate vector in (a) by setting $a_{13} = 0$. 
(iii) The complete network of five nodes is not a coexistence network; see also Figure~\ref{fig:ALVE_topological_stability}(b). Additional near-perfect matchings arise through the interior edges $(5, 3), (1, 4)$, and $(2, 5)$. 
}
\label{fig:coexistence_networks}
\end{figure*}

\textit{Perfect Matchings, near-perfect matchings, and factor-critical networks.} 
A \textit{matching} of a network is a subset of its edge set $E' \subseteq E$ such that no two edges in $E'$ share the same node. 
In other words, every node is covered by at most one edge of the matching $E'$.
A matching that covers all nodes of a network is referred to as a \textit{perfect matching} $\mu$ of a network~\cite{West2001}; see Figure~\ref{fig:Pfaffian}(b). 
Consequently, the number of nodes in a network with a perfect matching needs to be even.
Because a perfect matching is a subset of the network's edges such that every node is covered exactly once, it can be interpreted as a partition of the set $\{1,...,S=2n\}$ into pairs. 
For networks with an odd number of nodes, one introduces the notion of a \textit{near-perfect matching}, which is a matching that covers all but one node. 
Thus, a near-perfect matching of a network is a perfect matching of a subnetwork that is obtained by deleting one node from the network; see Figure~\ref{fig:coexistence_networks}(a). 
In case of odd $S$, a network is called \textit{factor-critical} if there exists a perfect matching for every subnetwork that is created by deleting one node from the network~\cite{West2001}.

%

\textit{Graph-theoretical definition of the Pfaffian.} 
The Pfaffian of an antisymmetric matrix $A$ can be calculated via all perfect matchings of the network $\mathcal{N}(A)$.
Here we follow the convention that a network's link $i\to j$ gives rise to (a) the pair $(i, j)$ in the partition corresponding to the perfect matching $\mu$ and to (b) the negative matrix entry $a_{ij}<0$. 
With these two conventions, the Pfaffian of the antisymmetric matrix $A$ is computed as:
\begin{align}\label{eq:Pfaffian_matchings}
\text{Pf} (A) = \sum_{\substack{\text{perf. match.}\\ \mu\in \mathcal{N}(A)}}\left(\text{sign}(\sigma_{\mu }) \prod_{(i\to j)\in \mu } a_{ij} \right) \ ,
\end{align}
where the sum runs over all perfect matchings of the network $\mathcal{N}(A)$; see Figure~\ref{fig:Pfaffian}. 
Thus, the Pfaffian is a sum over signed products of negative matrix entries (a link $i\to j$ contributes with $a_{ij}<0 $), which are determined by the edges of each perfect matching. 
The permutation $\sigma_{\mu}$ denotes the partition of the node set $\{1, 2, \dots, S=2n\}$ obtained from the edges in the perfect matching $\mu$:
\begin{align}\nonumber
\sigma_{\mu} = &\begin{pmatrix}
1 & 2 & 3 & 4& \dots & 2n-1 & 2n\\
(i_1, & j_1) & (i_2, & j_2)& \dots & (i_n, & j_n)
\end{pmatrix}\ ,\\\nonumber
\equiv &\ \big(i_1\ j_1\ i_2\  j_2\ \dots  i_n\ j_n\big)
\ .
\end{align}
$\mathrm{sign}(\sigma_{\mu })$ is determined by the number of transpositions needed to permute the partition $(i_1, j_1, i_2,  j_2, \dots,  i_n, j_n)$ into the partition $(1, 2, \dots, S=2n)$: $\text{sign}(\sigma_{\mu})= + 1$ if the number of transpositions is even, and $\text{sign}(\sigma_{\mu})= - 1$ if it is odd.
For simplicity, we also to refer to $\mathrm{sign}(\sigma_{\mu })$ as the sign of the perfect matching $\mu$.
In 
Section~S2.b of the Supplementary Material \cite{Supplement}, we discuss why this graph-theoretical definition agrees with the ``standard'' (combinatorial) definition of the Pfaffian.

\textit{Example of the Pfaffian for a pretzel-like network.} To illustrate the graph-theoretical definitions of the Pfaffian, consider the pretzel-like interaction network sketched in Figure~\ref{fig:Pfaffian}, whose antisymmetric adjacency matrix is: 
\begin{align}\label{eq:pretzel}
A_\text{pretzel} = 
\begin{pmatrix}
0&-a_{21}&a_{13}& 0 \\
a_{21}&0&-a_{32}&a_{24} \\
-a_{13}&a_{32}&0&-a_{43} \\
0&-a_{24}&a_{43}&0
\end{pmatrix}\ .
\end{align}
There exist two perfect matchings of this network: $\mu_1 = \big((1\to 2),(3\to 4)\big)$ and $\mu_2 = \big((3\to 1), (4 \to 2)\big)$ and the Pfaffian of $A_\text{pretzel}$ is obtained via definition~\eqref{eq:Pfaffian_matchings} as:
\begin{align}\nonumber
\text{Pf}(A_\text{pretzel}) 
&=\text{sign}(1\ 2\ 3\ 4)a_{12}a_{34}+\text{sign}(3\ 1\ 4\ 2)a_{31}a_{42}\ , \\\nonumber
&=(+1) (-a_{21})(-a_{43})  + (-1) (-a_{13})(-a_{24})\ , \\\label{eq:Pfaffian_pretzel}
&=a_{21}a_{43}-a_{13}a_{24}\ .
\end{align}
The kernel of $A_\text{pretzel}$ is only nontrivial if the matrix entries of $A_\text{pretzel}$ fulfill $\text{Pf}(A_\text{pretzel}) = a_{21}a_{43}-a_{13}a_{24}=0$, that is, for a fine-tuned choice of weights. 
Thus, $\mathcal{N}(A_\text{pretzel})$ is not a coexistence network.

\subsection{Adjugate vector of an antisymmetric matrix and coexistence networks}
\label{sec:main_adjugate_vector}

By using the notion of the Pfaffian of an antisymmetric matrix $A$, an explicit analytical expression for the kernel of $A$ is obtained via the adjugate vector if the kernel dimension is 1; see Figure~\ref{fig:coexistence_networks} for an illustration. High-dimensional kernels are discussed in Section~\ref{sec:numerics}. In brief, we have neither found any even-sized coexistence networks nor odd-sized coexistence networks with a high-dimensional kernel thus far. 

\textit{The adjugate vector of A.} If $S$ is odd, the kernel of an antisymmetric matrix $A\in\mathbb{R}^{S\times S}$ is characterized by the \textit{adjugate vector} $\vec{r}\in\mathbb{R}^S$, which is defined as:
\begin{align}\label{eq:adjugate_vector}
r_i = (-1)^{i+1} \text{Pf}(A_{\hat{i}}) \quad ,\ i = 1, \dots, S\ .
\end{align}
Here, $A_{\hat{i}}$ denotes the matrix obtained by deleting the $i$th row and column from $A$. The computation of the adjugate vector~\eqref{eq:adjugate_vector} via the Pfaffians of all submatrices $A_{\hat{i}}$ is reminiscent of Cramer's rule~\cite{Mirsky1990} adjusted to antisymmetric matrices. Note that $A_{\hat{i}}$ corresponds to the weighted network obtained from $\mathcal{N}(A)$ by deleting node $i$. The adjugate vector is, thus, determined by all near-perfect matchings of $\mathcal{N}(A)$.
The adjugate vector is a kernel vector of $A$ if $\mathrm{dim}(\mathrm{Ker}(A)) = 1$ and the zero-vector $\vec{r} = 0$ if $\mathrm{dim}(\mathrm{Ker}(A)) = 3, 5, \cdots,S$~\cite{Cullis1913}. In any case, it is $A\vec{r} = 0$.

\textit{Algebraic characterization of coexistence networks via the adjugate vector.} 
In our work, we use the adjugate vector to characterize coexistence networks of odd size~$S$ as follows. 
A network $A$ whose kernel is one-dimensional for all choices of weights is a coexistence network if its adjugate vector is always strictly positive, that is, if all entries of the adjugate vector have the same sign independent of the weights (if $\vec{r}$ is a kernel vector of $A$, so is $-\vec{r}$). 
For further details on the influence of the network topology on the kernel dimension of the network's adjacency matrix see also 
Section~S2.e of the Supplementary Material \cite{Supplement}.

Figure~\ref{fig:coexistence_networks}(a) illustrates the connection between coexistence networks and the graph-theoretical interpretation of the adjugate vector for an exemplary cycle of five states with two additional interior edges. Upon identifying the near-perfect matchings of a network and their signs, the adjugate vector is computed. 
A necessary condition to obtain a strictly positive adjugate vector is that the network is factor-critical. 
Only if the network is factor-critical can the adjugate vector have non-vanishing values in all of its components. 
The sign of the $i$th component of the adjugate vector, $r_i$, is determined by the signs of the near-perfect matchings and by whether the index $i$ is even or odd; see Equations~\eqref{eq:Pfaffian_matchings} and~\eqref{eq:adjugate_vector}. 
Both contributions determine whether all summands occurring in the adjugate vector have the same sign or not and, thus, whether the one-dimensional kernel is strictly positive or not for all choices of weights.

\textit{Examples for the algebraic characterization.} 
As was shown in Section~\ref{sec:topological_robustness}, simple directed cycles of odd length are coexistence networks (see Figure~\ref{fig:coexistence_networks}(b)(i)), while cycles of even length are not. 
The directed cycle with one additional interior edge depicted in Figure~\ref{fig:coexistence_networks}(b)(ii) is a coexistence network as well; see also Figure~\ref{fig:ALVE_topological_stability}(a). The interior edge $(4\to 2)$ creates one new near-perfect matching $\big((4\to 2),(5\to 1)\big)$ that gives rise to an entry in the third component of the adjugate vector. The corresponding permutation has the same sign as the permutation of the near-perfect matching $\big((4\to 5),(1\to 2)\big)$. Therefore, the kernel of $A$ remains strictly positive for all choices of weights, and this network is a coexistence network. 
Note that if the direction of this edge is reversed to $(2\to 4)$, the sign of the near-perfect matching will be negative and the adjugate vector could have a negative entry in the third component upon choosing suitable weights; thus, not being a coexistence network.
Figure~\ref{fig:coexistence_networks}(b)(iii) shows the complete network of five nodes (see also Figure~\ref{fig:ALVE_topological_stability}(b)). 
This network is factor-critical, but not a coexistence network. As indicated above, factor-criticality is not sufficient to obtain a coexistence network; in addition to factor-criticality, the signs of all summands in all entries of the adjugate vector need to be the same to give rise to a coexistence network.

\section{Coexistence networks}
\label{sec:coexistence_networks}

\textit{Overview of this section.} 
In this section, we present graph-theoretical rules for how to construct coexistence networks. 
Recall that for coexistence networks, coexistence of all states in the ALVE~\eqref{eq:ALVE} is robust against arbitrary changes of the weights (the defining interaction strengths) on the given network topology; see Section~\ref{sec:ALVE}. 
We begin our analysis with coexistence networks that have a one-dimensional kernel for all choices of weights.
For such networks, the vector of steady state concentrations of the ALVE~\eqref{eq:ALVE} is given by the unique normalized kernel vector as described in Section~\ref{sec:ALVE_long_time}.

First, we introduce Hamiltonian coexistence networks, which are coexistence networks with a directed cycle through all nodes and, thus, generalize directed cycles of odd size.
In Section~\ref{sec:hamiltonian} we present the two \textit{coexistence conditions} on the network topology (cycle condition~ \eqref{eq:conditions1} and crossing condition~\eqref{eq:conditions2}) with which all Hamiltonian coexistence networks are identified. The proof of these conditions exploits the connection between the adjugate vector~\eqref{eq:adjugate_vector} of an antisymmetric matrix and near-perfect matchings of its network topology; details are deferred to 
Section~S4 of the Supplementary Material \cite{Supplement}. 
In Section~\ref{sec:non-hamiltonian}, we show how network topologies without a Hamiltonian cycle can be obtained from Hamiltonian coexistence networks by deleting suitable links from the cycle. 
For networks with up to 9 nodes, we numerically verify that all coexistence networks are obtained from Hamiltonian coexistence networks this way; see Section~\ref{sec:numerics}. 
At present, our numerical enumerations are limited to networks of 9 nodes because of the vastly growing number of network topologies with more nodes.
We also briefly discuss the possibility of coexistence networks with a degenerate kernel ($\mathrm{dim}(\mathrm{Ker}(A)) = 2, 3, \dots$), even though we did not find any for $S\leq 9$. 
We present examples for the construction of coexistence networks in Section~\ref{sec:examples}, applications of our results are discussed in Section~\ref{sec:discussion}.

\subsection{Hamiltonian coexistence networks}\label{sec:hamiltonian}

\begin{figure*}[htb!]
\centering
\includegraphics{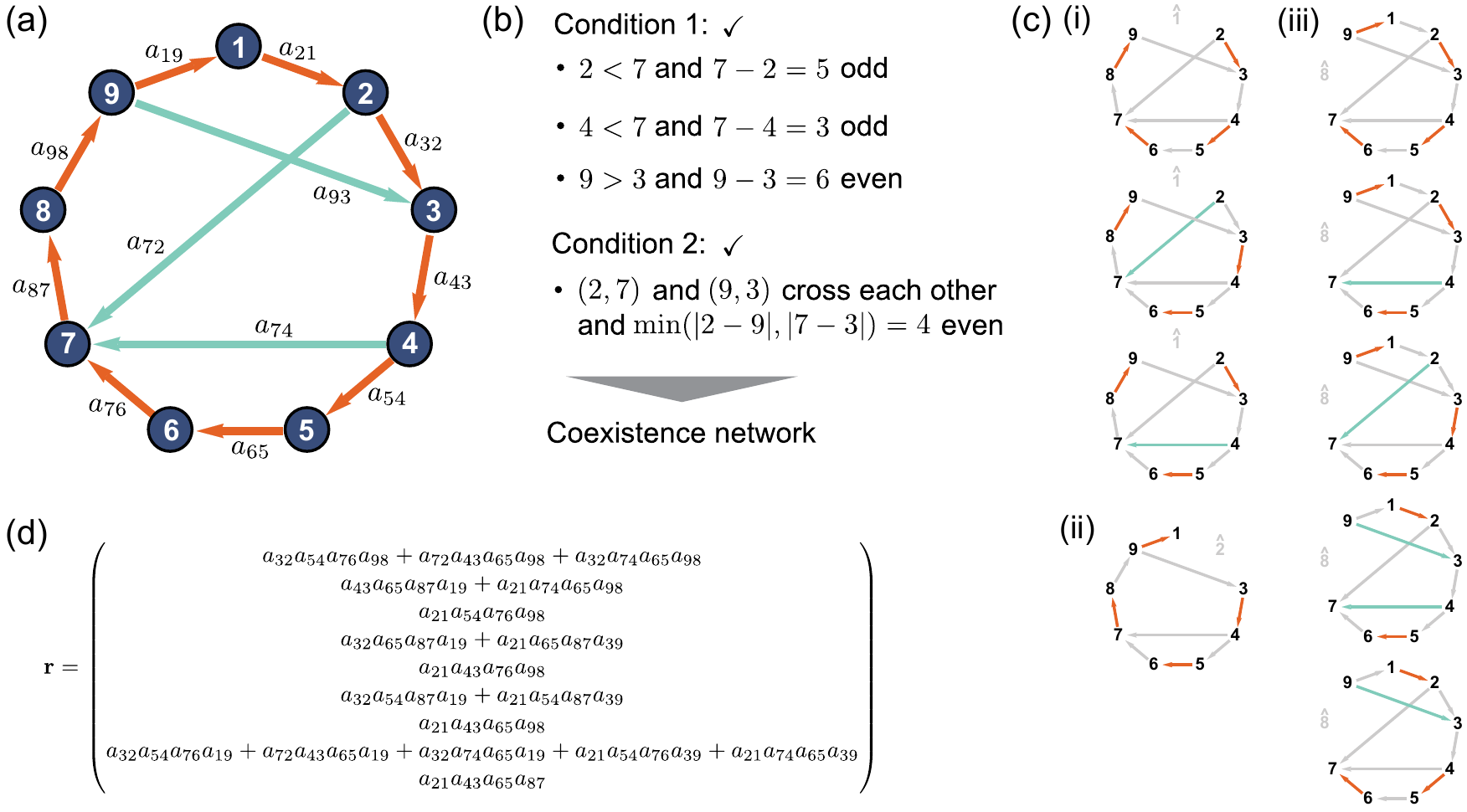}
\caption{(Color online)
\textbf{Graph-theoretical conditions for Hamiltonian coexistence networks. } 
(a) Hamiltonian network $\mathcal{N}^{(9)}$ of 9 nodes consisting of edges from the Hamiltonian cycle (orange) and interior edges $(2,7) ,(4,7)$, and $(9,3)$ (cyan). 
(b) We identified the cycle condition~\eqref{eq:conditions1} and the crossing condition~\eqref{eq:conditions2} to check whether a network topology is a coexistence network.
These conditions are both necessary and sufficient.
The cycle condition ensures that only cycles of odd length are created within the Hamiltonian cycle through any interior edge.
The crossing condition ensures that two crossing cycles share only an odd number of nodes connected by an even number of edges.
It follows that $\mathcal{N}^{(9)}$ is a coexistence network.
(c) Near-perfect matchings contributing to the first, second, and eighth component of the adjugate vector~\eqref{eq:adjugate_vector}. Crossing edges do not contribute to the same near-perfect matching.
(d) The component-wise calculation of the adjugate vector $\mathbf{r}$ confirms that the network topology $\mathcal{N}^{(9)}$ in (a) is a coexistence network because all vector components are strictly positive for all choices of weights.
}
\label{fig:ALVE_9}
\end{figure*}

\textit{Conditions for Hamiltonian coexistence networks.}
We now present a scheme to identify Hamiltonian coexistence networks of odd size $S = 2n-1$ ($n= 2,3, \dots$).
A Hamiltonian network $\mathcal{N}(A)$ contains at least one directed cycle, that is, a directed closed path passing exactly once through all nodes. 
Thus, its edge-set $E(\mathcal{N})$ can be split into the edges that constitute one such Hamiltonian cycle, $E_{\text{cycle}}$, and all other interior edges in the cycle, $E_{\text{in}}$, such that $E(\mathcal{N}) = E_{\text{cycle}} \cup E_{\text{in}}$ with $E_{\text{cycle}}  \cap E_{\text{in}} =\emptyset $. An ascending labeling of the network can be chosen such that $E_{\text{cycle}} = \{ (1,2),(2,3),...,(i,i+1),...,(S,1)\}$; see Figure~\ref{fig:ALVE_9}.

With this assignment of the edges of the Hamiltonian network into cycle edges and interior edges, we identified the following necessary and sufficient conditions for a Hamiltonian network to be a coexistence network. The coexistence conditions are stated first, before we illustrate, discuss, and prove them. 
A Hamiltonian network with the chosen ascending labeling of the cycle $E_{\text{cycle}} = \{ (1,2),(2,3),...,(i,i+1),...,(S,1)\}$ is a coexistence network \textit{if and only if} its interior edges fulfill the following two \textit{coexistence conditions}
\begin{enumerate}
\item \textit{Cycle condition:} For every interior edge $(i,j) \in E_{\text{in}}$ it holds that:
\begin{equation}
\begin{aligned}\label{eq:conditions1}
&(i,j) \text{ is ascending, and } j-i \text{ is odd,}\\
&(i,j) \text{ is descending, and } i-j \text{ is even.}
\end{aligned}
\end{equation}

An edge $(i,j)\in E_{\text{in}}$ is called \textit{ascending} (with respect to the labeling of the Hamiltonian cycle) if $i<j$, and \textit{descending} if $i>j$; see Figure~\ref{fig:ALVE_9}.

\item \textit{Crossing condition:} For every pair of crossing interior edges $\{(i,j), (k,l) \}\subseteq E_{\text{in}}$ it holds that: 
\begin{equation}
\begin{aligned}\label{eq:conditions2}
&(i,j) \text{ and } (k,l) \text{ cross each other,}\\
&\text{and } \min (|i-k|,|j-l|) \text{ is even.}
\end{aligned}
\end{equation}
Two interior edges $(i,j), (k,l) \in E_{\text{in}}$ are called \textit{crossing} if $\min (k,l )<i < \max (k,l)$ or $\min (k,l )<j < \max (k,l)$. If the Hamiltonian network is drawn in the two-dimensional plane, crossing edges cross in the interior of the cycle; see Figure~\ref{fig:ALVE_9}.
\end{enumerate}

%

\textit{Illustration of the conditions.}
The cycle condition~\eqref{eq:conditions1} governs the relation of interior edges with the Hamiltonian cycle, while the crossing condition~\eqref{eq:conditions2} governs the relation of interior edges to each other.
fulfilllment of the cycle condition~\eqref{eq:conditions1} ensures that interior edges only create directed subcycles of odd length (due to $S$ being odd).
If the difference between start and end node of an interior edge is odd, the direction of the interior edge needs to be ascending with respect to the Hamiltonian cycle to ensure a strictly positive kernel. 
In contrast, if the difference is even, the edge needs to be descending.
Taken together, Hamiltonian coexistence networks do not have even cycles.
The crossing condition~\eqref{eq:conditions2} ensures that no two crossing edges occur in the same near-perfect matching. This is the case if and only if the minimal difference between end and start nodes of crossing edges is even. 
In other words, the two directed cycles of odd length created by every pair of interior edges share an odd number of nodes that are connected by an even number of edges.
In 
Section~S4 of the Supplementary Material \cite{Supplement}, we prove that the coexistence conditions~\eqref{eq:conditions1} and~\eqref{eq:conditions2} are both sufficient and necessary for Hamiltonian coexistence networks.

\textit{Example of a coexistence network.} 
To illustrate the above conditions and ideas of the proof, we consider an exemplary Hamiltonian network $\mathcal{N}^{(9)}= E_{\text{cycle}}^{(9)}\cup E_{\text{in}}^{(9)}$ of 9 nodes, constituted by the Hamiltonian cycle $E_{\text{cycle}}^{(9)} = \{(1,2),(2,3),\dots (9, 1)\}$ and three interior edges $E_{\text{in}}^{(9)} = \{(2,7),(4,7),(9,3)\}$; see Figure~\ref{fig:ALVE_9}(a) for a sketch.

First, the coexistence conditions~\eqref{eq:conditions1} and ~\eqref{eq:conditions2} for a Hamiltonian coexistence network are fulfilled (Figure~\ref{fig:ALVE_9}(b)).
The cycle condition~\eqref{eq:conditions1} is fulfilled for all interior edges: 
(i) the interior edge $(2,7)$ is ascending because $2<7$ and their difference is odd, 
(ii) the interior edge $(4,7)$ is ascending because $4<7$ and the difference is odd,
(iii) the interior edge $(9,3)$ is descending because $9>3$ and the difference is even.
The crossing condition~\eqref{eq:conditions2} is fulfilled for the two crossing edges $(9,3)$ and $(2,7)$ (no other interior edges cross each other) because $\min (|9-2|,|3-7|) =4$, which is even.
Thus, we conclude from the coexistence conditions that the Hamiltonian network $\mathcal{N}^{(9)}$ is a coexistence network. 

\textit{Component-wise calculation of the adjugate vector confirms coexistence network.} 
To verify that the kernel of the antisymmetric matrix of $\mathcal{N}^{(9)}$ is indeed strictly positive as claimed 
above, we now explicitly calculate the adjugate vector~\eqref{eq:adjugate_vector} and check the sign of all entries. 
This algebraic check illustrates the main ideas underlying the proof presented in 
Section~S4 of the Supplementary Material \cite{Supplement}.
For illustration, we write out the near-perfect matchings for the first, second, and eighth component of the adjugate vector~\eqref{eq:adjugate_vector} and discuss their sign; see Figure~\ref{fig:ALVE_9}(c).

The first component $r_1$ is calculated as the Pfaffian of the submatrix $A_{\hat{1}}$. 
The summands contributing to $\mathrm{Pf}(A_{\hat{1}})$ originate from the near-perfect matchings of $\mathcal{N}^{(9)}$ for which the first node is removed, that is, the perfect matchings of the subnetwork $\mathcal{N}(A_{\hat{1}})$.
Three such near-perfect matchings exist (see Figure~\ref{fig:ALVE_9}(c)(i)):
\begin{align}\nonumber
 \mu_{\hat{1}, 1} &= \big(  (2\to 3), (4\to 5), (6\to 7), (8\to 9) \big)\ ,\\\nonumber
\mu_{\hat{1}, 2} &= \big(  (2\to 7), (3\to 4), (5\to 6),(8\to 9)  \big)\ , \\ \nonumber
 \mu_{\hat{1}, 3} &= \big( (2\to 3), (4\to 7), (5\to 6), (8\to 9) \big)\ .
\end{align}
The first of the above near-perfect matchings, $\mu_{\hat{1}, 1}$, comprises only edges from the Hamiltonian cycle. The two other near-perfect matchings, $\mu_{\hat{1}, 2}$ and $\mu_{\hat{1}, 3}$, include contributions from the Hamiltonian cycle and, in addition, one interior edge.
Thus, $r_1$ is computed as:
\begin{align}\nonumber
r_1 &=(-1)^{1+1}\mathrm{Pf}(A_{\hat{1}}) , \\\nonumber
&= 
\left[
\begin{aligned}
  \phantom{+ } \text{sign} \left( 
2\  3\ 4\  5\ 6 \ 7\ 8 \ 9
\right) a_{23} a_{45} a_{67} a_{89} \\
 +\text{sign} \left( 
2\ 7\ 3\ 4\ 5\ 6\ 8\ 9
\right) a_{2 7}a_{3 4}a_{5 6}a_{89}\\
 +\text{sign} \left( 
2\ 3\ 4\ 7\ 5\ 6\ 8\ 9
\right) a_{2 3}a_{4 7}a_{5 6}a_{89}\\
\end{aligned}\ \right],\\\nonumber
&= 
(+1)  a_{3 2} a_{5 4} a_{7 6} a_{9 8}  
+ (+1) a_{7 2}a_{4 3}a_{6 5}a_{9 8} \\
&\phantom{=(}+ (+1) a_{3 2}a_{7 4}a_{6 5}a_{9 8}\ .
\end{align}
The sign of the near-perfect matchings is calculated via the number of transpositions to order its elements in size. 

The permutation $\mu_{\hat{1},1}$ is already ordered in size and, thus, equals the identity permutation $\mu_{\hat{1},1} = (2\ 3\ \dots \ 8 \ 9 ) = \sigma_\mathds{1}$. 
For odd components of the adjugate vector, the permutation containing only edges from $E_{\text{cycle}}$ always equals the identity permutation and, thus, has the sign $+1$.
The first component of the adjugate vector is strictly positive because the permutations corresponding to $\mu_{\hat{1}, 2}$ and $\mu_{\hat{1}, 3}$ are ordered by an even number of transpositions, such that $r_1$ contains three positive summands.

For the second component, $r_2$, there exists only one perfect matching of $\mathcal{N} (A_{\hat{2}})$, which contains only edges from the Hamiltonian cycle; see Figure~\ref{fig:ALVE_9}(c)(ii): 
\begin{align}\nonumber
\mu_{\hat{2}, 1} =  
\big((3\to 4), (5\to 6), (7\to 8), (9\to 1)\big)\ .
\end{align}
Interior edges do not contribute to any near-perfect matching of $\mathcal{N}^{(9)}$ for which node 2 is removed. In general, it is both the placement of the interior edges relative to each other and their placement relative to the deleted node that determines whether additional near-perfect matching arise or not.
For $r_2$, it follows: 
\begin{align}\nonumber
r_2  &=(-1)^{1+2}\mathrm{Pf}(A_{\hat{2}})\ ,\\\nonumber
&=
(-1)\ \text{sign} \left(3\ 4\ 5\ 6\ 7\ 8\ 9\ 1 \right) a_{3 4}a_{5 6} a_{7 8} a_{9 1}\ ,\\
&=   a_{4 3}a_{6 5} a_{8 7} a_{1 9}\ .
\end{align}
For other even components of the adjugate vector, the permutation of the near-perfect matching containing only edges from $E_{\text{cycle}}$ is also not ordered in size in the same way as for $r_{2}$. The identity permutation $\sigma_{\mathbb{1}} = (1\ 2\ \dots i-1\ i+1\ \dots S)$ is achieved with an odd number of transpositions, resulting in the sign $-1$. This minus sign is balanced by the prefactor $(-1)^{1+i}$ in the adjugate vector. Therefore, all summands in all components, which arise from near-perfect matching with only edges from the Hamiltonian cycle, have the same sign.

Similarly, all other components of the adjugate vector are calculated from which $r_8$ as an instructive component is further discussed in the following. 
There exist five near-perfect matchings of $\mathcal{N}^{(9)}$ upon omitting node 8; see Figure~\ref{fig:ALVE_9}(c)(iii):
\begin{align}\nonumber
\mu_{\hat{8}, 1} = \big((2\to 3),  (4\to 5),  (6\to 7),  (9\to 1) \big)\ ,\\\nonumber
\mu_{\hat{8}, 2} = \big( (2\to 7),  (3\to 4),  (5\to 6),  (9\to 1) \big)\ ,\\\nonumber
\mu_{\hat{8}, 3} = \big(  (2\to 3),  (4\to 7),  (5\to 6),  (9\to 1) \big)\ ,\\\nonumber
\mu_{\hat{8}, 4} = \big( (1\to 2),  (4\to 5),  (6\to 7),  (9\to 3)\big)\ ,\\\nonumber
\mu_{\hat{8}, 5} = \big( (1\to 2),  (4\to 7),  (5\to 6),  (9\to 3) \big)\ .
\end{align}
Again, the first near-perfect matching $\mu_{\hat{8}, 1}$ comprises only edges stemming from the Hamiltonian cycle, while $\mu_{\hat{8}, 2}, \mu_{\hat{8}, 3}$ and $\mu_{\hat{8}, 4}$ involve contributions from the Hamiltonian cycle and exactly one interior edge. 
The corresponding permutations have the same sign because an even number of transpositions maps the partitions to the partition stemming from $\mu_{\hat{8}, 1}$; similarly as for the discussion of $r_1$.
The near-perfect matching $\mu_{\hat{8}, 5}$, however, contains the two non-crossing interior edges $(4, 7)$ and $(9,3)$. 
The sign of the permutation equals the sign of the identity permutation (which corresponds to the ordered set) because each interior edge fulfills the cycle condition~\eqref{eq:conditions1} and, thus, can be transferred to the partition stemming from $\mu_{\hat{8}, 1}$ by an even number of transpositions. 
This reasoning can be generalized to any number of interior, non-crossing edges occurring in the same near-perfect matching. 
The eighth component of the adjugate vector is, thus, obtained as:
\begin{align}\nonumber
r_8 &=(-1)^{1+8}\mathrm{Pf}(A_{\hat{8}})\ ,\\\nonumber
&= 
(-1)\cdot\left[
\begin{aligned}
\phantom{+}\text{sign} \left( 2\ 3\ 4\ 5\ 6\ 7\ 9\ 1 \right) a_{2 3} a_{4 5} a_{6 7} a_{9 1} \\
+\text{sign} \left( 2\ 7\ 3\ 4\ 5\ 6\ 9\ 1 \right) a_{2 7} a_{3 4} a_{5 6} a_{9 1} \\
+\text{sign} \left( 2\ 3\ 4\ 7\ 5\ 6\ 9\ 1 \right) a_{2 3} a_{4 7} a_{5 6} a_{9 1} \\
+\text{sign} \left( 1\ 2\ 4\ 5\ 6\ 7\ 9\ 3 \right) a_{1 2} a_{4 5} a_{6 7} a_{9 3} \\
+\text{sign} \left( 1\ 2\ 4\ 7\ 5\ 6\ 9\ 3 \right) a_{1 2} a_{4 7} a_{5 6} a_{9 3}  
\end{aligned}
\right]\ ,\\\nonumber
&=   a_{3 2} a_{5 4} a_{7 6} a_{1 9} + 
a_{7 2} a_{4 3} a_{6 5} a_{1 9} +
a_{3 2} a_{7 4} a_{6 5} a_{1 9} \\
&\phantom{=}+a_{2 1} a_{5 4} a_{7 6} a_{3 9} +
a_{2 1} a_{7 4} a_{6 5} a_{3 9} 
\ .
\end{align}

In total, the adjugate vector for the Hamiltonian network $\mathcal{N}^{(9)}$ is written out for all components in Figure~\ref{fig:ALVE_9}(d).
The adjugate vector of $\mathcal{N}^{(9)}$ is not the zero-vector for any choice of non-vanishing weights, such that the kernel of $A(\mathcal{N}^{(9)})$ is one-dimensional for all choices of weights; see Section~\ref{sec:main_adjugate_vector}.
The network topology of $\mathcal{N}^{(9)}$ determines the signs of all summands in all entries and components of the adjugate vector~\eqref{eq:adjugate_vector_9}. Because all these summands have the same sign, the kernel is strictly positive for all choices of weights and, thus, the network topology $\mathcal{N}^{(9)}$ is a coexistence network.

In summary, the decomposition of the edge-set into edges from the cycle and interior edges, $\mathcal{N}^{(9)} = E_{\text{cycle}}^{(9)}\cup E_{\text{in}}^{(9)}$, leads to a separation of all near-perfect matchings into two sets. The first set consists of all near-perfect matchings containing only edges of the Hamiltonian cycle $E_{\text{cycle}}$.
These near-perfect matchings are identical to the near-perfect matchings of a directed cycle of size $S$.
Analogous to the adjugate vector of odd cycles, the signs of these near-perfect matchings alternate such that the corresponding contributions to the adjugate vector~\eqref{eq:adjugate_vector} have the same sign ($\mathrm{sign} (\mathrm{Pf}(A_{\hat{i}})) = (-1)^{i-j} \mathrm{sign} (\mathrm{Pf}(A_{\hat{j}}))$), see Section~\ref{sec:topological_robustness}.
The second set consists of all near-perfect matchings containing one or several interior edges $E_{\text{in}}$. The cycle condition~\eqref{eq:conditions1} and the crossing condition~\eqref{eq:conditions2} together ensure that the signs of all near-perfect matchings excluding the same node are identical. Thus, $\mathcal{N}^{(9)}$ is a coexistence network.

\textit{Edge-wise decomposition of the adjugate vector.} 
To establish the coexistence conditions~\eqref{eq:conditions1} and~\eqref{eq:conditions2} for coexistence networks, it is also insightful to partition the adjugate vector into contributions arising from the presence of interior edges.
For example, the adjugate vector of the network $\mathcal{N}^{(9)}$ in Figure~\ref{fig:ALVE_9}(a) can be written as: 
\begin{widetext}
\begin{align}\nonumber
\vec{r} &= \vec{r}_\text{cycle} + \vec{r}_\text{in}\ ,\\\nonumber
&= \vec{r}_\text{cycle}+ \vec{r} _{2\to 7} + \vec{r}_{4 \to 7} + \vec{r}_{9 \to 3} + \vec{r}_{4\to 7,9\to 3}\ ,\\\label{eq:adjugate_vector_9}
&=
\begin{pmatrix}
a_{32}a_{54}a_{76}a_{98}\\
a_{43}a_{65}a_{87}a_{19}\\
a_{21}a_{54}a_{76}a_{98}\\
a_{32}a_{65}a_{87}a_{19}\\
a_{21}a_{43}a_{76}a_{98}\\
a_{32}a_{54}a_{87}a_{19}\\
a_{21}a_{43}a_{65}a_{98}\\
a_{32}a_{54}a_{76}a_{19}\\
a_{21}a_{43}a_{65}a_{87}
\end{pmatrix}
+
\begin{pmatrix}
a_{72}a_{43}a_{65}a_{98}\\
0\\
0\\
0\\
0\\
0\\
0\\
a_{72}a_{43}a_{65}a_{19}\\
0
\end{pmatrix}
+
\begin{pmatrix}
a_{32} a_{74}a_{65}a_{98}\\
0\\
a_{21}a_{74}a_{65}a_{98}\\
0\\
0\\
0\\
0\\
a_{32} a_{74} a_{65} a_{19}\\
0
\end{pmatrix}
+
\begin{pmatrix}
0\\
0\\
0\\
a_{21}a_{65}a_{87}a_{39}\\
0\\
a_{21}a_{54}a_{87}a_{39}\\
0\\
a_{21} a_{54}a_{76}a_{39}\\
0
\end{pmatrix}
+
\begin{pmatrix}
0\\
0\\
0\\
0\\
0\\
0\\
0\\
a_{21} a_{74} a_{65} a_{39} \\
0
\end{pmatrix}\ .
\end{align}
\end{widetext}
The contribution of the Hamiltonian cycle to the adjugate vector is denoted as $\vec{r}_\text{cycle}$, which is the adjugate vector of a directed cycle of 9 nodes in the ascending labeling. 
The contributions from near-perfect matchings of one interior edge and edges from the cycle are denoted as $\vec{r}_{2\to 7}, \vec{r}_{4 \to 7}$ and $\vec{r}_{9\to 3}$. The remaining contribution $\vec{r}_{4 \to 7, 9 \to 3}$ arises because these two non-crossing interior edges occur together in a near-perfect matching. 
$\vec{r}_\text{in}$ summarizes all contributions to the adjugate vector that stem from near-perfect matchings and include one or several interior edges.
In this edge-wise notation, the contribution of a single interior edge to the adjugate vector can be suitably discussed: whether it creates one or several near-perfect matchings, whether these matchings arise through combinations with edges of the Hamiltonian cycle only or in combination with further interior edges, and whether their sign agrees with the near-perfect matching stemming from the cycle.

\textit{Necessity of conditions.}
As demonstrated by the previous example, every Hamiltonian network in which all internal edges fulfill the coexistence conditions~\eqref{eq:conditions1} and~\eqref{eq:conditions2} is a coexistence network.
Furthermore, the coexistence conditions are not only sufficient, but also necessary to obtain a coexistence network. 
Reversing the direction of an edge changes the signs of the respective summands occurring in the adjugate vector because of an additional transposition in the permutation.
Thus, every Hamiltonian network not fulfilling conditions~\eqref{eq:conditions1} and~\eqref{eq:conditions2} is not a coexistence network; see 
Section~S4 of the Supplementary Material \cite{Supplement} for details.

\begin{figure*}[hptb!]
\centering
\includegraphics[width=0.95\textwidth]{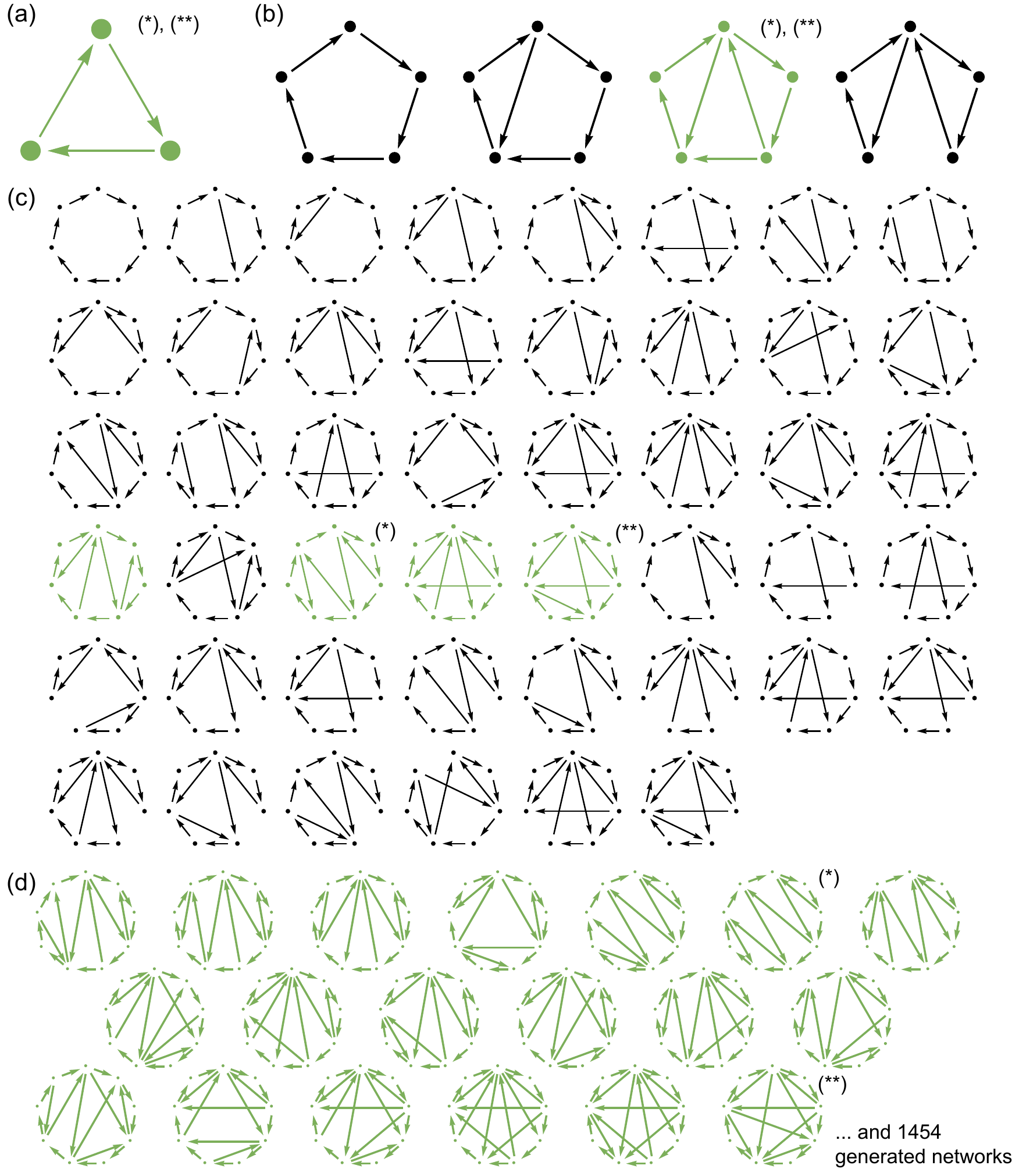}
\caption{(Color online)
\textbf{All coexistence networks with up to 9 nodes.}
We determined all coexistence networks for up to $S\leq 9$ by establishing all Hamiltonian coexistence networks via the graph-theoretical coexistence conditions~\eqref{eq:conditions1} and~\eqref{eq:conditions2} and deleting suitable edges; see Section~\ref{sec:numerics}.
(a)-(c)~All coexistence networks for $S=3, 5$, and 7 nodes. Coexistence networks are, for example, cycles, concatenations of smaller coexistence networks, and so-called generating coexistence networks (green color). Generating coexistence networks have a saturated number of edges: Upon adding any further edge to their network topology, they are no longer coexistence networks.
(*) denotes specific triangulations of cycles that are discussed in Section~\ref{sec:triangulations}. These are dilute networks, but the total number of near-perfect matchings grows exponentially fast with the number of nodes $S$ at a rate characterized by the golden ratio $\varphi = 1.6180...$; see Equation~\eqref{eq:growth_triangulation}.
(**) denotes cycles with complete subnetworks; see Section~\ref{sec:complete_subnetworks}. These networks are dense, but the total number of near-perfect matchings grows only polynomially as $\sim S^3$.
(d)~Generating coexistence networks with 9 nodes. Upon deleting suitable edges from these generating coexistence networks, all other 1454 coexistence networks for $S\leq 9$ are generated.
}
\label{fig:ListofCoexNW}
\end{figure*}

\subsection{Non-Hamiltonian coexistence networks}\label{sec:non-hamiltonian}

Non-Hamiltonian coexistence networks can be generated by deleting suitable links from Hamiltonian coexistence networks. To illustrate this approach, consider the network topology that is obtained by deleting the cycle edge $(2,3)$ from the above-studied Hamiltonian network $\mathcal{N}^{(9)}$ (equivalently, by setting $a_{32} = 0$ in $A(\mathcal{N}^{(9)})$). 
The resulting network $\mathcal{N}^{(9)}\backslash\{(2,3)\}$ is not Hamiltonian because no directed cycle passes through all of the network's nodes.
The adjugate vector of this non-Hamiltonian network is obtained from the adjugate vector~\eqref{eq:adjugate_vector_9} by setting all entries $a_{32} = 0$. 

Even though six previously existing near-perfect matchings vanish upon deleting the edge $(2,3)$ (four near-perfect matchings from the cycle, and two involving the interior edge $(4,7)$), all components of the resulting adjugate vector are different from 0 and all summands in all components have the same sign. 
In other words, the kernel of the corresponding antisymmetric matrix is still strictly positive for all choices of weights due to the contributions from the previously interior edges. 
Thus, the non-Hamiltonian network topology $\mathcal{N}^{(9)}\backslash\{(2,3)\}$ is a coexistence network. 

In general, deleting edges from a Hamiltonian coexistence network can only decrease the number of its near-perfect matchings.
Note that the resulting network does not have to be factor-critical or strongly connected (for example, upon deleting the edge $(5,6)$ from $\mathcal{N}^{(9)}$ $A_{\hat{4}}$ does not have a perfect matching).
However, as long as the network remains factor-critical upon removing edges, it remains a coexistence network: Removing edges from the network leads to removing summands in the adjugate vector. On the other hand, factor-criticality of the remaining network ensures that for every component at least one perfect matching and, thus, one summand in every component of the adjugate vector, remains.
Potentially non-Hamiltonian coexistence networks arise if edges are deleted from the defining Hamiltonian cycle of a Hamiltonian coexistence network as illustrated for the network topology $\mathcal{N}^{(9)}\backslash\{(2,3)\}$.
The smallest non-Hamiltonian coexistence network has five nodes; see Figure~\ref{fig:ListofCoexNW}(b). It is constituted of two 3-cycles that are trivially concatenated at one node. 
The smallest nontrivial non-Hamiltonian coexistence networks, which are \textit{not} obtained by concatenating smaller coexistence networks at single nodes, have seven nodes; see Figure~\ref{fig:ListofCoexNW}(c).

\subsection{All coexistence networks with up to 9 nodes}
\label{sec:numerics}

\textit{Numerical methods.} 
In order to support our theoretical findings from above, we numerically determined all coexistence networks for up to 9 nodes with two different methods. In our first method, we used the coexistence conditions~\eqref{eq:conditions1} and~\eqref{eq:conditions2} to successively build Hamiltonian coexistence networks, and deleted suitable edges to generate both all Hamiltonian and also non-Hamiltonian coexistence networks; see 
Section~S5.a of the Supplementary Material \cite{Supplement} for details.
Through a second method, we determined all coexistence networks in an algebraic manner. This approach explicitly exploits the notion of the adjugate vector~\eqref{eq:adjugate_vector} of an antisymmetric matrix; see 
Section~S5.b of the Supplementary Material \cite{Supplement}.
Crucially, our numerical results confirm that both methods yield the same coexistence networks for up to $9$ states, which also numerically confirms the validity of the coexistence conditions for Hamiltonian coexistence networks. 

\textit{All coexistence networks for $S\leq 9$.} The obtained list of coexistence networks for three, five, and seven nodes are shown in Figure~\ref{fig:ListofCoexNW}(a)-(c). 
Coexistence networks depicted in green indicate so-called \textit{generating coexistence networks}. A generating coexistence network is Hamiltonian and has a saturated number of edges: Upon adding any further edge to this network topology, it is not a coexistence network any longer. 
In general, every Hamiltonian coexistence network can be generated from a generating coexistence network by deleting suitable interior edges.
Our numerical enumerations show that for $S \leq 9$ also all non-Hamiltonian coexistence networks can be created from generating coexistence networks by deleting suitable interior and cycle edges.
For $S=7$ nodes, for example, four generating coexistence networks exist; see Figure~\ref{fig:ListofCoexNW}(c). All other Hamiltonian coexistence networks as well as all non-Hamiltonian are obtained from the four generating coexistence networks by deleting suitable edges. The specific form of two of these generating coexistence networks is further discussed in the next Section~\ref{sec:examples}.
Because of the large number of coexistence networks for $S=9$ (in total 1473 coexistence networks), only the generating coexistence networks are depicted in Figure~\ref{fig:ListofCoexNW}(d), from which again all coexistence networks are obtained. 
Whether also for $S\geq 10$ all non-Hamiltonian coexistence networks can be created from generating coexistence networks remains open at present.
Furthermore, it is an interesting question to us, whether all coexistence networks can be obtained by concatenating and fusing Hamiltonian coexistence networks of smaller size.

\textit{Do coexistence networks with a degenerate kernel exist?} 
Our numerical simulations of coexistence networks with $S\leq 9$ nodes did not yield any coexistence network with a kernel dimension other than $\mathrm{dim}(\mathrm{Ker}(A))=1$.
In other words, all coexistence networks, which we identified thus far, are odd-sized and have a strictly positive adjugate vector~\eqref{eq:adjugate_vector}. 
In 
Section~S3 of the Supplementary Material \cite{Supplement}, we show that coexistence networks with a two-dimensional kernel do not exist.
Whether coexistence networks with a degenerate kernel ($\mathrm{dim}(\mathrm{Ker}(A))\geq 2$) exist, remains an open question to us at present. 

\section{Specific generating coexistence networks }
\label{sec:examples}

\begin{figure}[th!]
\centering
\includegraphics{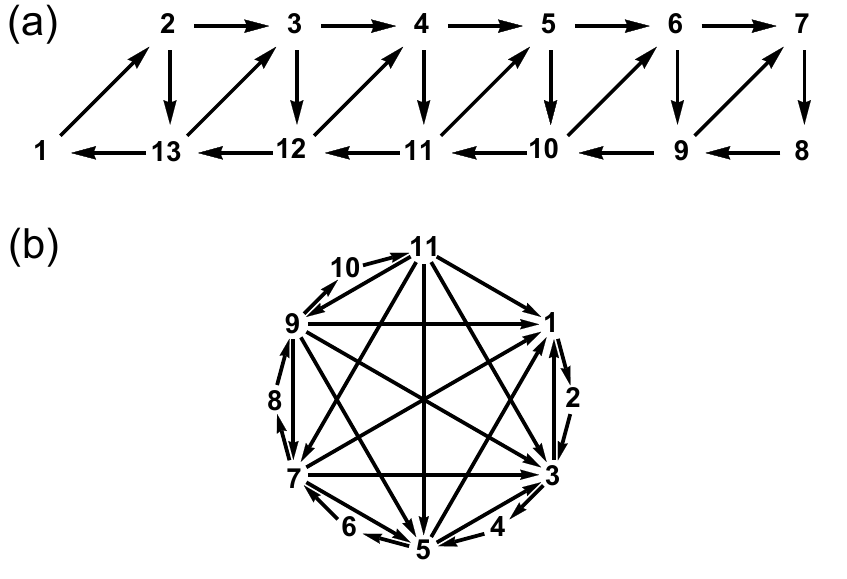}
\caption{
\textbf{Specific generating coexistence networks.}
(a) Triangulation of a cycle of size $S=13$. The orientation of the interior edges is chosen such that the network is a generating coexistence network. 
The adjugate vector is obtained as $\vec{r}^{(13)}= (13, 8, 5, 6, 6, 5, 8, 13, 8, 10, 9, 10, 8)$ if unit weights are chosen. 
Each entry equals the number of near-perfect matchings if the corresponding node is deleted from the network.
The ratio of its first two entries converges to the golden ratio $\varphi = 1.6180...$ as $S = 2n-1 \to\infty$. 
The total number of near-perfect matchings in the network is obtained from the sum over all entries of this adjugate vector and grows exponentially fast at a rate characterized by $\varphi$; see Equation~\eqref{eq:growth_triangulation}.
(b) Cycle of size $S=11$ with a complete subnetwork on the odd nodes. The adjugate vector for unit weights is obtained as $\vec{r}^{(11)}= (1, 5  , 1, 8, 1,9 , 1,8 , 1,5 , 1)$.
Such networks are dense ($1/4$ of all possible edges are realized as $S\gg 1$, as opposed to the triangulations for which only $\sim 4/S$ of all possible edges are realized). 
Even though these networks are dense, the number of near-perfect matchings grows only polynomially with the system size $S$.}
\label{fig:coexistence_examples}
\end{figure}

In the following, we discuss two examples of classes of generating coexistence networks.
These two exemplary classes are chosen because of their simple topological structure that can be constructed for arbitrary odd size. 
Furthermore, they illustrate the importance of topology for both the complexity and diversity of coexistence networks. Applications of these insights are further discussed in Section~\ref{sec:discussion}.
The first class of coexistence networks comprises specific triangulations of odd cycles; see Figure~\ref{fig:coexistence_examples}(a).
 As an application of the correspondence between the adjugate vector and the steady state concentrations of the ALVE~\eqref{eq:ALVE}, we propose a protocol to dynamically measure the golden ratio and the Fibonacci numbers by using these triangulations.
The second class comprises coexistence networks, which are fully connected on the subnetwork of all odd nodes; see Figure~\ref{fig:coexistence_examples}(b).
Applications of these and other coexistence networks are discussed in Section~\ref{sec:discussion}.

%

\subsection{Triangulations of cycles}
\label{sec:triangulations}
\textit{Triangulations of cycles are dilute networks.} 
A triangulation of a cycle is created by adding the maximal amount of internal edges such that no crossing edges occur. That is, the cycle is divided into triangles.
Every triangulation of an odd-sized cycle can be oriented to be a Hamiltonian coexistence network because interior edges in a triangulation do not cross each other and, thus, can always be oriented to fulfill the cycle condition~\eqref{eq:conditions1}. 
Here, we consider a specific class of triangulations of odd-sized cycles, which give rise to generating coexistence networks; see Figure~\ref{fig:coexistence_examples}(a) and the networks indicated with $(\text{*})$ in Figure~\ref{fig:ListofCoexNW}. 
These triangulations are created by adding to a cycle of odd size $S$ the ascending edges $(2,S), (3,S-1),(4,S-2),\dots ,\left((S-1)/2 ,(S-1)/2+3 \right)$ and the descending edges $(S,3),(S-1,4), \dots , \left( (S-1)/2+3, (S-1)/2+1\right)$. 
In other words, this triangulation is obtained from merging the two directed paths $S\to 1\to 2\to 3 \dots \to S-1 \to S $ (in total, $S$ edges that form the cycle) and $2\to S\to 3\to S-1 \to \dots ... (S-1)/2 \to (S-1)/2+3\to (S-1)/2+1$ (in total, $S-3$ interior edges that triangulate the cycle).
Note that the total number of edges in a triangulation grows as $\sim 2S$ for $S\gg 1$. Compared to a complete network of $S$ nodes (with $\binom{S}{2}\sim S^2/2$ edges), only $\sim 4/S$ of the possible edges are realized in a triangulation for $S\gg 1$. 
Heuristically speaking, triangulations are dilute (or sparse) network topologies.


\textit{These triangulations are generating coexistence networks.} 
Triangulations that are built in the above manner are coexistence networks because they are Hamiltonian networks, and all of their interior edges fulfill the coexistence conditions. The cycle condition~\eqref{eq:conditions1} is fulfilled because every interior edge does not create any even-sized cycle, and the crossing condition~\eqref{eq:conditions2} is trivially fulfilled because no two interior edges cross each other.
A network topology, in which an arbitrary interior edge is added to this triangulation, will not be a coexistence network any longer, which can be seen as follows. 
For every node $i$ of the network, there is a neighboring node ($i+1$ or $i-1$) that is both starting and end point of two different interior edges. 
Because of the way in which these triangulations are created, every additional interior edge starting or ending in $i$ crosses at least one neighboring interior edge such that the distance between starting and end points is 1. 
Therefore, the crossing condition is violated in triangulations with an additional interior edge.
Taken together, the presented triangulations of odd cycles are generating coexistence networks.

\textit{Counting the total number of near-perfect matchings by choosing unit rates.}
It is worth calculating the adjugate vector for these triangulations explicitly for unit weights. 
Upon setting all weights to 1 in a coexistence network, the $i$th component of the adjugate vector $r_i$ counts the number of near-perfect matchings for node~$i$, that is, the number of perfect matchings when node $i$ is removed from the coexistence network; see also Figure~\ref{fig:dimer_problem} for an illustration.
Because the constructed triangulations are coexistence networks, the adjugate vector is strictly positive for all choices of weights.
With unit rates, the sum over all entries of the adjugate vector, $|\sum_i r_i|$, counts the total number of near-perfect matchings in the coexistence network. 
Note that this procedure to count the number of near-perfect matchings can be applied to any undirected graph for which an orientation as a coexistence network can be found. We refer to such an orientation as a Pfaffian orientation of an odd-sized graph; see Section~\ref{sec:Pfaffian_orientation} for a detailed discussion of this application in the context of the dimer problem in statistical physics.

\textit{Number of near-perfect matchings for triangulations is characterized by the golden ratio.} We found that the adjugate vector of the triangulation of a cycle of odd size $S=2n-1$ ($n=2, 3, \dots$) with chosen unit weights is analytically computed as (see 
Section~S6.a of the Supplementary Material \cite{Supplement} for details):
\begin{align}\label{eq:triangulation_adjugate_vector}
	r_i =
\begin{cases} 
      F(n)& \text{, for } i=1\ , \\
      F(n+1-i)F(i-1) & \text{, for } i=2, \dots, n\ ,\\
      F(2n+1-i)F(i-n) & \text{, for } i=n+1, \dots, 2n-1\ ,
   \end{cases}
\end{align}
where $F(k)$ denotes the $k$th Fibonacci number with $F(0)=0, F(1)=1$, and $F(k+1)=F(k)+F(k-1)$. For example, the adjugate vector of the triangulation of the cycle of size $S=13$ ($n=7$) is obtained as $\vec{r}^{(13)}= (13, 8, 5, 6, 6, 5, 8, 13, 8, 10, 9, 10, 8)$. The Fibonacci numbers arise in this context because of the iterative decomposition of the triangulations into the so-called ladder graphs~\cite{Grimaldi12}; see 
Section~S6.a of the Supplementary Material \cite{Supplement}.
The total number of near-perfect matchings $\#(\text{npm})$ for the triangulation of an odd-sized cycle is calculated (by applying the convolution expansion for Fibonacci numbers~\cite{Grimaldi12, WeissteinFibonacci}) as:
\begin{align}
\label{eq:number_npm_triangulation}
	\#(\text{npm}) 
	= nF(n)\frac{1}{5}\left(3\frac{F(n+1)}{F(n)}+1+\frac{1}{n}\right)\ ,
\end{align}
which grows asymptotically as:
\begin{align}\label{eq:growth_triangulation}
	\#(\text{npm}) \sim \frac{3\varphi+1}{5\sqrt{5}}n \mathrm{e}^{\ln\varphi\cdot n}\quad \text{, for }n\gg 1.
\end{align}
Here, $\varphi = 1/2\left(1+\sqrt{5}\right)$ denotes the golden ratio. Note that the ratio of two consecutive Fibonacci numbers converges to the golden ratio ($F(n+1)/F(n)\to \varphi$ and $F(n)\sim \varphi^n/\sqrt{5}$ as $n\to \infty$).
In other words, the total number of near-perfect matchings grows exponentially fast with the system size $S = 2n-1$ at a rate determined by the golden ratio $\varphi$; see Section~\ref{sec:Pfaffian_orientation} for discussion.

\textit{Dynamical measurement of the Fibonacci numbers and the golden ratio.}
Interestingly, these triangulations of cycles of odd size $S=2n-1$ suggest a recipe to \textit{dynamically measure} the Fibonacci numbers and the golden ratio that we present in the following.
Recall from Section~\ref{sec:ALVE_long_time} that if the kernel of $A$ is one-dimensional, the adjugate vector determines the kernel element and, after normalization, equals the steady state concentrations $\langle \vec{x} \rangle_{t} = \frac{1}{t}\int_0^t \mathrm{d}s\ \vec{x}(s)$ in the ALVE~\eqref{eq:ALVE} as $t\to \infty$. 
Therefore, after normalization, the entries of the adjugate vector~\eqref{eq:triangulation_adjugate_vector} denote the steady state concentrations of the evolutionary zero-sum game that is defined by the triangulation of an odd-sized cycle as $\langle x_i \rangle_\infty \coloneqq \text{lim}_{t\to \infty}\langle x_i\rangle_t = r_i/\#(\text{npm})$.
Note also that, because all entries $r_i$ of the adjugate vector~\eqref{eq:triangulation_adjugate_vector} scale as $r_i\sim \mathcal{O}(\varphi^n)$ or $\sim \mathcal{O}(\varphi^{n+1})$ as $n\gg 1$ for all $i$, the trajectory defined by the ALVE~\eqref{eq:ALVE} remains in the center of the ($S-1$)-simplex $\Delta_{S-1}$.

As an application of the correspondence between the adjugate vector and the steady state concentrations, one may carry out the following protocol to dynamically measure the golden ratio and the Fibonacci numbers:
\begin{enumerate}
\item Pick a number $n=2, 3, \dots$. 
\item Construct the triangulation of the cycle $\mathcal{N}_\text{triang}$ of (odd) size $S=2n-1$ by merging the two directed paths $S\to 1\to 2\to 3 \dots \to S-1 \to S $ and $2\to S\to 3\to S-1 \to \dots ... (S-1)/2 \to (S-1)/2+3\to (S-1)/2+1$ as illustrated in Figure~\ref{fig:coexistence_examples}. Choose unit weights for every edge. $A(\mathcal{N}_\text{triang})$ denotes the antisymmetric adjacency matrix of the constructed weighted network. 
\item Simulate the evolutionary zero-sum game defined by $A(\mathcal{N}_\text{triang})$, that is, numerically integrate the ALVE~\eqref{eq:ALVE} specified by $A(\mathcal{N}_\text{triang})$.
\item Measure the long-time average of all state concentrations $\langle\vec{x}\rangle_t = \frac{1}{t}\int_0^t \mathrm{d}s\ \vec{x}(s)$ for $t\gg 1$.
\item Compute approximate values of:
\begin{itemize}
	\item the golden ratio $\varphi$ by computing the ratio $\langle x_1 \rangle_t/\langle x_2 \rangle_t$, which converges to $F(n)/F(n-1)$ as $t\to\infty$ (and $F(n)/F(n-1)\to \varphi$ as $n\to \infty)$;
	\item the Fibonacci number $F(k)$ by computing the ratio $\prod_{l=1}^k\langle x_{n+l} \rangle_t/\langle x_l \rangle_t$ for $k=1, 2, \dots, n-1$, which converges to $F(k)$ as $t\to \infty$. $F(n)$ is obtained by computing the ratio $\langle x_{n+1} \rangle_t/\langle x_n \rangle_t\prod_{l=1}^k\langle x_{n+l} \rangle_t/\langle x_l \rangle_t$ as $t\to\infty$. 
	
	The Fibonacci numbers can also be computed more efficiently by successively computing the ratios $\langle x_{n+1} \rangle_t/\langle x_1 \rangle_t$ (converging to $F(1)$ as $t\to \infty$), $F(1)\langle x_{n+2} \rangle_t/\langle x_2 \rangle_t $ (converging to $F(2)$ as $t\to \infty$), continuing with $F(k-1)\langle x_{n+k} \rangle_t/\langle x_k \rangle_t $ (converging to $F(k)$ as $t\to \infty$) for $k=1,2, \dots, n-1$, and finally computing $F(n-1)\langle x_{n+1} \rangle_t/\langle x_n \rangle_t $ (converging to $F(n)$ as $t\to \infty$).
\end{itemize}
\end{enumerate}
Even though we are not aware of any real-world application of the above protocol and the procedure is not numerically efficient, this measurement of the Fibonacci numbers and the golden ratio with a dynamical system is an interesting number-theoretical observation.

\subsection{Cycles with complete subnetworks}
\label{sec:complete_subnetworks}

\textit{Cycles with complete subnetworks are dense networks.}
Odd-sized cycles with complete subnetworks on the odd nodes are generating coexistence networks as well; see Figure~\ref{fig:coexistence_examples}(b) and the networks indicated with $(\text{**})$ in Figure \ref{fig:ListofCoexNW}.
These network topologies are built by starting from an odd-sized cycle of size $S = 2n-1$ in ascending labeling $S=2n-1 \to1\to2\to\dots\to 2n-2\to 2n-1 $ and adding descending edges between all pairs of odd nodes, that is, $E_\text{in} =\{(2n-1, 2n-3), (2n-1, 2n-5), \dots (2n-1, 1), (2n-3, 2n-5), \dots, (2n-3, 1),\dots, (3, 1) \}$.
Thus, the number of interior edges is given by $\binom{n}{2}-1$ (the complete network on the odd nodes with $\binom{n}{2}$ edges minus the cycle edge $(S, 1)$). 
The total number of edges in this network topology grows as $(S-1)(S+9)/8$ such that a ratio of $(1+9/S)/4$ of the possible edges are realized compared with a complete network in which all possible edges are realized. This ratio scales as $\sim 1/4$ for $S\gg 1$. Thus, these network topologies are dense network topologies (that is, with a macroscopic number of edges).

\textit{Cycles with complete subnetworks are generating coexistence networks.}
These cycles with complete subnetworks on the odd nodes are coexistence networks because they are Hamiltonian networks, and the coexistence conditions are fulfilled by construction: Every interior edge is descending and the difference between start and end node is even (the cycle condition~\eqref{eq:conditions1} is fulfilled), and the minimal distance between the starting nodes and end nodes of two crossing edges is always even (the crossing condition~\eqref{eq:conditions2} is fulfilled).
Upon adding another arbitrary edge, the network topology is not a coexistence network any longer, which can be seen as follows. Any further edge $(k,l)$ needs to either start or end in an even node (or both). 
Therefore, the added edge would cross the interior edge connecting the two neighboring odd nodes (connecting either nodes $k-1$ and $k+1$, or $l-1$ and $l+1$). Thus, the minimal distance between the start and end nodes of the added edge and the crossing edge is always 1, which is not even. 
In other words, every additional interior edge creates a network in which the crossing condition is violated. 
In total, the cycle with a complete subnetwork on the odd nodes is a generating coexistence network.

\textit{Number of near-perfect matchings for cycles with complete subnetworks grows polynomially slowly.} 
To further characterize cycles with complete subnetworks of size $S=2n-1$ ($n=2, 3, \dots$), we computed the total number of near-perfect matchings by setting all weights equal to 1 and calculating the adjugate vector~\eqref{eq:adjugate_vector} as described before. 
We found that the adjugate vector is obtained as (see 
Section~S6.b of the Supplementary Material \cite{Supplement} for details):
\begin{align}
\label{eq:adjugate_vector_CompleteSubnetwork}
	r_i =
\begin{cases} 
      1 & \text{, for } i \text{ odd}\ , \\
      \frac{i(n-i/2)}{2} & \text{, for } i \text{ even}\ .
   \end{cases}
\end{align}
For example, the adjugate vector of the network topology of size $S=11$ ($n=6$) is obtained as $\vec{r}^{(11)}= (1, 5  , 1, 8, 1,9 , 1,8 , 1,5 , 1)$; see Figure~\ref{fig:coexistence_examples}.
The total number of near-perfect matchings is computed as:
\begin{align}
	\#(\text{npm}) 
	= n + \frac{1}{6}(n-1)n(n+1)  \ ,
\end{align}
which grows polynomially as $\sim n^3/6$ for $n\gg 1$.

\textit{Summary.}
In this section we have investigated specific triangulations of odd-sized cycles that are dilute networks (with $\sim 4/S$ edges realized of all possible edges), but their number of near-perfect matchings grows exponentially fast with $S$ at a rate characterized by the golden ratio.
Additionally, we have shown that, even though odd-sized cycles with complete subnetworks on the odd nodes are dense network topologies ($1/4$ of all possible edges are realized as $S\gg 1$), the number of their near-perfect matchings grows only polynomially $\sim S^3$ with the system size.
This different scaling behavior between the two classes of network topologies underlines the importance of the topology of a network. The number of near-perfect matchings of a network, and thus the structure of the adjugate vector, mainly depends on the arrangement of internal edges, but only secondary on their number. Topology matters.

\section{Applications of coexistence networks}
\label{sec:discussion}

In the following we present applications of our results on coexistence networks in different contexts. We outline applications of coexistence networks for the ALVE, which was our initial motivation of this work. These applications include topologically robust zero-sum games in evolutionary game theory and topologically robust quantum networks for non-interacting bosons in driven-dissipative systems. 
Furthermore, our results on coexistence networks find also applications for symmetric zero-sum games in the field of game theory and for the dimer problem for odd-sized graphs in statistical physics.

\subsection{The ALVE and coexistence networks: Topologically robust coexistence in evolutionary game theory and driven-dissipative bosonic systems} 
\label{sec:discussion_ALVE}

\textit{The ALVE has applications in physics and biology.}
The ALVE was originally studied in the context of population biology by Volterra \cite{Volterra1931, May1972}, and has recently gained attention (i) as the replicator equation for zero-sum games (in the field of evolutionary game theory) and (ii) as the equation of motion for non-interacting bosons in driven-dissipative systems (in the field of open quantum systems) \cite{Akin1984, Hofbauer1998, Knebel2015,Vorberg2013, Vorberg2015, Reichenbach2006, Volterra1931, Goel1971, May1972, Sadeghi2018, Intoy2013, Feldager2017, Frey2010, Dobrinevski2012, Berr2009}.
Furthermore, the ALVE occurs in the fields of plasma physics and chemical kinetics as summarized in reference~\cite{Knebel2015}.
In the following, we outline applications of coexistence networks to evolutionary game theory and open quantum systems.

\textit{Replicator dynamics for symmetric zero-sum games.} 
In the context of evolutionary game theory, the ALVE is derived as the replicator equation of zero-sum games~\cite{Akin1984, Hofbauer1998, Knebel2015, Dobrinevski2012}. States correspond to pure strategies (labeled by $i = 1, \dots, S$) that are played by agents in a well-mixed population. Agents interact pairwisely with each other through a prescribed symmetric zero-sum game (antisymmetric payoff matrix $A$ such that the value of the game is zero) or, equivalently, by a weighted network.
The payoff gained or lost in each interaction translates to fitness and determines the rate at which agents of a certain strategy reproduce.
The ALVE describes the temporal evolution of the fraction of agents $x_i$ playing strategy $i$ in an infinitely large population. Depending on the entries of the payoff matrix, one typically observes the survival of certain strategies in the population and the extinction of others at long times~\cite{Chawanya2002, Knebel2013, Sadeghi2018, Intoy2013, Feldager2017}.
In other words, some of the strategies will not be played by the agents of the population at long times while other strategies survive and constitute the so-called evolutionary stable set of strategies \cite{Hofbauer1998}.

\textit{Topologically robust evolutionary zero-sum games.}
Our work was originally motivated by the observation of zero-sum games in evolutionary game theory for which all strategies coexist \textit{for all choices} of interaction strengths. The rock-paper-scissors zero-sum game~\cite{Hofbauer1998, Szabo2007, Frachebourg_2_1996, Reichenbach2006} and cycles with an odd number of strategies are coexistence networks~\cite{Zia2011,Durney2011}, and we asked whether other coexistence networks with a more complex interaction structure exist. 
Our results on how to determine these coexistence networks as described in Section~\ref{sec:coexistence_networks} are not trivial because strategies typically go extinct for some choice of interaction strengths~\cite{Knebel2013}.
Coexistence networks as determined in this work give rise to topologically robust zero-sum games in evolutionary game theory. Irrespective of the chosen interaction strengths, all strategies will be played in the population. No extinction can ever occur on these network topologies.

%

\textit{Condensation dynamics in driven-dissipative, bosonic systems.}
In the context of open quantum systems, the ALVE describes the condensation dynamics of non-interacting bosons in driven-dissipative systems~\cite{Vorberg2013, Knebel2015, Vorberg2015}.
In a theoretical model that was proposed only recently~\cite{Vorberg2013}, a system of non-interacting bosons is weakly coupled to a heat bath and driven by an external, time-periodic potential (a so-called Floquet system~\cite{Blumel1991,Kohler1997, Breuer2000}
On a coarse-grained time scale, the dynamics of this open quantum system becomes incoherent~\cite{Vorberg2013, Vorberg2015}. 
In other words, in this effective description the temporal evolution of such a driven-dissipative quantum system is captured by a classical stochastic process.
Each state $i = 1, \dots, S$ corresponds to a quantum Floquet state and the fraction of bosons in this state is given by the concentration~$x_i$.
Even though the bosons transition incoherently between the different quantum states, the transition rates still reflect the quantum statistics of the bosons: The more bosons occupy a quantum Floquet state, the higher is the rate for other bosons to jump into this state, reflecting the fact that bosons tend to congregate due to their indistinguishability. 
Furthermore, the differences of forward and backward jump rates between any two states are characterized by an antisymmetric matrix $A$, whose entries depend on microscopic properties of the system, the heat bath, and the coupling between the two.
Due to these dynamics, certain quantum Floquet states become condensates over time, that is, bosons congregate in a subset of the possible states, while other states become depleted. 
It was shown~\cite{Knebel2015} that the ALVE captures this condensation on the leading-order time scale. 
Whether a state becomes a condensate or a depleted state depends on the antisymmetric matrix $A$ alone.
This theoretical observation can be understood as a generalization of the Bose-Einstein condensation in thermodynamic equilibrium to a condensation of bosons in nonequilibrium~\cite{Vorberg2013}, which has stimulated further research recently~\cite{Choudhury2015, Leymann2017, Klembt2018, Schnell2018, Vidmar2015}. 

\textit{Topologically robust bosonic quantum networks.}
Coexistence networks as determined in this work give rise to topologically robust quantum networks, in which all states are condensates and no state becomes depleted, irrespective of how the individual jump rates are tuned.
However, coexistence networks cannot be complete networks (see the list of coexistence networks in Figure~\ref{fig:ListofCoexNW} and the networks in Figure~\ref{fig:coexistence_examples} for illustration). It is straightforward to show that some transitions between states need to be forbidden (or forward and backward jump rate need to be equal) in order to create a coexistence network.
Once it is possible to engineer the topology of such transition networks for non-interacting bosons in driven-dissipative systems, topologically robust quantum networks, on which none of the states becomes ever depleted, might be observable.

\subsection{Game theory and coexistence networks: Stability of optimal, totally mixed strategies in symmetric zero-sum games}
\label{sec:discussion_game_theory}

Our results on coexistence networks may also gain significance in the field of game theory for symmetric zero-sum games and so-called tournaments \cite{Fisher92,Brandl17,Brandt09,Bapat14} that we outline in the following.

\textit{Symmetric zero-sum games in game theory.} 
\textit{Tournaments} are typically introduced as symmetric two-player zero-sum games on fully connected networks of size~$S$ with unit payoff. In other words, all pairs of different nodes of the game's network (representing the $S$ actions that the two players pick from) are connected by a directed edge with weight $+1$ (representing the dominance relation between the actions; that is, the payoff matrix is antisymmetric, $A = -A^T$, and all off-diagonal entries are $\pm 1$) \cite{Fisher92,Brandl17,Brandt09,Bapat14}. 
Tournament games were introduced as a generalization of the rock-paper-scissors game with unit weights \cite{Fisher92}.
More generally, \textit{weighted tournaments} are symmetric zero-sum games on a fully connected network with arbitrary payoff (that is, the payoff matrix is antisymmetric and all off-diagonal entries are non-zero) \cite{DeDonder00}, while \textit{weak tournaments} are characterized by unit payoff on an arbitrary network topology (that is, the payoff matrix is antisymmetric, off-diagonal entries may be zero and non-zero entries are $\pm 1$) \cite{Brill16}.
If neither the game's underlying network topology nor the weights of the payoff are restricted, one simply refers to the game as a \textit{symmetric zero-sum game} characterized by the antisymmetric payoff matrix $A = -A^T\in \mathbb{R}^{S \times S}$. The results of our work apply to such symmetric zero-sum games.

\textit{Optimal strategies in symmetric zero-sum games.}
For symmetric zero-sum games, one is typically interested in so-called optimal sets of actions (or briefly an \textit{optimal strategy}), in which ``optimal'' may have different meanings depending on the context \cite{Brandt09,Aziz15}.
For example, an optimal set of actions may denote a probability vector ($\vec{p}^*\in \Delta_{S-1}$, that is, $\vec{p}^*\in \mathbb{R}^{S}$, $\vec{p}^*\geq 0$, and $\sum_{i=1}^S p_i = 1$), whose $i$th entry denotes the probability to play the $i$th action, and that maximizes the player's minimum expected payoff against all other sets of actions (a so-called \textit{mixed Nash equilibrium} of the symmetric game~\cite{Barron13,Gonzalez10}). 
For a symmetric zero-sum game, it is straightforward to show that a normalized positive kernel vector $\vec{p}^*\geq 0$ of the payoff matrix ($A\vec{p}^* = 0$) is an optimal strategy. 
A single action of this optimal strategy is referred to as essential if it is chosen with non-vanishing probability ($p^*_i>0$).
In other words, in an infinitely repeated game, it is reasonable to choose an essential action.
If all actions of the optimal strategy are essential ($\vec{p}^*> 0$), the optimal strategy is called \textit{totally mixed}~\cite{Brandl17}; that is, all actions are played. 
One central result obtained for odd-sized tournaments is that an optimal, totally mixed strategy always exists and is unique. In other words, there exists a unique way to optimize the player's payoff in odd-sized tournament games and all possible actions must be played \cite{Fisher92,Laffond93}. 
Recent research results further characterized optimal, totally mixed strategies and extended this concept to other types of games \cite{Michael06,DeDonder00,Kaplansky95,Brandt09,Brill16,Brandl17,Breton07,Roberts2006}.

\textit{Stability of optimal, totally mixed strategies in coexistence networks.}
Our results on coexistence networks contribute to this line of research by identifying the symmetric zero-sum games whose optimal, totally mixed strategies are stable.
An optimal, totally mixed strategy of a game is referred to as \textit{stable} if all actions remain essential for any change of the payoff on the game's network topology \cite{Arsham95}.
In other words, arbitrary changes of the payoff values do not change the fact that all actions are essential (even though their specific probability to be played may change).
In our work, we determined and characterized those antisymmetric matrices (defining the game's network topology) whose kernel remains one-dimensional and strictly positive for all choices of off-diagonal entries. 
Therefore, coexistence networks define symmetric zero-sum games for which the optimal strategy is totally mixed for any choice of payoff values; see Section~\ref{sec:coexistence_networks}. The list of coexistence networks in Figure~\ref{fig:ListofCoexNW} depicts the network topologies of such games for $S\leq 9$.

Note that, apart from the rock-paper-scissors game, none of the coexistence networks are complete graphs as already mentioned above. 
Instead, our discussion of coexistence networks in Section~\ref{sec:examples} exemplifies how symmetric zero-sum games with an arbitrary number of nodes can be constructed to have a unique optimal, totally mixed strategy for all choices of payoff values. 
These games include, for example, cycles with $S$ edges, triangulations of cycles with $\sim 2S$ edges, and cycles with complete subnetworks and $\sim \frac{1}{4} S^2$ edges; see~Sections~\ref{sec:topological_robustness} and~\ref{sec:examples}.
The stability of optimal strategies against arbitrary changes of payoff values in symmetric zero-sum games is, in general, an interesting topic for further research, for example, when the optimal strategy is not totally mixed or not unique. 

\subsection{Graph theory and coexistence networks: Pfaffian orientation and the dimer problem of odd-sized graphs}
\label{sec:Pfaffian_orientation}

\begin{figure}[t!]
\centering
\includegraphics{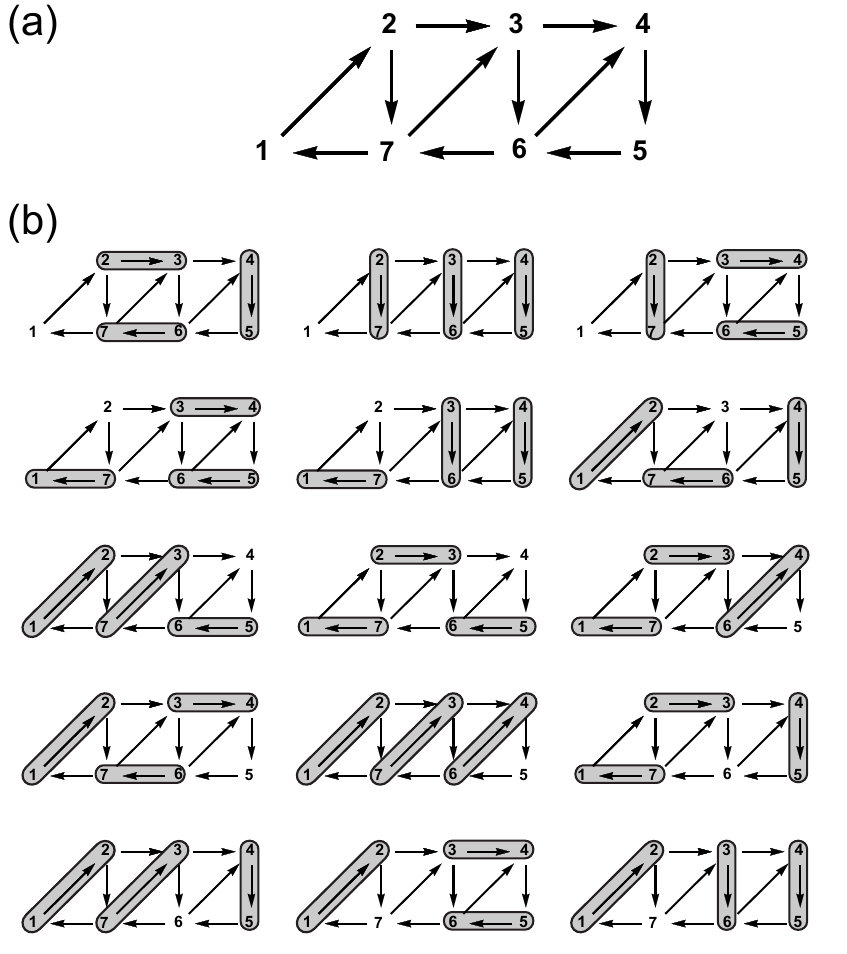}
\caption{
\textbf{The dimer problem for graphs of odd size.}
(a) What is the number of closest-packing configurations with dimers (so-called dimer coverings) for the triangulation of a cycle of size $S=7$? A closest-packing configuration is a covering of the graph that leaves only one node of the graph uncovered. Our results obtained in Section~\ref{sec:coexistence_networks} show that this question can be answered for all factor-critical coexistence networks.
(b) All possible dimer coverings of the triangulation in (a) are depicted. 
Each closest-packing configuration with one uncovered node corresponds to one near-perfect matching of the triangulation. In total, $15$ near-perfect matchings exist. 
The number of near-perfect matchings excluding node $i$ equals the $i$th component of the adjugate vector of the chosen directed graph (the Pfaffian orientation of the network topology) upon setting all weights to one. 
Here, the adjugate vector is obtained as $\vec{r}^{(7)}= (3,2,1,2,3,2,2)$ for unit weights; see Equation~\eqref{eq:triangulation_adjugate_vector}. The total number of dimer coverings for the triangulation of a cycle grows exponentially fast with the number of nodes $S$ at a rate characterized by the golden ratio $\varphi = 1.6180...$; see Section~\ref{sec:triangulations} for details.
}
\label{fig:dimer_problem}
\end{figure}

One interesting application of our results on coexistence networks relates to the so-called dimer problem in statistical physics: how many configurations exist that completely cover the edges of a lattice graph of even size with non-overlapping dimer molecules?
In graph theory and combinatorial mathematics, the dimer problem relates to counting the number of perfect matchings for a given even-sized graph, and motivated the introduction of the so-called Pfaffian orientation of even-sized graphs~\cite{Kasteleyn1961,Kasteleyn1963,Temperley1961,Fisher1961}.
As we explain in the following, our results obtained in this manuscript suggest a possibility to extend the concept of the Pfaffian orientation from even-sized to odd-sized graphs; see also Figure~\ref{fig:dimer_problem} for an illustration. This way, our results facilitate to study closed-packed dimer configurations on odd-sized graphs and may stimulate further research in this direction.

\textit{The dimer problem in statistical physics.}
Starting with the work of Kasteleyn, Fisher, and Temperley~\cite{Kasteleyn1961,Kasteleyn1963,Temperley1961,Fisher1961}, the graph-theoretical interpretation of the Pfaffian~\eqref{eq:Pfaffian_matchings} has been appreciated and applied in statistical physics to compute the entropy of dimer molecules adsorbed on lattices.
Such systems are typically defined by regular lattice graphs of even size on which dimer molecules cover the graph's vacancies. Every dimer molecule covers two connected nodes of the graph and dimers do not overlap. Of specific interest are closed-packing configurations (so-called dimer coverings), for which every node of the graph is covered exactly once by a dimer and the graph is completely filled. The partition function counts the total number of such dimer coverings and its computation is often referred to as the \textit{dimer problem}. The contribution of a dimer covering to the partition function may be weighted by introducing edge weights to the graph.

Counting the number of dimer coverings of regular lattice graphs was first motivated in statistical physics by the adsorption of dimer molecules on two-dimensional surfaces. Kasteleyn, Fisher, and co-workers~\cite{Kasteleyn1961,Kasteleyn1963,Fisher1966} also established a connection of the two-dimensional dimer problem to the two-dimensional Ising model. Notably, the dimer problem gained fresh attention through the work of Rokhsar and Kivelson~\cite{Rokhsar1988} to describe the so-called short-range resonating valence bond ground state in the field of superconductivity; see, for example, reference~\cite{Fendley2002} for a concise description of the physical background and on how the quantum dimer problem relates to the classical dimer problem described above.
All of these physical applications continue to stimulate the research of dimer statistics on lattices and networks in the fields of statistical physics and graph theory; see below.

\textit{Pfaffian orientation of even-sized graphs.}
From a mathematical point of view, counting the number of dimer configurations that cover the whole graph amounts to counting the number of perfect matchings of that graph.
As it was motivated in Section~\ref{sec:Pfaffian_graph}, the Pfaffian~\eqref{eq:Pfaffian_matchings} can be thought of as the sum of the signed perfect matchings of a graph. 
Therefore, to count the number of perfect matchings, it suffices to find an orientation of that graph (that is, assigning every edge a direction) such that every perfect matching has the same sign. 
Such an orientation of an even-sized graph is referred to as a \textit{Pfaffian orientation}~\cite{Lovasz86}.
Thus, upon assigning every directed edge the weight $+1$ on a Pfaffian orientation of the graph, the value of the Pfaffian of its antisymmetric adjacency matrix equals the number of perfect matchings. 
In other words, the Pfaffian of a graph's antisymmetric adjacency matrix in a chosen Pfaffian orientation equals the partition function for the dimer problem on that graph.
 
Two questions are of specific interest for the dimer problem: (i) Which graphs admit a Pfaffian orientation? (ii) If a Pfaffian orientation of a graph exists, how many perfect matchings exist on that graph?

Question (i). 
Kasteleyn showed that every \textit{planar graph} has a Pfaffian orientation~\cite{Kasteleyn1963}. A planar graph can be drawn in the two-dimensional plane such that no two edges intersect. 
To find a Pfaffian orientation of a planar graph, one can orient the graph's edges such that each face (regions bounded by the graph's edges) has an odd number of lines oriented clockwise. 
Such an orientation can be found in polynomial time~\cite{Kasteleyn1967}.
Ever since this result for planar graphs was established, progress has also been made for Pfaffian orientations of more general graphs~\cite{Little1974,Vazirani1989,Tesler2000,Norine2008b,Thomas2006}. For example, it was shown that a graph of even size has a Pfaffian orientation if and only if it can be drawn in the two-dimensional plane (possibly with crossings) such that every perfect matching intersects itself an even number of times~\cite{Norine2008b,Thomas2006}.

Question (ii). 
If a graph has a Pfaffian orientation, the number of perfect matchings can be efficiently calculated; see reference~\cite{Thomas2006} for a review and details on the computational complexity of this task. 
Analytical expressions for the partition function were first obtained for the square lattice by Kasteleyn, Temperley, and Fisher~\cite{Kasteleyn1961,Temperley1961,Fisher1961}, have been computed for other regular lattices~\cite{Wu1968,Wu2006,Fendley2002,Lu2010,Loh2008,Wu2008,Wang2007,Kenyon2000,Li2016,Li2015,Dong2013}, such as the honeycomb, triangular, and kagome lattice since then (see ~\cite{Wu2006} for a review), and for other geometries~\cite{Tesler2000,Yan2008,Lu2011} and scale-free networks~\cite{Zhang2015,Li2017}.


\textit{Pfaffian orientation and dimer problem of odd-sized, factor-critical graphs.} 
Our results on coexistence networks suggest that the above concept of a Pfaffian orientation can be generalized to odd-sized, factor-critical graphs.
For a given factor-critical graph of odd size one may consider closest-packing configurations with dimers (that is, a dimer covering), which leave only one node of the graph uncovered; see Figure~\ref{fig:dimer_problem}.
The total number of near-perfect matchings of that graph relates, as above, to the partition function of the dimer problem.

A natural generalization of the concept of a Pfaffian orientation from even-sized to odd-sized graphs is obtained in the following sense.
In Section~\ref{sec:coexistence_networks}, we constructed those networks $\mathcal{N}(A)$ for which (i) every subnetwork $\mathcal{N}(A_{\hat{i}})$ has a Pfaffian orientation for all $i$ and, moreover, (ii) the signs of the corresponding orientations fulfill the sign-condition $\mathrm{sign} (\mathrm{Pf}(A_{\hat{i}})) = (-1)^{i-j} \mathrm{sign} (\mathrm{Pf}(A_{\hat{j}}))$ for all $i$ and $j$. 
These two conditions originate from the notion of the adjugate vector $\vec{r}$~\eqref{eq:adjugate_vector} of an odd-sized antisymmetric matrix $A$. 
We emphasize that the sign-condition (ii) imposes a strong condition on the network topology: The signs of all summands in all near-perfect matchings of a network need to be aligned, and not just the sign of the summands of a single perfect matching as for the Pfaffian orientation of an even-sized graph. 
Thus, networks that fulfill conditions (i) and (ii) can be understood as odd-sized graphs with a Pfaffian orientation.
 
Factor-critical coexistence networks (see Section~\ref{sec:coexistence_networks}) are networks with a Pfaffian orientation in the above sense because perfect matchings (that is, closest-packing dimer configurations) exist for every node removed from the graph and all of their signs are aligned in the sense of the above conditions (i) and (ii). We note that all coexistence networks that we have found thus far are factor-critical; see Section~\ref{sec:numerics}.
The total number of near-perfect matchings can be computed by assigning unit weights to all edges of a factor-critical coexistence network and by computing the adjugate vector of the antisymmetric adjacency matrix. 
As was demonstrated in Section~\ref{sec:examples}, the $i$th component of the adjugate vector, $r_i$, then counts the number of perfect matchings when node $i$ is removed. The sum over all components of the adjugate vector, $\sum_i |r_i|$, counts the total number of near-perfect matchings in the factor-critical coexistence network.

As an example for a planar graph, we computed the number of near-perfect matchings for specific triangulations of an odd-sized cycle; see Section~\ref{sec:triangulations} and Figure~\ref{fig:coexistence_examples}(a). 
We showed that the number of near-perfect matchings~$\#\text{(npm)}$ grows as $\sim n\mathrm{e}^{\ln\varphi\cdot n}$ for $n\gg 1$. Consequently, the entropy of adsorption per dimer molecule on that triangulation of the cycle is given by:
\begin{align}
	s &= \lim_{D\to \infty} \frac{1}{D}\ln(\#\text{(npm)})\ ,\\
	& = \ln\varphi \approx 0.4812...\ .
\end{align}
Here, $D$ denotes the maximal number of dimers on the graph and is equal to $D=n-1=(S-1)/2$ in the notation of Section~\ref{sec:triangulations}; $\varphi = 1/2\left(1+\sqrt{5}\right)$ denotes the value of the golden ratio. Notably, the value of the entropy per dimer $s= \ln\varphi\approx 0.4812...$ for the triangulation of the cycle, which is an effectively one-dimensional lattice, lies above the value of the two-dimensional honeycomb lattice ($s\approx 0.3230$)~\cite{Wu1968}, but below the two-dimensional triangular lattice of even size ($s\approx 0.8571$)~\cite{Fendley2002}; see, for example, Table~1 in reference~\cite{Wu2006} for further comparison.

\textit{Outlook.} 
It will be interesting to extend our results on the Pfaffian orientation and the dimer problem of odd-sized graphs to two-dimensional lattices and non-planar graphs of odd size (see example in Section~\ref{sec:complete_subnetworks}), and to explore possible phase transitions that might occur upon choosing anisotropic dimer weights.

\section{Summary and conclusion}\label{sec:conclusion}

After having discussed applications of both coexistence networks and related concepts in Section~\ref{sec:discussion}, we provide a brief summary of our results and conclude.

\textit{Summary of coexistence networks.} 
In this work, we introduced the notion of coexistence networks, that is, networks which show coexistence of all states in the antisymmetric Lotka-Volterra equation (ALVE)~\eqref{eq:ALVE} as a consequence of the network topology alone.
We determined coexistence networks by mapping the question about the dynamical system of the ALVE to an algebraic question of the antisymmetric matrix $A$, which defines the interactions between the states.
By exploiting tools from graph theory related to antisymmetric matrices, we determined coexistence networks.

In detail, we showed that coexistence of all states in the ALVE is independent of the initial conditions and only depends on the antisymmetrix matrix $A$ defining the interactions between the states.
For matrices $A$ with a one-dimensional kernel, a non-trivial kernel element is computed as the adjugate vector via the Pfaffians of submatrices.
Exploiting the graph-theoretical interpretation of the Pfaffian, we characterized the kernel of $A$ in terms of network topological properties, namely by all near-perfect matchings of the network defined by $A$.
This approach enabled us, first, to construct Hamiltonian coexistence networks as generalizations of odd-sized, directed cycles, and, second, to generalize these results to construct also non-Hamiltonian coexistence networks. A numerical survey of coexistence networks with up to 9 nodes verifies our results; see Figure~\ref{fig:ListofCoexNW} for an overview.

\textit{Summary of applications of coexistence networks.} 
With respect to the ALVE, we outlined applications in the fields of evolutionary game theory as topologically robust zero-sum games. In the context of driven-dissipative systems of non-interacting bosons, topologically robust quantum networks might be an interesting application.
The algebraic results of our work include the characterization of antisymmetric matrices whose kernel remains strictly positive for all choices of weights that respect the sign structure of the matrix. 
We discussed the applications of our findings in the field of game theory for the stability of optimal, totally mixed strategies in symmetric zero-sum games. Furthermore, coexistence networks suggest the introduction of a Pfaffian orientation for odd-sized graphs to study the dimer problem on such graphs. 

\textit{Methodological approach and outlook.} 
Besides these specific applications of coexistence networks, we emphasize the methodological approach with which we studied the long-time behavior of the ALVE. 
With a suitable mapping from the dynamical system to an algebraic problem (via a Lyapunov function or a conserved quantity) and by solving the algebraic problem with a graph-theoretical approach, we characterized topologically robust coexistence in the ALVE. 
It might be possible to generalize this approach to study topologically stable attractors of general Lotka-Volterra systems on arbitrary networks~\cite{Haerter2016, Haerter2018, Tu2018} or in other population-dynamical models~\cite{Botta2014, West2018, Busiello2017}.

Overall, we believe that the results of this work will stimulate further research to investigate the interplay between interaction topologies and nonlinear dynamical systems. Ultimately, such studies will help to characterize the long-time behavior of nonequilibrium systems. 

\begin{acknowledgments}
It is a pleasure to thank Markus F. Weber, Felix Brandt, Florian Brandl, Isabella Graf, and Patrick Wilke for fruitful discussions.
This research was supported by the German Excellence Initiative via the program ``Nanosystems Initiative Munich'' (NIM). 
The authors declare no conflict of interest. P.M.G. and J.K. contributed equally to this work.
\end{acknowledgments}

%

\normalem
\providecommand{\noopsort}[1]{}\providecommand{\singleletter}[1]{#1}%

\clearpage
\newpage

\setcounter{page}{1}
\renewcommand\thesection{S\arabic{section}}
\renewcommand\thesubsection{\thesection.\alph{subsection}}
\renewcommand\theequation{S\arabic{equation}}
\renewcommand\thefigure{S\arabic{figure}}
\setcounter{section}{0}
\setcounter{equation}{0}
\setcounter{figure}{0}

\onecolumngrid

\section*{Supplementary Material to: \\ Topologically robust zero-sum games and Pfaffian orientation -- How network topology determines the long-time dynamics of the antisymmetric Lotka-Volterra equation}
In this Supplement we provide detailed derivations and calculations that were skipped in the main text, as well as additional information that is closely related to the material in the main text, but not essential for its understanding.  
In Section~\ref{sec:ALVE_Supplement} we provide further details on condensation in the antisymmetric Lotka-Volterra equation (ALVE). We start by completing the argument in 
Section~2.2 of the main text that coexistence of all states occurs in the ALVE if and only if the antisymmetric interaction matrix $A$ has a strictly positive kernel vector in Section~\ref{sec:ALVE_coexistence}. Then, we show that coexistence of all states in the ALVE is only possible in strongly connected networks.
In Section~\ref{sec:Pfaffian_appendix} we supplement further information on antisymmetric matrices and the Pfaffian.
We provide the combinatorial definition of the Pfaffian in Section~\ref{sec:Pfaffian_combinatorial}. The graph-theoretical definition of the Pfaffian agrees with the combinatorial definition, as we argue in Section~\ref{sec:Pfaffian_definitions_agree}. Some examples of the application of the combinatorial definition of the Pfaffian are provided in Section~\ref{sec:Pfaffian_examples}.
We introduce the adjugate matrix as a tool to calculate kernel elements of antisymmetric matrices with two-dimensional kernel in Section~\ref{sec:adjugate_vector_matrix}. 
With help of the graph theoretical definition of the Pfaffian we show in Section~\ref{sec:minimal_kernel_dimension} that the network topology constraints the minimal kernel dimension of an antisymmetric matrix.
In Section~\ref{sec:Proof2DimCoexNW} we prove that there are no coexistence networks with two dimensional kernel with help of the adjugate matrix.
In Section~\ref{sec:proof} we show that the coexistence conditions (Equations~(9) and~(10) in the main text)
are necessary and sufficient for coexistence networks. This proves that all Hamiltonian coexistence networks can be identified by these conditions.
In Secton~\ref{sec:numerical} we present the numerical methods used to generate the data presented in 
Section~4.3 of the main text.
We finish this Supplement with Section~\ref{sec:number_matchings}, where we present the calculation of the adjugate vectors given in 
Equations~(15) and~(18) of the main text.







\section{Coexistence in the ALVE}
\label{sec:ALVE_Supplement}

\subsection{Coexistence in the ALVE if the kernel of $A$ is strictly positive}
\label{sec:ALVE_coexistence}

Here we outline the steps to show that all states coexist in the ALVE (Equation~(1)
of the main text), if the kernel of the antisymmetric matrix $A$ is strictly positive. 
To show this statement, one may (i) exploit an algebraic property of antisymmetric matrices, and (ii) connect this algebraic property to the long-time dynamics of the ALVE via a suitable collective quantity that has the same form as the conserved quantity in Equation~(2).
Mathematical details can be found in Reference~\cite{Knebel2015}.\\

\textit{Part (i).} Given an antisymmetric matrix $A$ with real entries, there exist specific vectors $\vec{c}$, which we refer to as \textit{condensate vectors}, that fulfill the following properties for an unique index set $I\subseteq \{1, \dots, S\}$:
\begin{align}\label{eq:condensate_condition_1}
& c_i > 0 \text{ and } (A\vec{c})_i = 0\ ,\quad \text{ for all } i \in I \\\label{eq:condensate_condition_2}
& c_i = 0 \text{ and } (A\vec{c})_i < 0\ ,\quad \text{ for all } i \in \bar{I} = \{1,...,S\} \backslash I.
\end{align}
A proof of this algebraic property of antisymmetric matrices can be found in the book on linear programming theory by Kuhn and Tucker~\cite{Tucker1956}.
Condensate vectors can be thought of as generalized positive and attractive kernel elements: they are strictly positive kernel elements on the index set $I$ of the submatrix $A_I$ (matrix built from $A$ by only including rows and columns whose indices lie in $I$), and they are zero vectors on the index set $\bar{I}$ (that is, $\vec{c}_{\bar{I}} = 0$) and at the same time attractive in that $(A\vec{c})_{\bar{I}}<0$.
The existence of such condensate vectors is not intuitive at first sight and is, indeed, special to antisymmetric matrices~\cite{Tucker1956}.
There may exist linearly independent condensate vectors for a given antisymmetric matrix $A$ if the kernel of $A_I$ is degenerate. 
However, the index set of positive entries of $\vec{c}$, that is, the set $I = \{i \in \{1,...,S \} : c_i > 0 \}$, is unique to the antisymmetric matrix $A$. \\

\textit{Part (ii).} To connect condensate vectors with the long-time dynamics of the ALVE, a collective quantity is defined in the spirit of the conserved quantities~(2)
from above.
In particular, one defines the Kullback-Leibler divergence (or relative entropy) $D(\vec{c} || \vec{x})$ of an arbitrarily chosen condensate vector $\vec{c}$ of $A$ (fulfilling properties~\eqref{eq:condensate_condition_1} and \eqref{eq:condensate_condition_2}) to the state concentrations $\vec{x}(t)$ as:
\begin{align}\label{eq:ALVE_relative_entropy}
D(\vec{c} || \vec{x}(t)) = \sum_{\substack{i=1\\(c_i\neq 0)}}^S c_i \log \left( \frac{c_i}{x_i(t)}\right) = \sum_{i\in I} c_i \log \left( \frac{c_i}{x_i(t)}\right)\ .
\end{align}
Note the asymmetry in the definition of $D$: we consider the relative entropy of $\vec{c}$ to $\vec{x}$ and \textit{not} the relative entropy of $\vec{x}$ to $\vec{c}$ as one might naively try.
The collective quantity $D$ in Equation~\eqref{eq:ALVE_relative_entropy} is not conserved over time, but is a Lyapunov function of the ALVE~(1).
In other words, the value of $D$ decreases over time as one computes directly ($\frac{\text{d} }{\text{d} t} D(\vec{c}||\vec{x}) = \sum_{i\in \bar{I}} (A\vec{c})_ix_i<0$). 
Due to the definition of $D$ as a relative entropy and due to the Lyapunov property, $D$ is bounded as $0<D(\vec{c}||\vec{x})<D(\vec{c}||\vec{x}(0))<\infty$.
Similarly to above, one concludes that all states with index $i\in I$ remain bounded away from 0 for all times, that is, $x_i(t)\geq Const>0$ for all $i\in I$ and for all $t$ (otherwise, $D$ would diverge in contradiction to the boundedness of $D$).
With further arguments exploiting the boundedness of $D$, it is possible to show that all other states with index $i\in \bar{I}$ become depleted, that is, $x_i(t)\to 0$ as $t\to\infty$ for all $i\in \bar{I}$~\cite{Knebel2015}.\\
In total, condensation and depletion in the ALVE~(1)
are determined by an algebraic property of the antisymmetric matrix. 
The set of condensates $I$ is given by the antisymmetric matrix alone through its condensate vectors. 
All states with index $i\in I$ become condensates, all states with index $i\in\bar{I}$ become depleted for $t \to \infty$; no other cases can occur for long times.
This selection of condensates in the ALVE is, notably, independent of the initial conditions.
Furthermore, the Lyapunov function $D(\vec{c} || \vec{x})$ approaches a conserved quantity of the form $D(\vec{p} || \vec{x})$ at long times.
Finally, if all states coexist, the set of condensates is given by $I = \{1, \dots, S\}$, and all condensate vectors~$\vec{c}$ are strictly positive kernel elements of the antisymmetric matrix $A$, see Equation~\eqref{eq:condensate_condition_1}, which proves the assertion.

\subsection{Coexistence networks are strongly connected}
\label{sec:strongly_connected}

\textit{Statement.}
Here we show that coexistence of all states in the ALVE (Equation~(1)
of the main text) is only possible for strongly connected networks. In particular, every coexistence network is also strongly connected.
A network is strongly connected if for all pairs of nodes $i$ and $j$ there is a directed path connecting $i$ to $j$ and, vice versa, a directed path connecting $j$ to $i$. 
Pictorially speaking, all states can coexist for all times ($x_i\geq \epsilon>0$ for all $i$ for all times) if each state can (i) gain mass through a directed path from all other states and (ii) release mass through a directed path to all other states. 
The simplest example of a strongly connected network is a directed cycle.\\

\textit{Notation.}
To show that coexistence of all states requires a strongly connected network, we apply an argument by contradiction:
Assume that $A$ is the antisymmetric adjacency matrix of a network that is not strongly connected, but weakly connected (that is, there exists an undirected path connection all pairs of nodes). Furthermore, we assume that $A$ has a positive kernel and, thus, coexistence of all states in the ALVE~(1);
see Section~2
of the main text.
Because the network is not strongly connected, there exists a subset of nodes that is not connected to the rest of the network through a directed path. 
Therefore, the set of nodes $V$ can be divided into two disjoint sets of nodes $V_1$ and $V_2$ ($V = V_1 \cup V_2$ and $V_1 \cap V_2= \emptyset$), for which all edges between nodes of $V_1$ and nodes of $V_2$ are directed from $V_2$ to $V_1$. 
In other words, the network's antisymmetric adjacency matrix in a suitable labeling takes the form:
\begin{align}
A = 
\begin{pmatrix}
A_1 & T \\
-T^T & A_2
\end{pmatrix},
\end{align}
with $T_{ij} \geq 0$ for all $i,j$ denotes the weights of the edges that connect $V_2$ to $V_1$. Note also that $T \neq 0$ (if $T \equiv 0$, the network would be divided into two separated subnetworks without any connecting edges between $V_1$ and $V_2$, but we assumed a weakly connected network). The antisymmetric matrices $A_1$ and $A_2$ characterize the weights connecting nodes solely within $V_1$ and $V_2$, respectively.
Accordingly, we decompose the state vector into the chosen partitions $V_1$ and $V_2$ and write $\vec{x} = (\vec{x}_1 , \vec{x}_2)$.  \\

\textit{Proof.}
Let us now compute how the total mass in the states $V_2$ evolves in time:
\begin{align}
\frac{\mathrm{d}}{\mathrm{d} t} \sum_{i \in V_2} x_i 
&= \sum_{i\in V_2} x_i (A \vec{x})_i 
=\begin{pmatrix}
0^T & \vec{x}_2^T
\end{pmatrix} 
\begin{pmatrix}
A_1 & T \\
-T^T & A_2
\end{pmatrix}
\begin{pmatrix}
\vec{x}_1 \\ \vec{x}_2
\end{pmatrix}\ , \nonumber \\
&= - \vec{x}_2^T T^T \vec{x}_1 + \vec{x}_2^T A_2 \vec{x}_2 
= - \vec{x}_2^T T^T \vec{x}_1 \ .
\end{align}
Since we assumed coexistence of all states ($x_i\geq \epsilon>0$ for all $i$ for all times), the last line can be estimated with $T\geq 0$:
\begin{align}\label{eq:V_2_decrease}
\frac{\mathrm{d}}{\mathrm{d} t} \sum_{i \in V_2} x_i = - \vec{x}_2^T T^T \vec{x}_1 \leq - Const\cdot \epsilon<0\ ,
\end{align}
for some positive constant $Const>0$. 
Because all concentrations are positive $x_i \geq 0$, also $\sum_{i \in V_2} x_i \geq 0$. Together with equation~\eqref{eq:V_2_decrease} it follows that all states $V_2$ go extinct over time.
This result is intuitively understood because all edges between $V_1$ and $V_2$ are directed from $V_2$ to $V_1$, that is, mass can only flow from $V_2$ to $V_1$, but not in the opposite direction (note that the total concentration $\sum_{i =1}^S x_i = 1$ is a conserved quantity of the ALVE~(1);
see Section~2 of the main text.
However, extinction of $V_2$ is in contradiction with our assumption that all states coexist. Thus, if a network is not strongly connected, all states cannot coexist. Therefore, strongly connected networks are necessary to obtain coexistence of all states in the ALVE~(1).
The simplest strongly connected network is a directed cycle or, more generally, Hamiltonian networks that are discussed in the main text.\\

\textit{Implications.} 
The fact that coexistence networks are strongly connected has further implications for the underlying topology of coexistence networks.
An undirected graph only admits a strong orientation (that is, a choice of the direction of all edges such that the resulting directed network is strongly connected) if and only if it has an ear-decomposition~\cite{Robbins39}. 
Pictorially speaking, a graph has an ear-decomposition if it can be decomposed to a cycle by successively detaching paths, which are connected to the graph with both ends as 'ears' without crossing any other paths. 
Notably, Lovasz~\cite{Lovasz86} showed that all factor-critical graphs can always be oriented to be strongly connected. 
We employed this fact for our numerical search of coexistence networks starting from undirected graphs; see Section~\ref{sec:numerical2}.


\section{Pfaffian of an antisymmetric matrix and further algebraic properties}
\label{sec:Pfaffian_appendix}

\subsection{Combinatorial definition of the Pfaffian}
\label{sec:Pfaffian_combinatorial}

Because the Pfaffian of an antisymmetric matrix is central to our analysis, we present here its combinatorial definition for completeness.
Let $\Pi$ denote the set of all partitions of the set $\{1, 2, \dots ,S = 2n\}$ into ordered pairs. 
In other words, every partition $\alpha \in \Pi$ is pairwisely ordered in the form $\alpha = \big((i_1,j_1),(i_2,j_2),\dots,(i_n,j_n) \big)$ with $i_k< j_k$ for all $k$ and $i_k< i_l$ for all $ k<l$.
Note that there are $|\Pi| = (2n-1)\cdot(2n-3)\cdot\dots\cdot 3\cdot 1 = (2n-1)!!$ different pairwisely ordered partitions of the set $\{1,\dots , S \}$.
We define the permutation $\sigma_{\alpha}$ of such a partition $\alpha\in\Pi$ as:
\begin{align}
\sigma_{\alpha} \coloneqq &\begin{pmatrix}
1 & 2 & 3 & 4& \dots & 2n-1 & 2n\\
(i_1 & j_1) & (i_2 & j_2)& \dots & (i_n & j_n)
\end{pmatrix}\ , \\
\equiv &\ \big(i_1\ j_1\ i_2\  j_2\ \dots  i_n\ j_n\big)
\ .
\end{align}
With these notions, the Pfaffian of an antisymmetric matrix $A\in\mathbb{R}^{S\times S}$ of even size $S = 2n$ is defined as~\cite{Muir1882, Thomas2006}:
\begin{align}\label{eq:Pfaffian}
\text{Pf} (A) \coloneqq \sum_{\alpha \in \Pi} \left(\text{sign}(\sigma_{\alpha}) \prod_{k = 1}^n a_{\alpha_k} \right).
\end{align}
For an odd-sized antisymmetric matrix, the Pfaffian is 0. Because the elements of every partition $\alpha$ are ordered pairs ($i<j$ for every pair $(i,j) \in \alpha $), every summand $\prod_{k = 1}^n a_{\alpha_e}$ is a product of above-diagonal matrix entries of $A$. With this definition~\eqref{eq:Pfaffian}, one can show that $\text{Pf} (A)^2 = \text{Det}(A)$.


For illustration, consider the pretzel-like interaction network sketched in Figure~3
with adjacency matrix~(6)
given in the main text. 
Using the combinatorial definition~\eqref{eq:Pfaffian} to compute the Pfaffian of $A_\text{pretzel}$ yields:

\begin{align}
\text{Pf}(A_\text{pretzel}) 
&=
\left[
\begin{aligned}
&\phantom{+}\text{sign}(1\ 2\ 3\ 4) (-a_{21})(-a_{43}) \\
&+ \text{sign}(1\ 3\ 2\ 4) a_{13} a_{24} \\
&+ \text{sign}(1\ 4\ 2\ 3)\cdot 0  \cdot (-a_{32})
\end{aligned}\right]
\ ,  \\
&= (+1) (-a_{21})(-a_{43}) + (-1) a_{13}a_{24} \ , \nonumber \\
&=a_{21}a_{43} - a_{13}a_{24}\ ,
\end{align}

which agrees with the graph-theoretical~(5)
computation in Equation~(7)
in the main text.

\subsection{Graph-theoretical and combinatorial definition of the Pfaffian}
\label{sec:Pfaffian_definitions_agree}

\textit{Graph-theoretical and combinatorial definition agree.}
The graph-theoretical definition of the Pfaffian~(5)
agrees with its combinatorial definition~\eqref{eq:Pfaffian} for the following reason.
If the weighted network $\mathcal{N}(A)$ is built from the antisymmetric adjacency matrix $A$, every perfect matching of the network $\mathcal{N}(A)$ corresponds to a distinct non-zero summand in the Pfaffian of $A$ in the graph-theoretical definition~(5).
In the combinatorial definition~\eqref{eq:Pfaffian} all matrix elements for the computation of the Pfaffian are taken from above the diagonal of $A$, whereas in the graph-theoretical definition only negative matrix elements are used (that is, from above or below the matrix diagonal).
If a positive entry $a_{ji}>0$ appears in a summand of the combinatorial definition of the Pfaffian~\eqref{eq:Pfaffian}, the negative matrix entry $a_{ij} = -a_{ji}<0$ appears in the graph-theoretical definition~(5).
Additionally, the permutation of the corresponding partition changes by one transposition ($i$ and $j$ are swapped) such that the sign of the permutation in the combinatorial and graph-theoretical definition differs by a factor of $-1$. This compensates for the minus sign originating from the different sign of the matrix entry. 

\textit{The graph-theoretical definition is suitable for our work on coexistence networks.}
The graph-theoretical definition of the Pfaffian~(5)
has the advantage over the combinatorial definition~\eqref{eq:Pfaffian} in that it distinguishes between network topology and edge weights. The combinatorial definition~\eqref{eq:Pfaffian} includes only above-diagonal entries of the antisymmetric matrix $A$. When $A$ contains zero-entries, zero summands appear in the expression for the Pfaffian, which do not contribute to its value.
In contrast, the graph-theoretical definition of the Pfaffian includes by construction only non-zero entries of $A$ and, thus, contains only non-zero summands in the definition of the Pfaffian. 
In addition, all matrix elements of the summands occurring in the Pfaffian are negative such that the sign of each summand is determined only by the sign of the perfect matching alone. 
For these reasons, the graph-theoretical definition~(5)
distinguishes between edge-weights (negative matrix elements) and network topology (signs of matchings) and, thus, is suitable for our discussion of coexistence networks in the main text.

\subsection{The Pfaffian of exemplary antisymmetric matrices}
\label{sec:Pfaffian_examples}

\textit{A general $2\times 2$ antisymmetric matrix.} As an example, consider an arbitrary antisymmetric $2 \times 2$ matrix (with $a_{12}>0$),
\begin{align}
A_2 = 
\begin{pmatrix}
0 & a_{12}  \\
-a_{12} & 0 
\end{pmatrix}\ .
\end{align}
The set of all pairwisely ordered partitions of $\{1, 2\}$ is simply $\Pi = \{\big((1,2)\big)\}$. 
Therefore, the Pfaffian of $A_2$ is given by
\begin{align}
\text{Pf}(A_2) = a_{12}\ .
\end{align}
%

\textit{A general $4\times 4$ antisymmetric matrix.} For an arbitrary antisymmetric $4 \times 4$ matrix (all above-diagonal entries are assumed to be positive),
\begin{align}
A_4 = 
\begin{pmatrix}
0 & a_{12} & a_{13} & a_{14} \\
-a_{12} & 0 & a_{23} & a_{24} \\
-a_{13} & -a_{23} & 0 & a_{34} \\
-a_{14} & -a_{24} & -a_{34} & 0 
\end{pmatrix}\ ,
\end{align}
the set of all pairwisely ordered partitions of the set $\{1, 2, 3, 4\}$ is given by $\Pi = \{\big((1,2),(3,4)\big) ,\ \big((1,3),(2,4)\big),\ \big((1,4),(2,3)\big)\}$. 
The Pfaffian of $A_4$ is obtained as:
\begin{align}\label{eq:Pfaffian_4}
\text{Pf}(A_4) =(+1) a_{12}a_{34} + (-1) a_{13}a_{24} + (+1) a_{14}a_{23}\ .
\end{align}
The sign of the permutation was highlighted in front of the corresponding factor (for example, $\text{sign}(1\ 3\ 2\ 4) = -1$). \\

\textit{A general $6\times 6$ antisymmetric matrix.} For a general antisymmetric $6 \times 6$ matrix $A_6$ (again all above-diagonal entries are assumed to be positive),
\begin{align}
A_6 = 
\begin{pmatrix}
0 & a_{12} & a_{13} & a_{14} & a_{15}& a_{16}\\
-a_{12} & 0 & a_{23} & a_{24} & a_{25}& a_{26}\\
-a_{13} & -a_{23} & 0 & a_{34} & a_{35}& a_{36}\\
-a_{14} & -a_{24} & -a_{34} & 0 & a_{45}& a_{46}\\
-a_{15} & -a_{25} & -a_{35} & -a_{45} & 0& a_{56}\\
-a_{16} & -a_{26} & -a_{36} & -a_{46} & -a_{56}& 0
\end{pmatrix}\ ,
\end{align}
the Pfaffian is obtained as:
\begin{align}
\text{Pf}(A_6) =
\left[
\begin{aligned}
 &\phantom{+}   (+1)a_{12}a_{34}a_{56} + (-1)a_{12}a_{35}a_{46}\\
 &+ (+1)a_{12}a_{36}a_{45} + (-1)a_{13}a_{24}a_{56}\\
 &+ (+1)a_{13}a_{25}a_{46} + (-1)a_{13}a_{26}a_{45}\\
 &+ (+1)a_{14}a_{23}a_{56} + (-1)a_{14}a_{25}a_{36}\\
 &+ (+1)a_{14}a_{26}a_{35} + (-1)a_{15}a_{23}a_{46}\\
 &+ (+1)a_{15}a_{24}a_{36} + (-1)a_{15}a_{26}a_{34} \\
 &+ (+1)a_{16}a_{23}a_{45} + (-1)a_{16}a_{24}a_{35}\\
 &+ (+1)a_{16}a_{25}a_{34}
 \end{aligned}
 \right] \ .
\end{align}
%

\textit{Directed cycle of even size.} For further illustration of how the Pfaffian is computed and to complete the statements from the previous section, let us also consider a directed cycle of even length: $ S\to 1\to 2\to 3\to \dots \to S-1 \to S$ (such that $a_{1,S}, a_{21}, a_{32}, \dots, a_{S, S-1}>0$ with $S = 2n$ even for $n =  2,3, \dots$) with according antisymmetric adjacency matrix $A_\text{even-cycle}$: 
%
\begin{align}
A_\text{even-cycle} = 
\begin{pmatrix}
0 & -a_{21} & 0  &  \dots & a_{1, S}  \\
a_{21} & 0 & -a_{32}  &  \dots & 0\\
0 & a_{32} & 0  &  \dots & 0\\
\vdots & 0 & \ddots & \ddots &   \vdots\\
0 &  \dots & a_{S-1, S-2} & 0 &   -a_{S, S-1}\\
-a_{1,S} & 0 & \dots &  a_{S, S-1} &   0\\
\end{pmatrix}\ .
\end{align}

The Pfaffian of the even-sized cycle is obtained as: 
\begin{align}
\text{Pf}(A_\text{even-cycle}) =
\phantom{-}(-1)^{n\, \text{mod}\, 2}a_{21}a_{43}\cdots a_{S, S-1} -(-1)^{n\, \text{mod}\, 2} a_{32}a_{54}\cdots a_{1,S}
\ .
\end{align}

%

\textit{Laplace-like formula of the Pfaffian.} 
As can be seen from the above examples, the computation of the Pfaffian of an antisymmetric matrix proceeds in a similar manner as the computation of the determinant of an arbitrary matrix, but is tailored to the antisymmetry of the matrix through the notion of the Pfaffian. 
In this line of thought, it is not surprising that a recursive definition of the Pfaffian for an antisymmetric matrix of even size can be obtained (in the spirit of Laplace's formula for determinants) as~\cite{Cullis1913, Wimmer2011}:
\begin{align}\label{eq:Pfaffian_recursive}
\text{Pf} (A)= \sum_{i=2}^S (-1)^i a_{1i}\text{Pf}(A_{\hat{1}\hat{i}})\ ,
\end{align}
which we mention here for completeness. 

\subsection{Adjugate vector and adjugate matrix of an antisymmetric matrix}
\label{sec:adjugate_vector_matrix}

By using the notion of the Pfaffian of an antisymmetric matrix $A$, an explicit analytical expression for the kernel of $A$ is obtained for odd-sized matrices with kernel dimension 1 and even-sized matrices with kernel dimension 2 via the adjugate vector or the adjugate matrix of $A$, respectively~\cite{Cullis1913}. \\

\textit{The adjugate vector of an odd-sized antisymmetric matrix.}
If $S$ is odd, the kernel of an antisymmetric matrix $A$ is characterized by the \textit{adjugate vector} $\vec{r}\in\mathbb{R}^S$ as defined in the main text; see Equation~(8).
Recall that the adjugate vector is a kernel vector of $A$ if $\mathrm{dim}(\mathrm{Ker}(A)) = 1$; and it is the zero-vector $\vec{r} = 0$ if $\mathrm{dim}(\mathrm{Ker}(A)) = 3, 5, \cdots, S -2, S$~\cite{Cullis1913}. 
In any case, it holds that $A\vec{r} = 0$ and the computation of the adjugate vector in Equation~(8)
via the Pfaffians of all submatrices $A_{\hat{i}}$ is reminiscent of Cramer's rule adjusted to antisymmetric matrices.\\

\textit{The adjugate vector of an odd-sized cycle.}
With the graph-theoretical definition of the Pfaffian~(5)
the adjugate vector of the adjacency matrix of an odd cycle of size $S$, with antisymmetric adjacency matrix $A_\text{odd-cycle}$ given in Equation~(4),
can be computed as follows.
Every subnetwork of the odd cycle created by deleting one node is a path of length $S-1$ and thus has exactly one perfect matching. Consequently every component of the adjugate vector consists of one product of matrix elements. 
All edges of the odd cycle connect nodes with different parity, apart from the edge $S \to 1$. 
Only when this edge $S \to 1$ is part of a near-perfect matching, the elements in the permutation of the near-perfect matching are not ordered in size, which is the case whenever an even node is deleted.

(i) \textit{$i$ even.} 
For example, the near-perfect matching upon deleting node $2$ is $\mu_\text{even} = ((S \to 1), (3 \to 4), \dots , (S-2 \to S-1))$. To order the corresponding permutation in size, one may shift the index $S$ from the first to the last position by applying $S-2$ transpositions. Thus, $\text{sign} (\mu_\text{even}) = -1$. 
Therefore, for even $i$, the component of the adjugate vector, $r_i$, is obtained as (recall that $S = 2n+1$ here is odd):

\begin{align}
r_{i}& = (-1)^{1+i}\text{sign}(\mu_\text{even}) (-a_{1,S})(-a_{32})\dots (-a_{i-1,i-2})(-a_{i+2,i+1})\dots (-a_{S-1,S-2})\ , \nonumber \\
&= (-1)(-1)(-1)^{n\, \mathrm{mod}\, 2}a_{1,S}a_{32}\dots a_{i-1,i-2}a_{i+2,i+1}\dots a_{S-1,S-2}\ .
\end{align}

(ii) \textit{$i$ odd.}
If, on the other hand, an odd node is deleted, an equal number of even and odd nodes remain in the network such that every edge of a perfect matching connects an even and an odd node. 
The edge $S\to 1$ is not part of the perfect matching $\mu_\text{odd}$ and, thus, the elements of the corresponding permutation are ordered in size, yielding the sign $+1$ for the permutation. Consequently, for odd $i$ the adjugate vector is obtained as:
\begin{align}
r_{i} & = (-1)^{i+1} \text{sign}(\mu_\text{odd}) (-a_{21})(-a_{43})\dots (-a_{i-1,i-2})(-a_{i+2,i+1})\dots (-a_{S,S-1})\ , \nonumber \\
&= (+1)(+1)(-1)^{n\, \mathrm{mod}\, 2}a_{21}a_{43}\dots a_{i-1,i-2}a_{i+2,i+1}\dots a_{S,S-1}\ .
\end{align}

Therefore, the kernel of an odd-sized cycle is given by $\text{Ker}(A) = \{(a_{32}a_{54}\cdots a_{S, S-1}, a_{43}a_{65}\cdots a_{1, S}, \dots, a_{21}a_{43}\cdots a_{S-1, S-2})\}$, as claimed in the main text.\\

\textit{The adjugate matrix of an even-sized antisymmetric matrix.}
The kernel of an even-dimensional antisymmetric matrix is characterized in terms of Pfaffians of submatrices as well.
If $S$ is even, the kernel of $A$ is characterized by the \textit{adjugate matrix} $R\in\mathbb{R}^{S\times S}$, whose entries are defined as follows:
\begin{align}\label{eq:adjugate_matrix}
R_{ij} = \text{sign}(\sigma_{ij}) \text{Pf}(A_{\hat{i}\hat{j}}) \quad ,\ i,j = 1, \dots, S\ .
\end{align}
Here, $A_{\hat{i}\hat{j}}$ denotes the matrix obtained by deleting both the $i$th and $j$th row and column from $A$. In case $i=j$, $A_{\hat{i}\hat{i}}\coloneqq A_{\hat{i}}$ and, thus, $\mathrm{Pf}(A_{\hat{i}\hat{i}}) = 0$. Furthermore, $\text{sign}(\sigma_{ij})$ denotes the signum of the permutation,
\begin{align}
\sigma_{ij} \coloneqq 
&\left(
\begin{array}{*{16}c}
1 & 2 & 3 & 4& \dots & i-1 & i      & i+1 & i+2    &  \dots   & j-1  & j       & j+1  & \dots & 2n-1 &2n \\
i & j & 1 & 2& \dots   & i-3  & i-2  & i-1  & i+1    & \dots    &  j-2 & j-1    & j+1 &  \dots & 2n-1& 2n \\
\end{array}
\right)\ , \nonumber \\
\equiv &\ \big(i\ j\ 1\ 2\ \dots  2n-1\ 2n\big)
\ ,
\end{align}
in which $i$ and $j$ are taken out of and put in front of the sequence $( 1, 2, \dots, 2n)$.
The adjugate matrix comprises two linearly independent column vectors as kernel vectors if $\mathrm{dim}(\mathrm{Ker}(A)) = 2$; and it is the zero-matrix if $\mathrm{dim}(\mathrm{Ker}(A)) = 4, 6, \dots, S$~\cite{Cullis1913}. 
In general, the adjugate matrix can be thought of as the generalized inverse of the antisymmetric matrix $A$ having the property $A R = -\text{Pf}(A)\mathbb{1}_{S\times S}$ with $\mathbb{1}_{S\times S}$ denoting the unit matrix of size $S\times S$.
If the kernel is trivial ($\mathrm{dim}(\mathrm{Ker}(A)) = 0$, that is, $\text{Det}(A) > 0$), the antisymmetric matrix $A$ is invertible and the adjugate matrix $R$ is proportional to its inverse.\\


\textit{The adjugate matrix of an even-sized cycle.}
For completeness, we also denote the elements of the adjugate matrix of an even-sized cycle, which are obtained in a similar manner as for the odd-sized cycle, but deleting two nodes from the network:
\begin{align}
R_{ij} = \begin{cases}
0 \; &\ ,\text{for } i = j\\
0 \; &\ ,\text{for } j-i \text{ even}\\
(-1)^{n\, \mathrm{mod}\, 2}a_{21}\dots a_{i-1,i-2} a_{i+2,i+1} \dots a_{j-1,j-2} a_{j+2,j+1}\dots a_{S,S-1}&\ , \text{for } j-i \text{ odd and } i \text{ odd} \\
(-1)(-1)^{n\, \mathrm{mod}\, 2} a_{1,S}\dots a_{i-1,i-2} a_{i+2,i+1} \dots a_{j-1,j-2} a_{j+2,j+1}\dots a_{S-1,S-2}&\ , \text{for } j-i \text{ odd and } i \text{ even} \\
-R_{ji} \; &\ , \text{for } j<i 
\end{cases} \ .
\end{align}
%

\subsection{Minimal kernel dimension of a network topology determined by perfect matchings of subnetworks}
\label{sec:minimal_kernel_dimension}

\textit{Minimal kernel dimension of a network topology.} Given a network topology, that is, a directed graph for which the magnitudes of the edge weights can be arbitrarily chosen, one may ask the following question: what is the lower bound for the kernel dimension of the antisymmetric adjacency matrices corresponding to that network topology? 
The lower bound for the kernel dimension in the set of all antisymmetric matrices that respect the specified network topology is referred to as the \textit{minimal kernel dimension} of this network topology.
Note that upon tuning the weights of a network topology, the Pfaffian of the antisymmetric adjacency matrix $A$ and Pfaffian of submatrices of $A$ may vanish and, thus, the dimension of the kernel may increase. In other words, tuning the weights on a given network topology can only increase the kernel dimension compared with the minimal kernel dimension of that network topology.\\

\textit{Factor-critical networks have minimal kernel dimension 1.}
First, consider a factor-critical network as depicted in Figure~5(a)
of the main text. Because the adjugate vector is not the zero-vector for some choice of weights (Figure~(5)d),
the minimal kernel dimension of this network topology is 1. 
In general, the minimal kernel dimension of factor-critical networks is always 1; see below.\\

\begin{figure*}[th!]
\centering
\includegraphics[width=1.0\textwidth]{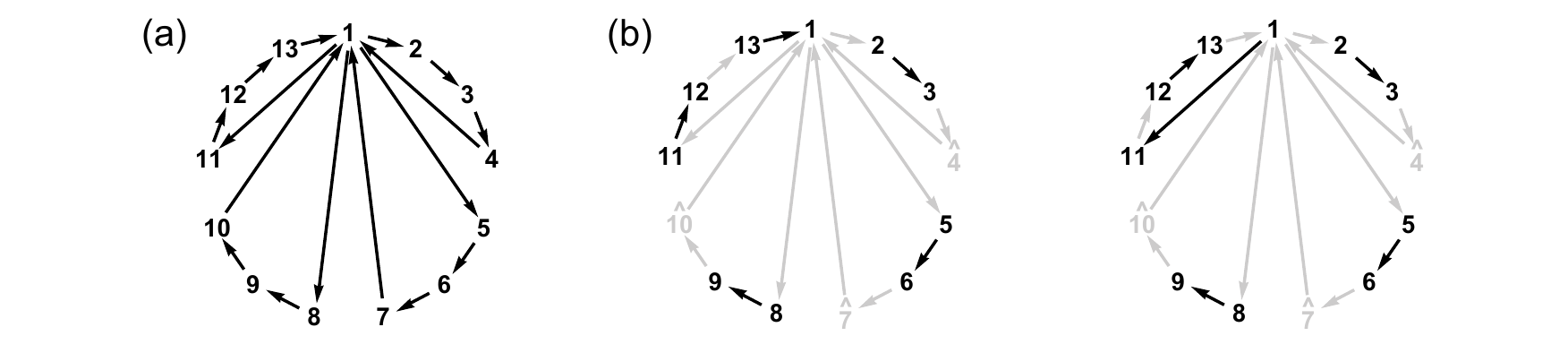}
\caption{
\textbf{Example of a network topology with minimal kernel dimension 3.}
(a) Example of a network topology of $13$ nodes, for which the minimal kernel dimension is $3$. The network consists of four cycles of length four connected at one single center node (node 1). 
(b) Two exemplary perfect matchings, which arise after deleting three nodes. No perfect matching can be identified after removal of no, one, or two arbitrary node(s).
 As a consequence, the kernel dimension of the network's adjacency matrix for an arbitrary choice of weights is at least 3. No choice of weights can yield a one-dimensional kernel. In total, the minimal kernel dimension of this network topology is 3.
}
\label{fig:ExampleHighDimKernel}
\end{figure*}

\textit{Example of a network topology with minimal kernel dimension 3.}
Consider now the exemplary network topology depicted in Figure~\ref{fig:ExampleHighDimKernel}(a), which is built from four 4-cycles connected at one single center node. For a generic choice of weights, the kernel of $A$ has dimension $3$, as validated by the three linearly independent kernel vectors:

\begin{align*}
v_1 = 
\begin{pmatrix}
0\\
 a_{4,3} (a_{11,1} a_{13,12} - a_{12,11} a_{1,13}) \\
 0\\ 
 a_{3,2} (a_{11,1} a_{13,12} - a_{12,11} a_{1,13}) \\ 
 0\\ 
 0\\ 
 0\\ 
 0\\ 
 0\\ 
 0\\ 
-a_{13,12} (a_{2,1} a_{4,3} - a_{3,2} a_{1,4}) \\ 
 0\\
 -a_{12,11} (a_{2,1} a_{4,3} - a_{3,2} a_{1,4}) 
\end{pmatrix}\ , \
v_2 = 
\begin{pmatrix}
0\\ 
-a_{4,3} (a_{1,10} a_{9,8} - a_{8,1} a_{10,9})\\ 
0\\
 -a_{3,2} (a_{1,10} a_{9,8} - a_{8,1} a_{10,9})\\
 0\\ 
 0\\ 
 0\\ 
 -a_{10,9} (a_{2,1} a_{4,3} - a_{3,2} a_{1,4})\\ 
 0\\ 
-a_{9,8}  (a_{2,1} a_{4,3} - a_{3,2} a_{1,4}) \\ 
 0\\ 
 0\\ 
 0
\end{pmatrix}\ ,\
 v_3 = 
\begin{pmatrix}
0\\
 a_{4,3} (a_{5,1} a_{7,6} - a_{6,5} a_{1,7}) \\ 
 0\\ 
 a_{3,2} (a_{5,1} a_{7,6} - a_{6,5} a_{1,7}) \\
 -a_{7,6}(a_{2,1} a_{4,3} - a_{3,2} a_{1,4})  \\ 
  0\\ 
-a_{6,5} (a_{2,1} a_{4,3} - a_{3,2} a_{1,4})  \\ 
  0\\ 
  0\\ 
  0\\
   0\\ 
   0\\ 
   0
\end{pmatrix}\; .
\end{align*}

Therefore, the minimal kernel dimension of the network topology in Figure~\ref{fig:ExampleHighDimKernel}(a) is 3.
However, all three kernel vectors $v_1, v_2$, and $v_3$ have zero entries at index $1,3,6,9$ and $12$. No linear combination of these vectors is strictly positive for any choice of weights and, thus, this network is not a coexistence network.
In other words, no choice of weights on this network topology can yield an antisymmetric adjacency matrix with kernel dimension smaller than 3.
In fact, when all weights are chosen to be equal, the three kernel vectors vanish. In this case, the kernel dimension is $5$. 

%

\textit{Minimal kernel dimension is determined by occurrences of perfect matchings in subnetworks.}
In the following, we show that the minimal kernel dimension of a network topology equals the minimal number of nodes that need to be deleted such that a perfect matching exists in the remaining subnetwork. 
In other words, the kernel of $A$ on the specified network topology is at least $K$-dimensional if it is necessary to delete at least $K$ nodes from the network $\mathcal{N}(A)$ to obtain a perfect matching in the remaining subnetwork. 

For example, a factor-critical network has a perfect matching upon removing any single node (the so-called near-perfect matchings). On the other hand, because factor-critical network are of odd size, they do not have a perfect matching.
Thus, the minimal kernel dimension of a factor-critical network is 1. 
The above example of four 4-cycles connected at a center node (Figure~\ref{fig:ExampleHighDimKernel}(a)) neither has a perfect matching nor does it have a perfect matching upon removing one arbitrary node or two arbitrary nodes. However, this network topology has perfect matchings after removal of three nodes as shown in Figure~\ref{fig:ExampleHighDimKernel}(b). 
Thus, the minimal kernel dimension of this network topology is~3. 

To briefly rationalize the graph-theoretical characterization of the minimal kernel dimension, we express the characteristic polynomial of an arbitrary antisymmetric matrix $A$ that respects a given network topology in terms of its principal minors. 
The characteristic polynomial of a matrix can be written as:
\begin{align}
\det (A - \lambda \mathbb{1}_S) =  \sum_{i = 0}^{S} c_{i} \lambda^i\ ,
\end{align}
where the coefficients $c_i\in \mathbb{R}$ in the expansion are given by~\cite{Mirsky1990}:
\begin{align}
c_0 &= \det(A)\ ,\\
c_i &= \sum_{\substack{J \subseteq \{1,\dots,S\},\\ |J| = i}}  (-1)^i\det (A_{\hat{J}}) \quad ,\ i = 1, \dots, S-1\ ,\\
c_S &= 1\ .
\end{align}
Here, $A_{\hat{J}}$ denotes the submatrix of $A$ (a principal minor) that is created by deleting the rows and columns with label $j_1, \dots, j_{|J|}$ (such that $J = \{j_1, \dots, j_{|J|}\}$) from $A$.
In a graph-theoretical interpretation, $A_{\hat{J}}$ is the antisymmetric adjacency matrix of the subnetwork that is created by deleting the nodes $J$ and their attached links from $\mathcal{N}(A)$. 
If it is necessary to delete at least $K$ nodes from the network $\mathcal{N}(A)$ to obtain a perfect matching in a remaining subnetwork, the first $K$ expansion coefficients $c_0, c_1, \dots, c_{K-1}$ are zero (because for antisymmetric matrices, the determinant is 0 if and only if the Pfaffian is 0). 
Upon tuning the weights of the network, also further expansion coefficients $c_i$ with $i\geq K$ may vanish and, thus, the kernel dimension can only be greater than $K$. 
Therefore, the kernel of $A$ on the specified network topology is at least $K$-dimensional if it is necessary to delete at least $K$ nodes from the network $\mathcal{N}(A)$ to obtain a perfect matching in a remaining subnetwork.

\section{No coexistence networks with a two-dimensional kernel}
\label{sec:Proof2DimCoexNW}

\textit{Overview.}
If the dimension of the kernel of a network's antisymmetric adjacency matrix is zero, the network does not have non-trivial kernel elements and, thus, no strictly positive kernel vectors. Consequently, this network topology cannot be a coexistence network.
In case of an one-dimensional kernel, we identified coexistence networks through the adjugate vector~(8).
Sufficient and necessary conditions for Hamiltonian coexistence networks of odd size were determined and all coexistence networks with up to 9 nodes were identified; see Sections~4.1-4.3
of the main text. 
Here we supplement the statements from the main text (see Section~4.3)
by showing that there are no coexistence networks with a two-dimensional kernel $\text{dim} (\text{Ker}( A)) = 2$. We employ the adjugate matrix~\eqref{eq:adjugate_matrix} to show this result.
Whether coexistence networks with a high-dimensional kernel $\text{dim} (\text{Ker}( A)) \geq 3$ exist, remains open at present.\\

\textit{Proof: No coexistence networks with a two-dimensional kernel.}
We use an argument by contradiction to show that coexistence networks cannot have a two-dimensional kernel.
Assume that $\mathcal{N}(A)$ is a coexistence network with a two-dimensional kernel (its minimal kernel dimension is 2).
Thus, $A$ has a strictly positive kernel vector for every choice of weights. We choose the rates such that indeed $\text{dim} (\text{Ker}( A)) = 2$ and it is possible to choose two strictly positive kernel vectors $u$ and $v$ that are linear independent and form a basis of the kernel of $A$.

For antisymmetric matrices with a two-dimensional kernel, the kernel vectors can be calculated with help of the adjugate matrix~\eqref{eq:adjugate_matrix}. 
If the kernel of $A$ is two-dimensional, the antisymmetric adjugate matrix has nonzero entries, its rank is 2, and two linearly independent column vectors form a basis of the kernel space.
We denote the column vectors of the adjugate matrix $R$ as $\vec{r}^{(j)}$ ($j = 1, \dots, S$) with elements $r_i^{(j)} = R_{ij}$ for $i, j = 1, \dots, S$.

Because the column vector of the adjugate matrix are kernel vectors of $A$ (see Section~\ref{sec:adjugate_vector_matrix}), they can be expressed as linear combinations of $u$ and $v$: $r_i^{j}= \mu^{(j)} u^{(j)}_i + \nu^{(j)} v^{(j)}_i$. 
From the antisymmetry of the adjugate matrix ($r_{jj}=0$), it follows that: 
\begin{align}\label{eq:R_diagonal}
\nu^{(j)} =-\mu^{(j)}\frac{v_j}{u_j}\ , \text{ for all }j = 1, \dots, S\ .
\end{align}
By assumption, the network has a two-dimensional kernel (or higher dimension) for all choices of weights. Thus, the Pfaffian of $A$ is always zero and the network has no perfect matching; see Section~3.2.

Because the network is a coexistence network, it is strongly connected; see Section~\ref{sec:strongly_connected}. In particular, for every node $k$, one finds another node $l$ that is connected to $k$. In the following we show that $R_{kl} = R_{lk} = 0$ follows. 
Because nodes $k$ and $l$ are connected, the network created by deleting these nodes, $A_{\hat{k}\hat{l}}$, has no perfect matching. If $A_{\hat{k}\hat{l}}$ had a perfect matching, $A$ would have a perfect matching as well (by combining the perfect matching of $A_{\hat{k}\hat{l}}$ together with the edge connecting $k$ and $l$); which is in contradiction to the assumption that the minimal kernel dimension is 2.
Therefore, $ \text{Pf}(A_{\hat{k}\hat{l}}) =0$ and $R_{kl} = R_{lk} = 0$ as claimed.
Furthermore, it follows that $R_{lk} = r_l^{(k)} = \mu^{(k)} u_i + \nu^{(k)} v_i = 0$, and with Equation~\eqref{eq:R_diagonal} for index $k$: $0 = \mu^{(k)} u_i - \mu^{(k)}\frac{v_k}{u_k} v_i = 0$.
If $\mu^{(k)}=0$, it also follows that $\vec{r}^{(k)} = \vec{0}$, which is in contradiction to $\text{dim} (\text{Ker}( A)) = 2$ because $k$ is arbitrary. As a consequence, if node $k$ and $l$ are connected, it follows that:
\begin{align}
	u_k u_l - v_l v_k =0\ .
\end{align}
This argument can be carried out iteratively for all pairs of connected nodes. This way, it follows that for any two nodes $k$ and $l$, which are connected via a path of arbitrary length, it holds $u_k u_l - v_l v_k =0$.

Because the network is strongly connected, one finds indeed for every pair of nodes $k$ and $l$ a path in the network that connects the two nodes, that is, $u_k u_l - v_l v_k =0$ for all $k,l$. Therefore, the two vectors $\vec{u}$ and $\vec{v}$ are \textit{not} linearly independent, which is in contradiction to the assumption at the beginning. 
In other words, the network $A$ cannot be a coexistence network and have a two-dimensional kernel.
This proves that coexistence networks with a two-dimensional kernel do not exist.\\

\textit{Outlook: Do coexistence networks with minimal kernel dimension $\geq 3$ exist?}
Our numerical simulations of coexistence networks with up to 9 nodes (see Section~4.3)
show that network topologies that give rise to high-dimensional kernels ($\mathrm{dim}(\mathrm{Ker}(A))\geq 3$) are not coexistence networks.
In other words, we have not found any coexistence network with minimal kernel dimension different from 1 thus far.
At present, it remains to us an open question for future research, whether this numerical observation generalizes to networks with $S\geq 10$: Do coexistence networks exist whose minimal kernel dimension is $\geq 3$?


\section{Proof of the conditions for Hamiltonian coexistence networks}
\label{sec:proof}

In the following, we supplement the proof for the main results stated in Section~4.1: A Hamiltonian network is a coexistence network if and only if in an ascending labeling all interior edges fulfill the coexistence conditions, that is the cycle condition~(9)
and the crossing condition~(10).
For convenience, we restate the coexistence conditions:
\begin{enumerate}
\item \textbf{Cycle condition:} For every interior edge $(i,j) \in E_{\text{in}}$ it holds that: 
\begin{equation*}
\begin{aligned}
&(i,j) \text{ is ascending, and } j-i \text{ is odd,}\\
&(i,j) \text{ is descending, and } i-j \text{ is even.}
\end{aligned}
\end{equation*}
An edge $(i,j)\in E_{\text{in}}$ is called \textit{ascending} (with respect to the labeling of the Hamiltonian cycle) if $i<j$, and \textit{descending} if $i>j$.

\item \textbf{Crossing condition:} For every pair of crossing interior edges $\{(i,j), (k,l) \}\subseteq E_{\text{in}}$ it holds that: 
\begin{equation*}
\begin{aligned}
&(i,j) \text{ and } (k,l) \text{ cross each other,}\\
&\text{and } \min (|i-k|,|j-l|) \text{ is even.}
\end{aligned}
\end{equation*}
Two interior edges $(i,j), (k,l) \in E_{\text{in}}$ are called \textit{crossing} if $\min (k,l )<i < \max (k,l)$ or $\min (k,l )<j < \max (k,l)$. If the Hamiltonian network is drawn in the two-dimensional plane, crossing edges cross in the interior of the cycle.
\end{enumerate}

\textit{Background and bridge to the main text.}
Because Hamiltonian networks of odd size are factor-critical the kernel is generically one-dimensional and determined by the adjugate vector~$\vec{r}$~(8);
see Section~3.3.
A Hamiltonian network is a coexistence network if all summands of all entries of the adjugate vector have the same sign. 
In this case, the kernel vector is strictly positive for all choices of weights.
Through the Pfaffian, each summand occurring in the adjugate vector stems from a near-perfect matching of the network; see graph-theoretical definition of the Pfaffian~(5).
The sign of each summand is determined by both the sign of the permutation of the corresponding near-perfect matching and the index of the vector-component, that is, the index of the node that is not part of the matching; see Figure~4
of the main text.

\subsection{Set-up of the proof}
%

\textit{Notation: adjugate vector $\vec{r} = \vec{r}_\text{cycle} + \vec{r}_\text{in}$.}
Recall that we label the nodes of a Hamiltonian network $E(\mathcal{N}) = E_{\text{cycle}} \cup E_{\text{in}}$ with $E_{\text{cycle}}  \cap E_{\text{in}} =\emptyset $ in an ascending manner such that the Hamiltonian cycle is given by the edge-set $E_{\text{cycle} }= \{(1,2), (2,3), \dots, (S-1,S), (S,1)\}$. Similar to the example in Section~4.1,
we separate the adjugate vector of this network into two components, such that $\vec{r} = \vec{r}_\text{cycle} + \vec{r}_\text{in}$.
In this notation, $\vec{r}_\text{cycle}$ contains only contributions from the edges that form the Hamiltonian cycle $E_{\text{cycle}}$. 
The vector $\vec{r}_{\text{in}}$ contains all contributions stemming from the interior edges $E_{\text{in}}$.
This separation is always possible because the presence of (near-)perfect matchings containing only cycle edges is not altered by interior edges. Adding interior edges only leads to additional near-perfect matchings.\\

\textit{$\vec{r}_\text{cycle}$ as the reference for the comparison of signs.}
Since an odd cycle is a coexistence network (see also Section~2.3),
all components of $\vec{r}_{\text{cycle}}$ have the same sign. 
As a consequence, a Hamiltonian network is a coexistence network if for every deleted node $i$ all permutations of all near-perfect matchings have the same sign as the permutation of the near-perfect matching containing only edges of the cycle. 
This way, we compare all summands in $r_{\text{in}, i}$ with the one summand in $r_{\text{cycle}, i}$. 
If these signs agree for every component, the signs of all summands agree because they do so for $\vec{r}_{\text{cycle}}$. Effectively, the adjugate vector of the cycle $\vec{r}_\text{cycle}$ serves as a reference for the signs that are induced by the interior edges through near-perfect matchings. \\

\textit{Notation: relabel nodes such that $\hat{i}\to\hat{S}$.}
Note that the property of being a coexistence network is independent of the order in which the nodes of the network are labeled. In algebraic terms, a relabeling of the nodes is achieved by multiplying the adjacency matrix with permutation matrices, which does not change the spectral properties of the network and the structural properties of the kernel.
Therefore, if all signs in the adjugate vector are equal in one selected labeling, the signs are equal in every labeling. 
For convenience, we relabel the nodes of the system such that the node, which is not part of the near-perfect matching, has the index $S$ by shifting the node labels as $i \to S, i+1 \to 1, \dots$. 
This shift facilitates an easier comparison of signs because the permutation corresponding to the perfect matching of the cycle after relabeling is the identity (that is, $(1 \ 2 \ \dots \ S-1)$) and, thus, has the sign $(+1)$.\\

 \textit{Directed paths contribute to near-perfect matchings if they are even-sized.}
Furthermore, we denote the set of edges of the network, from which the $i$th node was deleted, as $E_{\text{cycle}, \hat{i}}$ and after relabeling as $E_{\text{cycle}, \hat{S}}$. After relabeling, the remaining edges of $E_{\text{cycle}, \hat{S}}$ form a directed path from $1$ to $S-1$.
In the following, we also denote such a directed path as $P[1,S-1] = \{ (1, 2), (2, 3), \dots, (S-2, S-1)\}$. 
In general, a directed path $P[i,j] = \{(i, i+1), (i+1, i+2), \dots, (j-1,j) \}$ has a perfect matching if the number of nodes in the path $|P| = j-i+1$ is even and, thus, $j-i$ is odd. Thus, only directed paths of even length contribute to perfect matchings.\\

\textit{Structure of the proof.} 
First, we prove that a Hamiltonian coexistence network is a coexistence network if the coexistence conditions~(9) and~(10)
are fulfilled.
We show that the permutations all near-perfect matchings which contain (one or several) interior edges have the same sign as the near-perfect matching of the cycle. 
This step proves that the coexistence conditions are sufficient; see Section~\ref{sec:proof1}.
Second, we consider all cases in which interior edges do not fulfill the coexistence conditions. We show that, under this assumption, near-perfect matchings arise that have a different sign than the near-perfect matching of the cycle. This step proves that the coexistence conditions are necessary; see Section~\ref{sec:proof2}.\\


%
\begin{figure*}[th!]
\centering
\includegraphics[width=0.95\textwidth]{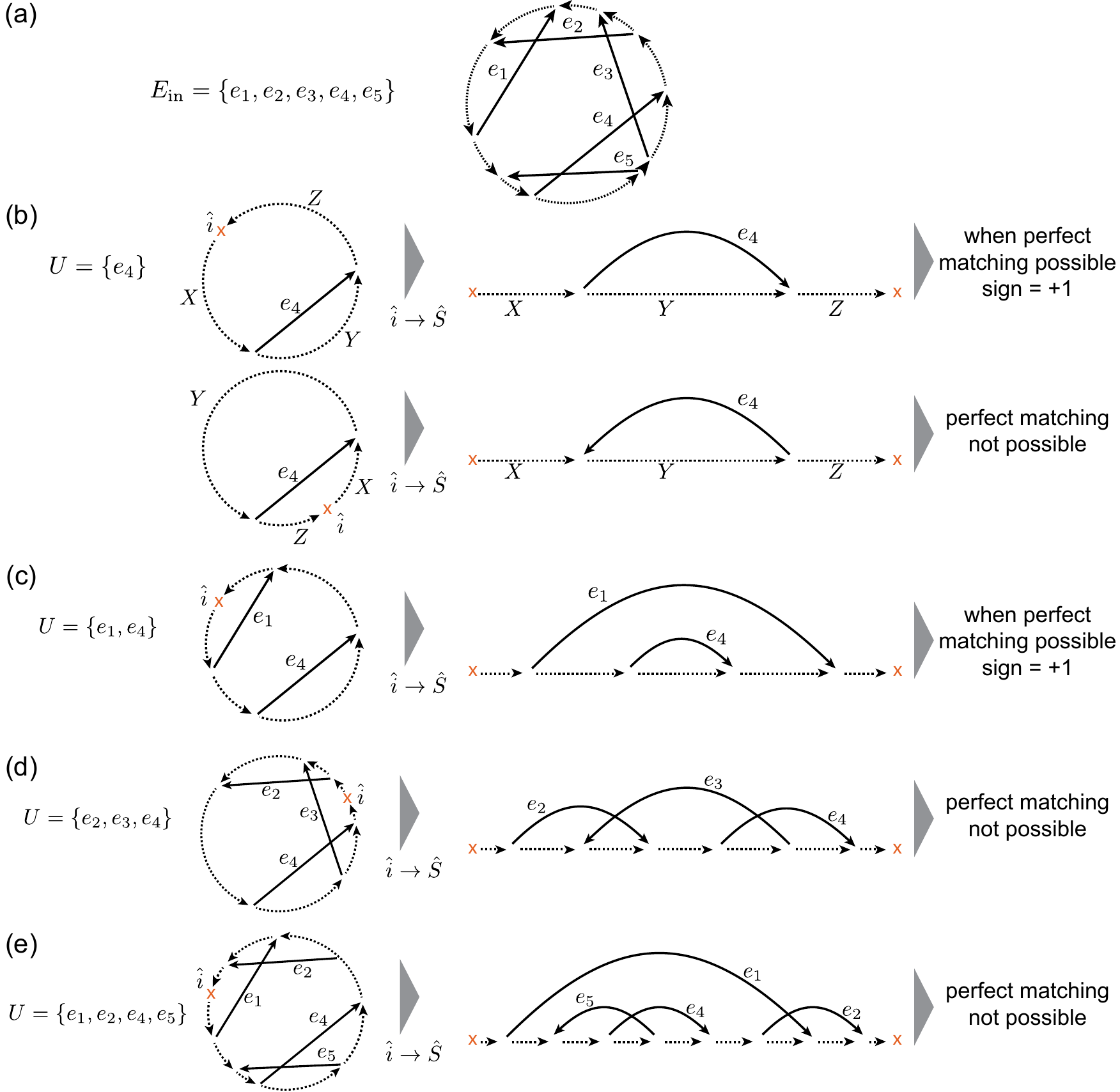}
\caption{(Color online)
\textbf{Illustration of important steps of the proof for Hamiltonian coexistence networks.}
(a) Hamiltonian network with internal edges $e_1,...,e_5$. This network is a coexistence network if for every subset $U\subseteq E_{\text{in}}$ and relative to every deleted vertex $\hat{i} \to \hat{S}$ all near-perfect matchings have a positive sign.
(b) Case of a single edge, $|U|=1$: The edge $e_4$ is either ascending or descending, depending on its position relative to the deleted vertex.
(c) Case of covering edges: Both vertices of the edge $e_4$ lie between start- and end-vertex of $e_1$, such that edge $e_1$ covers the edge $e_4$. Since both edges are ascending, a perfect matching exists with sign $+1$.
(d) Case of crossing edges: The edges $e_2, e_3$, and $e_4$ form a crossing set. Thus, a perfect matching does not exist. The lowest descending edge is $e_3$ because it does not cover any other descending edges. 
(e) Case of two crossing sets: The crossing set consisting of $e_1$ and $e_2$ covers the crossing set of $e_4$ and $e_5$. 
}
\label{fig:ExplanationsProof}
\end{figure*}
%

\subsection{Sufficiency of the conditions~(9) and~(10) for coexistence networks}
\label{sec:proof1}

First, we show that a Hamiltonian network is a coexistence network if all of its interior edges $E_{\text{in}}$ fulfill the coexistence conditions~(9) and~(10).
To show sufficiency of the conditions, we assume a Hamiltonian network in which all interior edges fulfill the coexistence conditions.
We investigate all near-perfect matchings that can arise through the presence of interior edges and discuss their sign as follows. 
For the subnetwork created by deleting node $i$ (denoted by $A_{\hat{i}}$) we 
(i) relabel the nodes in the manner mentioned above, such that $\hat{i}\to \hat{S}$,
(ii) analyze for every subset $U\subseteq E_{\text{in}}$ whether it can be completed to a perfect matching of $A_{\hat{S}}$ using only edges from $E_{\text{cycle},\hat{S}}$ (that is, the edges of the cycle that remain after deleting node $i$ and after relabeling $\hat{i}\to \hat{S}$), 
(iii) then determine the sign of the resulting perfect matching, 
(iv) and compare the sign with the sign of the perfect matching of $A_{\hat{i}}$ that includes only edges from $E_{\text{cycle},\hat{i}}$ (that is, the near-perfect matching of the cycle; its sign is referred to as $\text{sign} (\sigma_\mathds{1})$).
We discuss all cases for possible subsets of interior edges $U\subseteq E_\text{in}$ in the following.\\

\textit{Trivial cases: $U = \emptyset$, $\hat{S} \in U$, and edges starting or ending in the same node.} 
For the case $U = \emptyset$, there are no contributions from interior edges to any near-perfect matching of the network. When the deleted vertex $\hat{S}$ is part of an edge in $U$, a near-perfect matching including all edges in $U$ and excluding $S$ does not exist. Furthermore, when two or more edges in $U$ share the same node, they cannot be part of the same perfect matching. In the following, these trivial cases are disregarded.\\

\textit{$U$ contains a single edge ($|U| = 1$).}
The simplest case, for which interior edges can contribute to the network's near-perfect matchings is $|U| =1$. 
In this case, only the cycle condition~(9)
applies to the edge in $U$, but not the crossing condition~(10).
For this case, we denote $U = \{e\}$ with $e = (e_{\text{start}},e_{\text{end}})$ after relabeling of the cycle ($\hat{i}\to \hat{S}$) and discuss the contribution of edge $e$ to a perfect matching of $A_{\hat{S}}$.

If the nodes of the interior edge $e$ contribute to the perfect matching of $A_{\hat{S}}$, the unmatched nodes are $\{1, 2,\dots, S-1 \} \backslash \{ e_{\text{start}}, e_{\text{end}} \}$. 
After deleting $e$, the remaining edges of $E_{\text{cycle}}$ form three paths: 
$X = P[1, \min(e_{\text{start}},e_{\text{end}})-1], 
Y = P[\min( e_{\text{start}},e_{\text{end}})+1,\max( e_{\text{start}},e_{\text{end}})-1]$, and 
$Z = P[\max( e_{\text{start}},e_{\text{end}})+1, S-1]$; see Figure~\ref{fig:ExplanationsProof}(b).
A perfect matching of $A_{\hat{S}}$ including $e$ is possible if and only if all three created paths have a perfect matching on their own, that is, the difference between highest and lowest node is odd for each of the three paths (the number of vertices in the paths, $|X|, |Y|$, and $|Z|$, are even; that is, the paths $X, Y$, and $Z$ are of even length).
We distinguish two cases in the following: the edge $e$ is ascending or descending (again, meant after relabeling $\hat{i}\to \hat{S}$).

\begin{itemize}
\item \textit{Case (i): the edge $e$ is ascending: all near-perfect matchings have the same sign as the near-perfect matching stemming from the cycle.}
 In this case, $e_{\text{start}} < e_{\text{end}}$ and $e_{\text{end}}- e_{\text{start}}$ is odd (by the cycle condition~(9)).
Therefore, $(e_{\text{end}}-1)  - ( e_{\text{start}}+1)$ is odd and $Y$ has a perfect matching.
The perfect matching of $A_{\hat{S}}$ including the edge $e$ does indeed exist if, in addition, $|X|$ and $|Z|$ are even.
 
If this perfect matching exists, its corresponding permutation has the same sign as the perfect matching stemming from the cycle. If the paths $X, Y$ and $Z$ are of even length, $e_{\text{start}}$ is odd and $e_{\text{end}}$ is even. The permutation corresponding to a perfect matching induced by $e$ is given by:

\begin{align*}
\sigma_{\alpha_e} &= (e_{\text{start}}\ e_{\text{end}}\  1 \  2\ \dots\ e_{\text{start}}-1\ e_{\text{start}}+1\ \dots \  e_{\text{end}}-1 \ e_{\text{end}} +1 \  \dots \ S-1)\ .
\end{align*}
The sign of this perfect matching of $A_{\hat{S}}$ is determined by the number of transpositions necessary to transform the permutation into the identity permutation $(\sigma_\mathds{1})$, and is computed as:
\begin{align*}
\text{sign} (\sigma_{\alpha_e} ) &= \text{sign} (e_{\text{start}}\ e_{\text{end}}\  1\ \dots \ e_\text{start}-1 \ e_\text{start}+1 \ \dots \ e_\text{end}-1\ e_\text{end}+1\ \dots \ S-1)\ , \\
&=(-1)^{(2 \cdot ((e_\text{start}-1)+1))} \text{sign} (1\ \dots \ e_\text{start}-1 \  e_{\text{start}}\ e_{\text{end}} \  e_\text{start}+1 \ \dots \ e_\text{end}-1\ e_\text{end}+1\ \dots \ S-1)\ , \\
&=(-1)^{( (e_\text{end}-1)-(e_\text{start}+1)+1)} \text{sign} (1\ \dots \ e_\text{start}-1 \ e_{\text{start}}\ e_\text{start}+1 \ \dots \ e_\text{end}-1 \ e_{\text{end}} \ e_\text{end}+1\ \dots \ S-1)\ , \\
& = (+1) \text{sign} (\sigma_\mathds{1}) = +1\ .
\end{align*}

In other words, all near-perfect matchings arising from a single ascending interior edge have the same sign as the near-perfect matching that is constituted by edges from $E_{\text{cycle}}$ only.

%
\item \textit{Case (ii): the edge $e$ is descending: no contribution to a near-perfect matching.}
In this case, $e_{\text{end}} < e_{\text{start}}$ and $e_{\text{start}}-e_{\text{end}}$ is even (according to the cycle condition~(9)).
Thus, $Y$ is of odd length, and $e$ cannot contribute to a perfect matching of $A_{\hat{S}}$ as a single interior edge. For later purposes, note that for a descending edge, start node, $e_{\text{start}}$, and end node, $e_{\text{end}}$, have the same parity.
\end{itemize}

\textit{$U$ contains several edges ($|U| > 1$).}
Next, we consider the case that the set $U\subseteq E_\text{in}$ contains several edges fulfilling the coexistence conditions~(9) and~(10).
First, we study the possibility that all interior edges in $U$ do not cross each other. 
In this case the cycle condition~(9)
ensures that all near-perfect matchings that can possibly arise have the same sign as the near-perfect matching stemming from the edges of the cycle.
Second, we study the cases in which interior edges cross each other and the crossing condition~(10)
becomes relevant. In particular, we introduce the notion of a crossing set $E_{\times} \subseteq E_{\text{in}}$.
We show that no near-perfect matchings are created if $U$ contains exactly two crossing edges.
This result is generalized to any $U\subseteq E_\text{in}$ that contains a crossing set. 

\begin{itemize}
\item \textit{None of the edges in $U$ cross each other.}
In generalization of the reasoning for $|U|=1$, a near-perfect matching can only arise from all non-crossing edges in $U$ and edges of the cycle if all created paths are of even length. Two different cases can occur. 

\begin{itemize}
\item
\textit{Case (i): All non-crossing edges in $U$ are ascending: all near-perfect matchings have the same sign as the near-perfect matching stemming from the cycle.} 
If all edges in $U$ are ascending, for all of its edges, the path between start and end node is of even length (following the cycle condition~(9),
as above). When $U$ contains $k$ ascending edges, after deleting all edges of $U$, the remaining edges of $E_{\text{cycle}}$ form $2k+1$ paths (possibly of zero length), see Figure~\ref{fig:ExplanationsProof}(c)). Since the edges do not cross, the length of every path is either determined by the vertices of an ascending edge (similar to $Y$ in the case $|U|=1$) or by the relative placement of two ascending edges (similar to $X$ and $Z$ in the case $|U|=1$). As argued before, ascending edges lead to a path of even length, such that only the relative placement of edges determines whether a perfect matching of $A_{\hat{S}}$ exists.

The sign of the permutation corresponding to such a perfect matching is equal to the sign of the perfect matching stemming from $E_\text{cycle}$ because an even number of transpositions is needed to order its elements in size (all paths are of even length). 
In particular, all nodes of interior ascending edges can be ordered in size by pairwisely ordering them with an even number of transpositions. 
For example, if $e_1 = (e_{1,\text{start}}\ e_{1, \text{end}})$ and $e_2 = (e_{2,\text{start}}\ e_{2, \text{end}})$ with $e_{2, \text{end}}<e_{1,\text{start}}$, four transpositions are needed to order these indices in size:
\begin{align*}
\hspace{1cm} \text{sign} (e_{2,\text{start}}\ e_{2, \text{end}} \  e_{1,\text{start}} \ e_{1,\text{end}}\ 1 \ \dots \ S-1)
= 
(-1)^{(2\cdot2)}\text{sign} (e_{1,\text{start}} \ e_{1,\text{end}}\  e_{2,\text{start}}\ e_{2, \text{end}}\ 1 \ \dots \ S-1 )\ .
\end{align*}
%
If, on the other hand, $e_{2, \text{start}}<e_{1,\text{start}}$ and $e_{1, \text{end}}<e_{2,\text{end}}$ (the vertices of $e_1$ lie between the vertices of $e_2$, $e_2$ covers $e_1$), two transpositions are necessary to order these nodes in size (see Figure~\ref{fig:ExplanationsProof}(c)):
\begin{align*}
\hspace{1cm} \text{sign} (e_{2,\text{start}}\  e_{1,\text{start}}\ e_{1, \text{end}}\  \ e_{2,\text{end}}\ 1 \ \dots \ S-1)
= 
(-1)^{(2\cdot1)}\text{sign} (e_{1,\text{start}}\ e_{1, \text{end}} \  e_{2,\text{start}} \ e_{2,\text{end}}\ 1 \ \dots \ S-1)\ .
\end{align*}
%
%
This way, all nodes of the interior ascending edges can be ordered in size by pairwisely ordering them with an even number of transpositions.
Because all paths stemming from cycle edges are of even length, further ordering the ordered nodes from interior edges within the nodes of the cycle edges involves an even number of transpositions.
Thus, an overall even number of transpositions is needed to order the elements of the partition corresponding to the perfect matching in size.

\item
\textit{Case (ii): $U$ contains at least one non-crossing descending edge: no contribution to near-perfect matchings.}
If at least one non-crossing descending edge is contained in $U$, one path of odd length is created for every arrangement of non-crossing edges. This path of odd length cannot contribute to a near-perfect matching of the network. 
\end{itemize}

\item \textit{Definition of a crossing set $E_\times$.} 
All edges in $U$ that are related by condition~(10)
form a so-called crossing set $E_{\times}$. 
Pictorially speaking, in a crossing set all edges are related by the crossing condition.
More precisely, all pairs of edges in a crossing set $f,g \in E_{\times}$ either cross each other (their relative placement is explicitly constrained by the crossing condition~(10),
or there exists a series of edges $f,e_1,...,e_l,g \in E_{\times}$ that consecutively cross each other (the relative placement of $f$ and $g$ is implicitly constrained by the intermediate edges $e_1,...,e_l$).
For example, in Figure~\ref{fig:ExplanationsProof}(d) the edges $e_2$ and $e_4$ are contained in one crossing set because both edges cross the edge $e_3$. Thus, $E_{\times} = \{ e_2,e_3,e_4\}$ is a crossing set. 
In Figure~\ref{fig:ExplanationsProof}(e), the set $U$ consists of two disjoint crossing sets $U = E_{\times}^{(1)} \cup E_{\times}^{(2)}$ with $E_{\times}^{(1)} = \{e_1,e_2\}$ and $E_{\times}^{(2)} = \{e_4,e_5\}$. 
Note also that the edges $e_2$ and $e_4$ belong to one crossing set for the choice of $U$ as depicted in Figure~\ref{fig:ExplanationsProof}(d), but not for the choice of $U$ in Figure~\ref{fig:ExplanationsProof}(e). Hence, both the occurrence and the elements of crossing sets do not only depend on the interior edges of a network, but also on the choice of $U$ from which near-perfect matchings of the network are constructed.

\begin{figure*}[th!]
\centering
\includegraphics[width=0.95\textwidth]{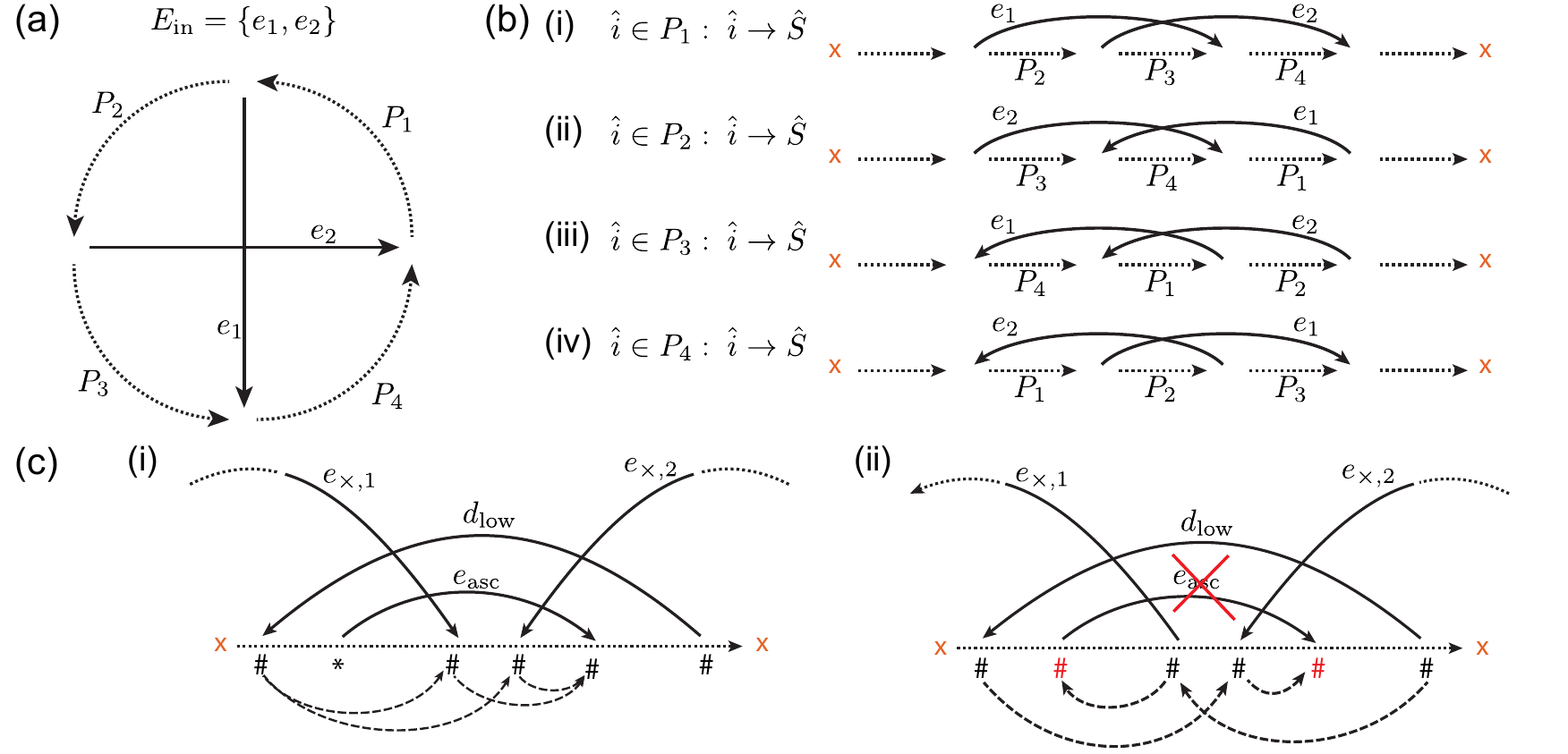}
\caption{(Color online)
(a) Schematic illustration of Hamiltonian coexistence network with two crossing edges $e_1$ and $e_2$. $P_1, \dots, P_4$ are the paths constituted by the edges of $E_{\text{cycle}}$ that do not include any vertices of $e_1$ and $e_2$.
(b) Because the network is a coexistence network, the crossing condition~(10)
is fulfilled for every ascending labeling. Thus, for the deletion of every vertex $i$ that is not part of $e_1$ and $e_2$, after relabeling $i \to S$, fulfilllment of the crossing condition constraints the length of the paths $P_1, \dots, P_4$. Hence, $|P_1|, |P_2|$ and $|P_4|$ are odd and $P_3$ is even.
(c) The lowest descending edge $d_{low}$ does not cover other descending edges. The edges $e_{\times,1}$ and $e_{\times,4}$ cross both $d_{low}$ and $e_{\text{asc}}$, such that all edges form one crossing set.
On the left, (i), the edges $e_{\times,1}$ and $e_{\times,2}$ both end between the vertices of a covered ascending edge and fulfill the crossing condition pairwisely, indicated by $\#$ and $\ast$ as placeholders for the vertices' parity (odd and even). 
On the right, (ii), the edge $e_{\times,1}$ starts and the edge $e_{\times,2}$ ends between the vertices of a covered ascending edge. In this arrangement, the crossing condition cannot be pairwisely fulfilled.
}
\label{fig:RelativePlacement}
\end{figure*}


\textit{Consequences of the crossing condition~(10)
for the relative placement of two crossing edges.}
We briefly discuss the consequences of the crossing condition for the relative placement of two crossing edges.
One pair of crossing edges divides a cycle into four paths $P_1, \dots, P_4$, as illustrated in Figure~\ref{fig:RelativePlacement}. 
Because the crossing condition is fulfilled in the chosen ascending labeling of the cycle, either $|P_2|$ or $|P_4|$ is the shortest path length either between the start points or between the end points.
Thus, either $|P_2|$ or $|P_4|$ is odd, or both are odd. 
Furthermore, because every interior edge also fulfillls the cycle condition~(9),
only directed cycles of odd length are created when combining every single interior edge with the Hamiltonian cycle. Thus, both $|P_4| + |P_1| + 3$ (the cycle consists of the start and end node of $e_1$, the nodes constituting $P_4$, the end node of $e_2$ and the nodes constituting $P_1$) and $|P_1| + |P_2| + 3$ are odd; see Figure~\ref{fig:RelativePlacement}(a).
Consequently, the numbers of nodes in each of the three paths $P_1, P_2$, and $P_4$ are odd.
It follows that $|P_3|$ is even because the overall number of nodes in the system $|P_1| + |P_2| + |P_3| + |P_4| + 4$ is odd. 
In total, two crossing interior edges create two cycles of odd length that share an odd number of nodes.

The same consequences for the placement of crossing edges are, of course, obtained if one considers an explicit labeling of the cycle with reference to the deleted index $\hat{i}\to \hat{S}$ as depicted in Figure~\ref{fig:RelativePlacement}(b)(i)-(iv).
Depending on the position of the deleted node $\hat{S}$, the two crossing edges $e_1$ and $e_2$ are arranged differently with respect to each other. 
In line with the above arguments, it follows that the node of edge $e_2$ that lies between the nodes of $e_1$ has the same parity as its counterpart of $e_1$ irrespective of the labeling.

\textit{$U$ contains only two crossing edges ($ U = E_\times$ and $|E_\times| = 2$): no contribution to near-perfect matchings.}
Now, we discuss whether near-perfect matchings arise in case $U$ contains exactly two crossing edges; see Figure~\ref{fig:RelativePlacement}(a).
Because the crossing edges fulfill the coexistence conditions~(9) and~(10),
either $P_2$ or $P_4$ build up the minimal distance between the two edges. 
After deleting the nodes of the edges $e_1$ and $e_2$ (as contributions to the near-perfect matching of the network $\mathcal{N}(A_{\hat{S}})$), either path $P_2$ remains or path $P_4$ remains, or both paths remain to be matched, Figure~\ref{fig:RelativePlacement}(b). 
Because both $P_2$ and $P_4$ have an odd number of nodes (see above), they do not have a perfect matching. 
Therefore, there does not exist any near-perfect matching that contains a single pair of crossing edges. 


%
%
%
%
%
%

\textit{$U$ contains only one crossing set of several ascending edges ($U=E_\times$ and $|U| \geq 3$ and $e_{i,\text{start}}<e_{i,\text{end}}$ for all $i = 1, \dots |U|$): no contribution to near-perfect matchings.}
When all edges in $U$ are ascending (again, ascending is meant after relabeling as above) and form a single crossing set, the above reasoning can be readily generalized. 
In particular, all start nodes have the same parity (for example, even) and all end nodes have the opposite parity (in this example, odd). 
The two lowest nodes of these ascending edges are start nodes and, thus, have the same parity. Therefore, these two start nodes enclose a path with an odd number of nodes, which does not have a perfect matching. 
Thus, a near-perfect matching containing a crossing set with only ascending edges does not exist.

\textit{$U$ contains only one crossing set with at least one descending edge ($U=E_\times$ and $|U| \geq 3$ and $e_{i,\text{start}}>e_{i,\text{end}}$ for at least one $i \in \{ 1, \dots |E_\times|\}$): no contribution to near-perfect matchings.}
First, we discuss some implications of the coexistence conditions~(9) and~(10)
for relative placement of an arbitrary edge crossing a descending edge.
For a descending edge $d$ the cycle condition~(9)
implies that both of its nodes have the same parity; see above. 
From the crossing condition~(10)
it follows that, if an arbitrary edge $e$ crosses a descending edge $d$, the node of $e$ that is placed between end and start node of $d$ has the same parity as the nodes of $d$. 
For example, in Figure~\ref{fig:RelativePlacement}(b)(ii), the end node of $e_2$ has the same parity as both nodes of $e_1$. 
For the configuration of crossing descending edges (as depicted in Figure~\ref{fig:RelativePlacement}(b)(iii)), the coexistence conditions imply that all nodes of the two edges have the same parity because the start-node of $e_1$ lies between end and start node of $e_2$, and because $e_1$ is descending. 

We now use these arguments to show that a crossing set with at least one descending edge is not part of a near-perfect matching.
In any crossing set $E_{\times}$ with descending edges we find (at least) one descending edge, to which we refer as $d_\text{low}$, that does not cover any other descending edges. In other words, there is no other descending edge in $E_{\times}$ for which both nodes lie between $d_{\text{low}, \text{end}}$ and $d_{\text{low},\text{start}}$.
Two cases can occur for this descending edge $d_\text{low}$. 

\begin{itemize}

\item \textit{Case (i): The descending edge $d_\text{low}$ does not cover any ascending edges: no contribution to near-perfect matchings.}
In this case $d_\text{low}$ does not cover any descending or ascending edges (see Figure~\ref{fig:ExplanationsProof}). Thus, any node that is covered by the edge $d_\text{low}$ and that is part of an edge of $E_{\times}$ belongs to an edge that crosses $d_\text{low}$.
As a consequence of the crossing condition~(10),
all of these nodes have the same parity and, thus, enclose paths of odd length. Therefore, this arrangement of crossing edges cannot be completed to form a perfect matching.

\item \textit{Case (ii): The descending edge $d_\text{low}$ covers ascending edges: no contribution to near-perfect matchings.}
In the following we show that every crossing set $U$ in which a descending edge covers ascending edges leads to at least one odd path and thus cannot be part of a near-perfect matching.

First, we consider the case that $d_{low}$ covers only non-crossing ascending edges, see Figure~\ref{fig:RelativePlacement}. 
Recall that $U$ contains only one crossing set, such that there are edges $e_{\times,i}$ that cross both $d_{low}$ and the covered ascending edge $e_{\text{asc}}$. A pairwise fulfilllment of the crossing condition~(10)
enforces that either all start or all end vertices of the edges $e_{\times,i}$ lie between the vertices of $e_{\text{asc}}$, see Figure~\ref{fig:RelativePlacement} (c)(i). When the start vertex of $e_{\times,1}$ and the end vertex of $e_{\times,2}$ lie between the vertices of $e_{\text{asc}}$ an arrangement consistent with the crossing condition cannot exist, see Figure~\ref{fig:RelativePlacement} (c)(ii).
Thus, either the two highest vertices of $U$ that lie between the vertices of $d_{\text{low}}$ (in case all $e_{\times,i}$ end between the vertices of $e_{\text{asc}}$) or the two lowest vertices (in case all $e_{\times,i}$ start between the vertices of $e_{\text{asc}}$) have the same parity. The arrangement leads to paths of odd length and cannot give rise to a perfect matching.

The same argument holds when $d_{\text{low}}$ covers a crossing set of ascending edges $E_{\times,\text{asc}}$.
As stated above, in a crossing set consisting only of ascending edges all start vertices have the same parity, while all end-vertices have the other parity. As above, in an arrangement consistent with the crossing condition the edges $e_{\times,i}$ that cross both $d_{\text{low}}$ and $E_{\times,\text{asc}}$ either all start or all end between the vertices of $E_{\times,\text{asc}}$. Hence, either the two highest vertices of $E_{\times,\text{asc}} \cup e_{\times,i} $, or the two lowest vertices of $E_{\times,\text{asc}} \cup e_{\times,i}$ (or both) have the same parity. 

These arguments show that every crossing set containing at least one descending edge enclose at least one path of odd length. Therefore, this arrangement of edges cannot be completed to form a perfect matching. 
\end{itemize}

Thus, for all cases of a single crossing set $E_{\times}\subseteq U$ for which all interior edges fulfill the coexistence conditions~(9) and~(10),
no contributions to near-perfect matchings occur. Therefore, no contributions to the adjugate vector arise through a single crossing set.

\item \textit{$U$ contains several crossing sets and single edges: no contribution to near-perfect matchings.} 
The above result for one crossing set readily generalizes to a set $U$ that consists of several crossing sets and further non-crossing edges.
Note that every $U$ can be decomposed into pairwisely disjoint sets of crossing sets and sets of single edges.
If a crossing set is contained in $U$, there exists a path of odd length between two nodes of edges contained in the crossing set. 
Thus, this path does not have a perfect matching, neither when its nodes are ordered (that is, if the path consists of edges from the cycle), nor when it its nodes are permuted (that is, if edges from the crossing set cover other edges).
We conclude that a near-perfect matching cannot contain arbitrary combinations crossing edges.


\end{itemize}

\textit{If the coexistence conditions~(9) and~(10)
are fulfilled, the network is a coexistence network.} 
In total, we have shown that for a network in which all interior edges satisfy the coexistence conditions, the adjugate vector consists of summands all of which have the same sign. 
Thus, the network topology is a coexistence network.
In particular, it was shown that through the cycle condition~(9)
only ascending single edges contribute to near-perfect matchings, but not descending edges. The resulting near-perfect matchings have the same sign as the near-perfect matching stemming from the Hamiltonian cycle. 
The crossing condition~(10)
implies that crossing edges do not occur in a near-perfect matching. 

\subsection{Necessity of the conditions~(9) and~(10) for coexistence networks}
\label{sec:proof2}

In the second part of this proof, we show that the conditions~(9) and~(10)
are also necessary for coexistence networks. 
To this end, we show that a Hamiltonian network $\mathcal{N}$ with edge-set $E(\mathcal{N}) = E_{\text{cycle}} \cup E_{\text{in}}$ is not a coexistence network if either condition~(9)
or condition~(10)
is not fulfilled.\\

\textit{Condition~(9)
is not fulfilled for one interior edge: at least one induced near-perfect matching has a different sign than the near-perfect matching stemming from the cycle.}
Assume that there is an edge $e \in E_{\text{in}}$ that violates condition~(9).
We choose the component of the adjugate vector such that after relabeling $\hat{i}\to \hat{S}$ the edge has the form $e = (e_{\text{start}},e_{\text{end}}) = (e_{\text{start}}, 1)$ with $2 < e_{\text{start}} <S -1$, and consider perfect matchings of $A_{\hat{S}}$. 
In this labeling, $e$ is descending. Thus, violating the cycle condition~(9)
implies that $e_{\text{start}}$ is even.
Deleting the nodes $e_{\text{start}}$ and $e_{\text{end}} = 1$ from $E_{\text{cycle},\hat{S}}$ creates the two separated paths $P[2, e_{\text{start}}-1]$ and $P[e_{\text{start}}+1,S-1]$. 
Both paths are of even length and, thus, have perfect matchings.
Therefore, a summand that includes the edge $e$ contributes to the $S$th component of the adjugate vector~(8).

However, this summand has the opposite sign than the summand stemming from the Hamiltonian cycle alone, as we show in the following.
To determine the sign of the perfect matching arising from the interior edge $e$ for the $S$th node deleted, we compute the number of transpositions needed to obtain the identity permutation:

\begin{align*}
\text{sign}[\sigma_{\alpha_{e}}] &= \text{sign} (e_{\text{start}}\ 1\ 2\ \dots \ S-1) = - \text{sign}(1\ e_{\text{start}} \ 2\ \dots\  S-1 )\ ,  \\
&=- (-1)^{((e_{\text{start}}-1)-1+1)} \text{sign}(1\  2\ \dots\ e_{\text{start}}-1 \ e_{\text{start}}\ e_{\text{start}}-1 \ \dots \ S-1)\ , \\
&= -\text{sign}(\sigma_\mathds{1}) = -1\ .
\end{align*} 

Thus, the interior edge $e$ contributes with a summand to the adjugate vector that has a different sign than the identity permutation. 
Therefore, if a network contains at least one interior edge violating the cycle condition, at least one summand in one component of the adjugate vector has the opposite sign compared to the contribution stemming from the cycle in that component of the adjugate vector~(8).
Such a network is not a coexistence network.\\

\textit{The crossing condition~(10)
is not fulfilled for two crossing edges: at least one induced near-perfect matching has a different sign than the near-perfect matching stemming from the cycle.}
Next, we consider a network with interior edges $E_{\text{in}}$ that fulfill the cycle condition, but violate the crossing condition. 
In other words, there exist two edges $e_1, e_2 \in E_{\text{in}}$ that do not fulfill the crossing condition. 
We choose the component of the adjugate vector such that after relabeling $\hat{i}\to \hat{S}$ the edges take the form $e_{1,\text{end}}=1 < e_{2,\text{end}}< e_{1,\text{start}} < e_{2,\text{start}}<S$. 
fulfilling the cycle condition and violating the crossing condition implies that $e_{1,\text{end}}$ is odd, whereas $e_{2,\text{start}}$ and $e_{2,\text{end}}$ are even. 
In this labeling, we consider now the edges of $E_{\text{cycle}, \hat{S}}$, that is, perfect matchings of $A_{\hat{S}}$.
By deleting all edges that contain the nodes of $e_1$ and $e_2$ from $E_{\text{cycle}, \hat{S}}$, the paths $P[2,e_{2,\text{end}}-1], P[e_{2,\text{end}}+1,e_{1,\text{start}}-1], P[e_{1,\text{start}}+1,e_{2,\text{start}}-1]$, and $P[e_{2,\text{start}}+1, S-1]$ remain. 
Because for every path start node and end node have different parity, all paths contain an even number of nodes. In other words, a near-perfect matching of $A_{\hat{S}}$ is possible.

The sign of the permutation of that near-perfect matching is obtained as follows:

\begin{align*}
\text{sign}(\sigma_{\alpha_{e_1,e_2}}) &= \text{sign} ( e_{1,\text{start}}\  1\ e_{2,\text{start}}\ e_{2,\text{end}}\  2\ \dots\ S-1)
= \text{sign}(1,e_{1,\text{start}}\ e_{2,\text{end}},e_{2,\text{start}}\ 2\  \dots\ S-1)\ , \\
&= (-1)^{3}\ \text{sign} (1\ e_{2,\text{end}}\  e_{1,\text{start}}\ e_{2,\text{start}}\ 2\ \dots\ S-1)\ ,\\
&= - (-1)^{((e_{2,\text{start}}-1) -3) } \\
& \hspace{1cm}  \text{sign} (1\ e_{2,\text{end}}\  e_{1,\text{start}}\ 2\ \dots\ e_{2,\text{end}}-1\  e_{2,\text{end}}+1\  \dots\ e_{1,\text{start}}-1\ e_{1,\text{start}}+1\ \dots\ e_{2,\text{start}}\ \dots\ S-1)\ ,
\\
&= - (-1)^{((e_{1,\text{start}}-1) -2 )} \text{sign} (1\ e_{2,\text{end}}\ 2\ \dots\ e_{2,\text{end}}-1\  e_{2,\text{end}}+1\  \dots\  e_{1,\text{start}} \  \dots\ S-1)\ ,\\
&= - (-1)^{((e_{2,\text{end}}-1) -1) } \text{sign}(1\   2\  \dots  \  e_{2,\text{end}} \ \dots\ S-1)\ ,\\
& =- \text{sign}(\sigma_\mathds{1}) = -1\ .
\end{align*} 

Thus, if two interior edges of a given network do not fulfill the crossing condition, at least one near-perfect matching exists that has a different sign than the corresponding near-perfect matching of the cycle. 
Thus, such a network is not a coexistence network.\\

\textit{If the coexistence conditions~(9) or~(10)
are not fulfilled, the network is not a coexistence network.} 
In total, we have shown that the existence of (i) one interior edge that does not fulfill the cycle condition~(9)
or (ii) two interior edges that fulfill the cycle condition, but not the crossing condition~(10)
implies that the network is not a coexistence network. 
Thus, the coexistence conditions are necessary for a Hamiltonian network to be a coexistence network.

\section{Methods: Numerical determination of coexistence networks}\label{sec:numerical}

We numerically determined all coexistence networks for $S\leq 9$ nodes with two different methods. 
In our first approach, we used conditions~(9) and~(10)
to constructively generate all coexistence networks.
Through our second approach, we determined algebraically all coexistence networks to verify our first approach.

\subsection{Method 1: Construction of coexistence networks via conditions~(9) and~(10)}\label{sec:numerical1}

To generate coexistence networks, we implemented conditions~(9) and~(10)
as a constructive algorithm. By successively adding edges fulfilling both conditions to a cycle of odd size $(S\to 1\to 2\to\dots\to S)$, we constructed an exhaustive list of Hamiltonian coexistence networks. 
From that list, we deleted network duplicates and isomorphic network topologies. 
Only networks with up to $9$ nodes were considered because of the limiting computing time needed to identify the vast number of both possible network duplicates (naively there are $S!$ ways to label $S$ nodes) and network isomorphisms.

Note that in our understanding two networks are isomorphic if one network is equal to the other after reversing all of its edges and/or suitably relabeling its nodes. In other words, a graph isomorphism is expressed in terms of multiplication with $-1$ and/or simultaneous reordering of row and column vectors of the antisymmetric adjacency matrix.
This notion of isomorphism of network topologies is justified by the algebraic properties of the antisymmetric adjacency matrix. A relabeling of all nodes is achieved by multiplying the antisymmetric matrix with permutation matrices, whose eigenvalues are $\pm 1$ only; see \cite{Brualdi1991}. Therefore, algebraic characteristics such as positivity of the kernel and spectrum do not change for an antisymmetric matrix. 

\subsection{Method 2: Algebraic determination of coexistence networks via the adjugate vector~(8)}\label{sec:numerical2}

We also determined all coexistence networks for up to networks of 9 nodes through a second, algebraic approach. We examined all possible orientations of all connected, undirected graphs whether a strictly positive kernel of the antisymmetric adjacency matrix is obtained for all choices of weights. 
Starting from databases containing all undirected graphs, we exploited, first, necessary conditions for coexistence networks with an one-dimensional kernel and, second, the notion of the adjugate vector~(8)
to find all orientations of undirected graphs that form coexistence networks.

We started with a list of all connected, undirected graphs with $S\leq 9$ nodes. In reference~\cite{McKay2017}, complete lists are available for up to $S  = 10$ nodes; the number of undirected graphs grows super-exponentially for larger $S$. 
For every undirected graph in that list, we examined whether it admits an orientation yielding a coexistence network through a sieve of necessary conditions that is outlined in the following.
\begin{itemize}
\item[(i)] First, we checked whether the graph can be oriented such that it is strongly connected (in the graph-theoretical literature referred to as \textit{2-edge-connected}~\cite{Lovasz86, Robbins39}). 
Every network that is not strongly connected cannot admit an orientation yielding a coexistence network and, thus, does not have to be considered further; see Section~\ref{sec:strongly_connected}.
\item[(ii)] If the (undirected) graph can be oriented to be strongly connected, we checked whether it admits at least one near-perfect matching. For graphs having at least one near-perfect matching, the kernel of the antisymmetric adjacency matrix is one-dimensional and is given by the adjugate vector (see Section~3).
Thus far, we have not found any coexistence network whose kernel is not one-dimensional; see Section~4.3.
\item[(iii)] Next, we checked whether the (undirected) graph is factor-critical.
An undirected graph having a near-perfect matching, but not being factor-critical has at least one zero-entry in its adjugate vector~(8)
irrespective of the orientation of that graph. Thus, such a graph cannot be oriented to be a coexistence network.
For factor-critical graphs, however, every component of the adjugate vector~(8)
has at least one non-vanishing summand irrespective of the orientation; see Section~3.
\item[(iv)] Only for factor-critical graphs did we search for an orientation such that all summands occurring in the adjugate vector~(8)
have the same sign; see Section~\ref{sec:strongly_connected}.
To this end, we computed all near-perfect matchings from the above-diagonal matrix elements. By choosing $+1,-1$ as matrix entries, the signs of all summands for all orientations were tested, and all coexistence networks were determined.
\end{itemize}



Carrying out the described procedure for all connected, undirected graphs with $S\leq 9$ nodes yields all coexistence networks (up to graph isomorphisms; see above). 
The list obtained through our algebraic approach agrees with the list of coexistence networks that was constructed as described in Section~\ref{sec:numerical1} (see Figure~6
of the main text).


\section{Calculation of the number of near-perfect matchings for selected coexistence networks}
\label{sec:number_matchings}

Here we supplement the derivation of the adjugate vectors for the generating coexistence networks with unit rates given in~(15) and~(18),
respectively, as stated in Section~5
of the main text. 
More explicitly, for both networks we derive the number of near-perfect matchings. Exploiting that both examples suffice the coexistence conditions~(9) and~(10),
the number of near-perfect matchings excluding vertex $i$ equals the $i$th entry of the adjugate vector for unit rates.

\begin{figure*}[th!]
\centering
\includegraphics[width=0.95\textwidth]{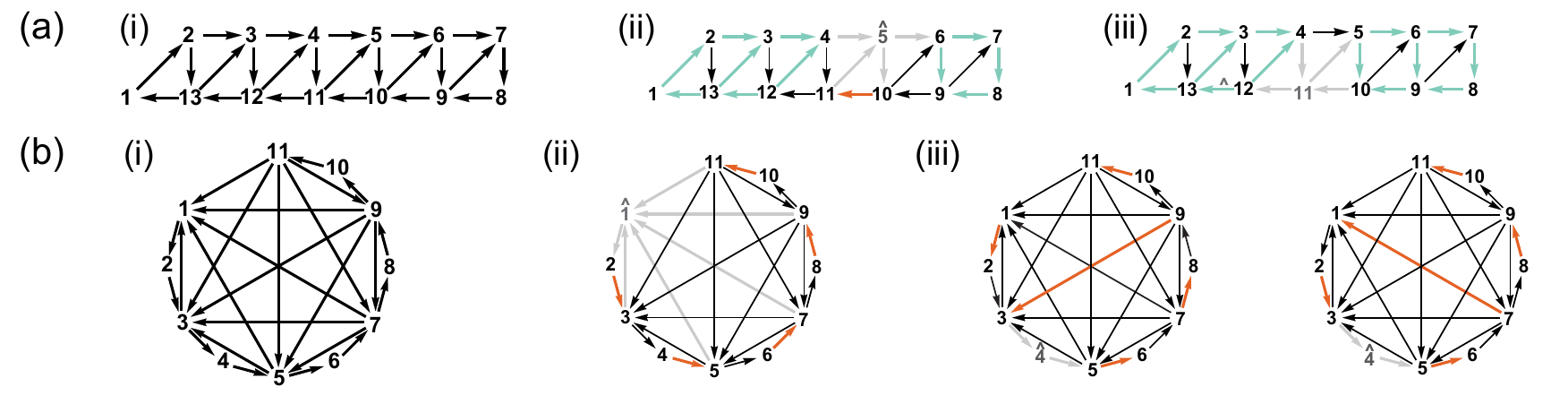}
\caption{(Color online)
\textbf{Exemplary sketches for the calculation of the number of near-perfect matchings for (a) the triangulation of the cycle, and (b) the cycles with complete subnetwork.}
(a) The triangulation of the cycle with $S=11$ nodes is shown in (a)(i). Depending on the position of the deleted node, the whole network is separated either such that two ladder-graphs (highlighted in blue) and one additional edge (orange) contribute to perfect matchings ((a)(ii), for the nodes $\{2,3,4,5,6,7\}$), or such that only two ladder graphs (blue) contribute ((a)(iii), for the nodes $\{8,9,10,11,1\}$).
(b)(i) The cycle of $S=11$ nodes with a complete subnetwork on the odd nodes is shown. When deleting odd nodes, only one near-perfect matching exists consisting only of edges of the cycle (ii). When deleting an even node, every near-perfect matching contains one edge between odd nodes (iii). The deleted node effectively divides the subnetwork into two partitions; in the depicted example $\{1,2,3\}$ and $\{5,6,7,8,9,10,11\}$. Every pair of one odd node from each partition gives rise to one near-perfect matching. Thus, in the depicted example, there are $2 \cdot 4 = 8$ near-perfect matchings.
}
\label{fig:AdjugateVectorofGeneratingNetworks}
\end{figure*}

\subsection{Near-perfect matchings for the triangulation of the cycle}
\label{sec:triangulation}

In the following we derive the number of near-perfect matchings of the triangulation of an odd cycle as specified in Equation~(15)
in Section~5.1.

For this calculation, it is helpful to introduce so-called ladder graphs and their number of perfect matchings. The ladder graphs with $2,4$, and $6$ vertices are a single edge, a rectangle, and a domino tile, respectively. In general, a ladder graph with $2T$ vertices consists of $T$ rungs and $2(T-1)$ rails. The number of perfect matchings of a
A ladder graph with $2T$ vertices has $F(T+1)$ perfect matchings, where $F(n)$ is the $n$th Fibonacchi number. This can be shown inductively by considering the additional perfect matchings arising when augmenting the graph by one rung and two rails~\cite{Grimaldi12}.

Note that the triangulation of an odd cycle presented in Section~5.1
can be thought of as a superposition of two ladder graphs, see Figure~\ref{fig:AdjugateVectorofGeneratingNetworks}(a)(i).
One of the two ladder graphs consists of the vertices $2, \dots, S$ and the ascending edges on these vertices (that is, $2 \to S, 3 \to S-1, \dots$), the other ladder graph is diagonally placed over the first one and contains the vertices $1, \dots , \frac{(S+1)}{2},  \frac{(S+1)}{2} + 2, \dots, S$ and all edges of the cycle together with all descending edges on these vertices.
Additionally, note that for a ladder graph with diagonal edges (for example, the subnetwork formed by the vertices $2,\dots, 13$ in Figure~\ref{fig:AdjugateVectorofGeneratingNetworks}(a)(i)), the diagonal edges do not contribute to any perfect matchings. In other words, the number of perfect matchings of a ladder graph with diagonal edges with $2T$ vertices is $F(T+1)$ as well.

Using these observations, the components of the adjugate vector for the triangulation of the cycle with unit rates are calculated as the number of perfect matchings of the subnetworks created by deleting each node.
Upon deleting node $1$ from the network, a network with $S-1 = 2n-2$ vertices remains. This graph is a ladder graph with $2(n-1)$ vertices and additional diagonal edges. Thus, it has $F(n)$ perfect matchings. Because the network is a coexistence network, the number of perfect matchings equals the entry of the first component of the adjugate vector with unit rates, that is, $r_1 = F(n)$.\\
In case one of the vertices $i = 2, \dots, n$ is deleted, the network is divided into two parts that can be viewed as two ladder graphs with $2(i-2)$ and $2(n-i)$ vertices (ignoring diagonal edges that cannot contribute to perfect matchings). 
The two ladder graphs are connected by three edges, see Figure~\ref{fig:AdjugateVectorofGeneratingNetworks}(a)(ii). Every combination of perfect matchings of the two separated ladder graphs is a perfect matching of the full graph, such that $r_i = F(i-1)F(n-i+1)$. \\
The remaining matrix entries $i=n+1, \dots, 2n-1$ are calculated in a similar way. Deleting one of these vertices divides the remaining network into two ladder graphs with $2(2n-i)$ and $2(i-n-1)$ vertices (see Figure~\ref{fig:AdjugateVectorofGeneratingNetworks}(a)(iii)), such that the corresponding components of the adjugate vector are $r_i = F(2n-i+1)F(i-n)$.
Taken together, the adjugate vector for the triangulation of a cycle with unit rates is given by Equation~(15).

Using the convolution formula for Fibonacci numbers~\cite{WeissteinFibonacci},
\begin{align*}
\sum_{k = 0}^{n} F(k)F(n-k) = \frac{1}{5} (n(F(n-1)+ F(n+1))- F(n)),
\end{align*}
the total number of near-perfect matchings is
\begin{align*}
 \sum_{i=2}^{n} F(i-1)&F(n-(i-1))+\left(\sum_{i = n+1}^{2n-1}  F(n-(i-n)+1)F(i-n)+F(n)\right) = \\
 &= \sum_{i=0}^{n} F(i)F(n-i) +  \sum_{i = 1}^{n+1}  F(n+1-i)F(i)\ , \\
 &= \frac{1}{5} (n (F(n-1)+ F(n+1))- F(n)) + \frac{1}{5} ((n+1) (F(n)+ F(n+2)) - F(n+1))\ ,\\
 &=\frac{1}{5} \left( n (3 F(n+1)+F(n)) + F(n)\right) = \frac{1}{5} n F(n) \left(3  \frac{F(n+1)}{F(n)} + 1 + \frac{1}{n} \right),
\end{align*}

as stated in Equation~(16)
the main text.

\subsection{Number of near-perfect matchings in a cycle with complete subnetwork on odd nodes}
\label{sec:SI_complete_subnetworks}

Here we supplement the calculation of the form of the adjugate vector given in Equation~(18)
in Section~5.2.
The form of the adjugate vector of an odd cycle with a complete subnetwork on the odd nodes can be understood as follows; see Figure~\ref{fig:AdjugateVectorofGeneratingNetworks}(b)(i) for illustration. 

Upon deleting an odd node $2k-1$ ($k=1, \dots, n$) from the network, the only perfect matching that covers all remaining nodes consists of edges from the Hamiltonian cycle. In detail, the perfect matching
is $\mu_{\widehat{2k-1}} = \big((2k\to 2k+1), (2k+2\to 2k+3), \dots, (S-1\to S), (1\to 2), (3\to 4), \dots (2k-3\to 2k-2)\big)$, see Figure~\ref{fig:AdjugateVectorofGeneratingNetworks}(b)(ii). Thus, with unit rates, for odd $i$ the components of the adjugate vector are $r_i = 1$.\\

When deleting an even node $i = 2k$ ($k=1, \dots, n-1$) from the network topology, the remaining network contains the edges of the cycle $1 \to \dots \to i-1$, $i+1 \to \dots \to S$, and edges connecting all pairs of odd vertices. In total, the remaining subnetwork has $n-1$ odd nodes and $n-3$ even nodes. Each perfect matching thus contains one edge connecting two odd vertices. There are $\frac{i-1+1}{2} \cdot \frac{S-i+1}{2} = \frac{i(n-i/2)}{2}$ possibilities for choosing odd edges such that the remaining paths consisting of edges from the cycle have a perfect matching; two such possibilities are shown in Figure~\ref{fig:AdjugateVectorofGeneratingNetworks}(b)(iii). Thus, for unit weights, for even $i$ the component of the adjugate vector are $r_i = \frac{i(n-i/2)}{2}$. \\
Taken together, the adjugate vector for the cylce with complete subnetwork on the odd nodes with unit rates is given by Equation~(18).

\end{document}